\documentclass[reprint,amsfonts, amssymb, amsmath,  showkeys,prx, superscriptaddress, twocolumn,longbibliography,nofootinbib,dvipsnames]{revtex4-2}

\usepackage{enumerate}
\usepackage{amsmath}
\usepackage{graphicx}
\usepackage{comment}

\usepackage{tikz} 
\usepackage{etoolbox} 

\def\HashTableNumCells{5}
\def\HashTableCellSize{.6}
\def\HashTableFilledCellColour{white}
\def\HashTableEmptyCellColour{lightgray}
\def\HashTableInactiveOpacity{0.2}
\def\HashTableArrowVerticalSpace{0.15 *.6}
\def\HashTableVerticalSeparation{2}

\pgfmathsetmacro\SecondHashTableX{(1 + \HashTableNumCells)*\HashTableCellSize}

\newcommand{\DrawEmptyHashTableCell}[4]{%
    \pgfmathsetmacro\cellx{(#1 - 1) * \HashTableCellSize};
    \draw[fill=\HashTableEmptyCellColour, opacity=#4] 
        (#2+\cellx,#3) rectangle ++(\HashTableCellSize, \HashTableCellSize);
}

\newcommand{\DrawEmptyHashTable}[2]{%
    \foreach \i in {1, ..., \HashTableNumCells } {
        \DrawEmptyHashTableCell{\i}{#1}{#2}{1};
    }
}

\newcommand{\DrawUnusedHashTable}[2]{%
    \foreach \i in {1, ..., \HashTableNumCells } {
        \DrawEmptyHashTableCell{\i}{#1}{#2}{\HashTableInactiveOpacity};
    }
}

\newcommand{\DrawFilledHashTableCell}[8]{%
    \pgfmathsetmacro\cellx{#2 + (#1 - 1) * \HashTableCellSize};
    \pgfmathsetmacro\celly{#3}; 
    \draw[fill=\HashTableFilledCellColour]
        (\cellx,\celly) rectangle ++(\HashTableCellSize, \HashTableCellSize);
    \pgfmathsetmacro\labelx{\cellx+0.5*\HashTableCellSize};
    \pgfmathsetmacro\labely{\celly+0.5*\HashTableCellSize};
    \node at (\labelx,\labely) {$#4$};
    \node at (\labelx,\labely+\HashTableCellSize) {\small $#5$};
    \node at (\labelx,\labely-\HashTableCellSize-0.2) {\small $#6$};
    
    \node[coordinate] (#7) at (\labelx, \celly+\HashTableArrowVerticalSpace+\HashTableCellSize) {};
    \node[coordinate] (#8) at (\labelx, \celly-\HashTableArrowVerticalSpace) {};
}

\newcommand{\LabelHashTables}{ {
        \color{gray}
	\node at (1.5,1.25) {\textit{enumerated}};
	\node at (5,1.235) {\textit{auxiliary}};
} }

\def\HashTableDiagramSingle{

    \def\HashTableNumCells{11}

      \node[text=gray,align=left] at (-1.5+.6,1.6*\HashTableCellSize) {address:};
      \node[text=gray,align=left] at (-1.34+.6,0.4*\HashTableCellSize) {array:};

    \def\y{0};
    \def\x{0};
    
        \DrawEmptyHashTable{\x}{\y};
        
        \DrawFilledHashTableCell
            {2}{\x}{\y}
            {c_{P_1}}
            {h(P_1)}{}
            {}{nodeTop1}
        \DrawFilledHashTableCell
            {5}{\x}{\y}
            {c_{P_2}}
            {h(P_2)}{}
            {}{nodeTop2}
        \DrawFilledHashTableCell
            {9}{\x}{\y}
            {c_{P_3}}
            {h(P_3)}{}
            {}{nodeTop3}

}

\def\HashTableDiagramOverviewGate{

    \def\HashTableNumCells{11}

    \node[text=gray,align=left] at (-1.2+.7,.5*\HashTableCellSize) {$S$:};
    \node[text=gray,align=left] at (-1.2+.7,-2*\HashTableCellSize) {$S'$:};
    
    \def\y{0};
    \def\x{0};
    
        \DrawEmptyHashTable{\x}{\y};
        
        \DrawFilledHashTableCell
            {2}{\x}{\y}
            {c_{P_1}}
            {h(P_1)}{}
            {}{nodeTop1}
        \DrawFilledHashTableCell
            {5}{\x}{\y}
            {c_{P_2}}
            {h(P_2)}{}
            {}{nodeTop2}
        \DrawFilledHashTableCell
            {9}{\x}{\y}
            {c_{P_3}}
            {h(P_3)}{}
            {}{nodeTop3}
        
    \def\y{-1.5};
    \def\x{0};
    
        \DrawEmptyHashTable{\x}{\y};
        
        \DrawFilledHashTableCell
            {2}{\x}{\y}
            {c_{P_1'}}
            {}{h(P_1)}
            {nodeBot1}{}
        \DrawFilledHashTableCell
            {4}{\x}{\y}
            {c_{P_4}}
            {}{h(P_4)}
            {nodeBot2}{}
        \DrawFilledHashTableCell
            {7}{\x}{\y}
            {c_{P_5}}
            {}{h(P_5)}
            {nodeBot3}{}
        \DrawFilledHashTableCell
            {9}{\x}{\y}
            {c_{P_6}}
            {}{h(P_6)}
            {nodeBot4}{}

    \draw[->] (nodeTop1) -- (nodeBot1);
    \draw[->] (nodeTop1) -- (nodeBot2);
    \draw[->] (nodeTop2) -- (nodeBot3);
    \draw[->] (nodeTop3) -- (nodeBot4);

}

\def\HashTableDiagramPauliNoise{

    \def\HashTableNumCells{5}

    \LabelHashTables{}
			
    \def\y{0};
    
        \def\x{0};
        \DrawEmptyHashTable{\x}{\y};
        \DrawFilledHashTableCell
            {1}{\x}{\y}
            {c_{P_1}}{}{}
            {}{top1}
        \DrawFilledHashTableCell
            {4}{\x}{\y}
            {c_{P_2}}{}{}
            {}{top2}
        
        \def\x{\SecondHashTableX};
        \DrawUnusedHashTable{\x}{\y};

    \def\y{-\HashTableVerticalSeparation};
    
        \def\x{0};
        \DrawEmptyHashTable{\x}{\y};
        \DrawFilledHashTableCell
            {1}{\x}{\y}
            {c_{P_1}}{}{}
            {bot1}{}
            \DrawFilledHashTableCell
            {4}{\x}{\y}
            {c_{P_2'}}{}{}
            {bot2}{}

        \def\x{\SecondHashTableX};
        \DrawUnusedHashTable{\x}{\y};
        
    \draw[->] (top1) -- (bot1) node[midway,right] {retain};
    \draw[->] (top2) -- (bot2) node[midway,right] {mutate};

}

\def\HashTableDiagramClifford{

    \def\HashTableNumCells{5}

    \LabelHashTables{}

    \def\y{0};
    
        \def\x{0};
        \DrawEmptyHashTable{\x}{\y};
        \DrawFilledHashTableCell
            {3}{\x}{\y}
            {c_{P_1}}{}{}
            {}{top1}
        \DrawFilledHashTableCell
            {5}{\x}{\y}
            {c_{P_2}}{}{}
            {}{top2}
        
        \def\x{\SecondHashTableX};
        \DrawEmptyHashTable{\x}{\y};
    
    \def\y{-\HashTableVerticalSeparation};
    
        \def\x{0};
        \DrawEmptyHashTable{\x}{\y};

        \node at (\x+\HashTableCellSize,\y+1.6*\HashTableCellSize) {clear};
        
        \def\x{\SecondHashTableX};
        \DrawEmptyHashTable{\x}{\y};
        \DrawFilledHashTableCell
            {1}{\x}{\y}
            {c_{P_3}}{}{}
            {mid1}{}
        \DrawFilledHashTableCell
            {5}{\x}{\y}
            {c_{P_2}}{}{}
            {mid2}{}

    \draw[->] (top1) -- (mid1); 
    \draw[->] (top2) -- (mid2) node[midway,right] {\hphantom{ins} insert};

    \def\swaplabelxoffset{.6};
    \pgfmathsetmacro\leftarrx{\HashTableCellSize*\HashTableNumCells/2+\swaplabelxoffset};
    \pgfmathsetmacro\rightarrx{\leftarrx + (\HashTableNumCells+1)*\HashTableCellSize-2*\swaplabelxoffset};
    \pgfmathsetmacro\midarrx{(\leftarrx + \rightarrx)/2)};

    \def\yarrgap{.3};

    \draw[<->, bend right=45, color=gray, dashed] (\leftarrx,\y-\yarrgap) to (\rightarrx,\y-\yarrgap);

    \node[color=gray] at (\midarrx,\y-1.2) {relabel};
    
}

\def\HashTableDiagramPauliRotation{

    \def\HashTableNumCells{5}

    \LabelHashTables{}

    \def\y{0};
    
        \def\x{0};
        \DrawEmptyHashTable{\x}{\y};
        \DrawFilledHashTableCell
            {1}{\x}{\y}
            {c_{P_1}}{}{}
            {}{top1}
        \DrawFilledHashTableCell
            {5}{\x}{\y}
            {c_{P_3}}{}{}
            {}{top2}

            \DrawFilledHashTableCell
                {4}{\x}{\y}
                {c_{P_2}}{}{}
                {}{newabove}
        
        \def\x{\SecondHashTableX};
        \DrawEmptyHashTable{\x}{\y};
    
    \def\y{-\HashTableVerticalSeparation};
    
        \def\x{0};
        \DrawEmptyHashTable{\x}{\y};
        \DrawFilledHashTableCell
            {1}{\x}{\y}
            {c_{P_1}}{}{}
            {mid1}{}
        \DrawFilledHashTableCell
            {5}{\x}{\y}
            {c_{P_3'}}{}{}
            {mid2}{}

            \DrawFilledHashTableCell
                {4}{\x}{\y}
                {c_{P_2'}}{}{}
                {newmidlefttop}{}
        
        \def\x{\SecondHashTableX};
        \DrawEmptyHashTable{\x}{\y};
        \DrawFilledHashTableCell
            {3}{\x}{\y}
            {c_{P_4}}{}{}
            {mid3}{undermid3}

            \DrawFilledHashTableCell
                {1}{\x}{\y}
                {c_{P_1'}}{}{}
                {newmidrighttop}{newmidrightbot}

    \def\y{-2*\HashTableVerticalSeparation};
    
        \def\x{0};
        \DrawEmptyHashTable{\x}{\y};
        \DrawFilledHashTableCell
            {3}{\x}{\y}
            {c_{P_4}}{}{}
            {bot1}{}
        \DrawFilledHashTableCell
            {1}{\x}{\y}
            {c_{P_1''}}{}{}
            {newmergedbot}{}
        \DrawFilledHashTableCell
            {5}{\x}{\y}
            {c_{P_3'}}{}{}
            {}{}

            \DrawFilledHashTableCell
                {4}{\x}{\y}
                {c_{P_2'}}{}{}
                {}{}
        
        \def\x{\SecondHashTableX};
        \DrawEmptyHashTable{\x}{\y};

        \node at (\x+2.5*\HashTableCellSize,\y+1.6*\HashTableCellSize) {clear};

    \draw[->] (top1) -- (mid1) node[midway,left] {retain};
    
    \draw[->] (newabove) -- (newmidlefttop) node[midway,left] {mutate};
    \draw[->] (top2) -- (mid2) node[midway,left] {}; 

    \draw[->] (newabove) -- (newmidrighttop) node[midway,right] {}; 
    \draw[->] (top2) -- (mid3) node[midway,right] {\hphantom{in}insert};
    
    \draw[->] (undermid3) -- (bot1) node[midway,right] {}; 

    \draw[->] (newmidrightbot) -- (newmergedbot) node[midway,left] {merge\hphantom{xx}};

}

\usepackage{dcolumn}
\usepackage{bm,bbm}
\usepackage{physics}
\usepackage{amsmath}
\usepackage{amssymb}
\usepackage{amsthm}
\usepackage{amsfonts}
\usepackage{MnSymbol}
\usepackage[colorlinks=true,citecolor=blue,linkcolor=magenta]{hyperref}
\usepackage{tikz}
\graphicspath{{Images/}}

\usepackage[export]{adjustbox}

\usepackage{listings}
\newcommand{\code}[1]{\texttt{#1}}

\usepackage{algorithm}
\usepackage{algorithmicx}
\usepackage{algpseudocode} 

\definecolor{keywordblue}{RGB}{0,0,180}
\lstdefinelanguage{Pseudocode}{
  keywords={for, remove, insert, apply, truncate, overlap
  },
  sensitive=false,
  comment=[l]{\#},
  keywordstyle=\color{keywordblue}\bfseries,
  commentstyle=\color{gray}\ttfamily\itshape
}

\newcommand{\EC}{\mathcal{E}}
\newcommand{\ECdag}{\mathcal{E}^\dagger}

\newcommand{\OC}{\mathcal{O}}
\newcommand{\PC}{\mathcal{P}}

\renewcommand{\geq}{\geqslant}
\renewcommand{\leq}{\leqslant}

\renewcommand{\vec}[1]{\boldsymbol{#1}}  

\newcommand{\cP}{\ensuremath{\code{c}_\code{P}}}
\newcommand{\cQ}{\ensuremath{\code{c}_\code{Q}}}

\newcommand{\thv}{\vec{\theta}}

\graphicspath{{Images/}}

\newcommand{\be}{\begin{equation}}
\newcommand{\ee}{\end{equation}}

\renewcommand{\vec}[1]{\boldsymbol{#1}}  

\usepackage{mathtools}

\usepackage{amssymb}

\newcommand{\ppjl}{\code{PauliPropagation.jl}}
\newcommand{\ppjllink}{\href{https://github.com/MSRudolph/PauliPropagation.jl}{\code{\mbox{PauliPropagation.jl}}}}

\usepackage[normalem]{ulem}

\usepackage{cancel}

\renewcommand{\bold}[1]{\mathbf{#1}}

\usepackage[framemethod=TikZ]{mdframed}
\usepackage{tcolorbox}
\usepackage{xcolor} 

\usepackage{color, colortbl}

\definecolor{lightgray}{gray}{0.9}

\definecolor{theorycolour}{RGB}{148, 156, 168} 
\definecolor{softwarecolour}{RGB}{50, 136, 140}
\definecolor{datacolour}{RGB}{101, 143, 226} 
\definecolor{conventioncolour}{RGB}{194, 97, 186}
\definecolor{glossarycolour}{RGB}{255, 205, 105}

\newtcolorbox[auto counter]{pbbox}[2][]{fonttitle=\bfseries,
title=Theory Box~\thetcbcounter: #2,#1,colframe=gray}

\newtcolorbox[use counter from=pbbox]{theorybox}[2][]{
floatplacement=t,float,
colback=theorycolour!5!white,colframe=theorycolour!75!black,title= Theory Box~\thetcbcounter: #2,#1}

\newtcolorbox[auto counter]{pabox}[2][]{fonttitle=\bfseries,
title=Software Box~\thetcbcounter: #2,#1,colframe=gray}

\newtcolorbox[use counter from=pabox]{softwarebox}[2][]{
floatplacement=t,float,
colback=softwarecolour!5!white,colframe=softwarecolour!75!black,title= Software Box~\thetcbcounter: #2,#1}

\newtcolorbox[auto counter]{glossarybox}[2][]{
floatplacement=t,float,
colback=glossarycolour!5!white,colframe=glossarycolour!90!black,title= Glossary Box~\thetcbcounter: #2,#1}

\newtcolorbox[use counter from=pbbox]{theoryboxlarge}[2][]{
  float*=t, 
  colback=theorycolour!5!white,
  colframe=theorycolour!75!black,
  title= Theory Box~\thetcbcounter: #2,#1,
  width=\textwidth, 
}

\newtcolorbox[auto counter]{pcbox}[2][]{fonttitle=\bfseries,
title=Software Box~\thetcbcounter: #2,#1,colframe=gray}

\newtcolorbox[use counter from=pcbox]{datastructbox}[2][]{
floatplacement=t,
float, 
colback=datacolour!5!white,colframe=datacolour!75!black,title= Under the Hood Box~\thetcbcounter: #2,#1}

\newtcolorbox[use counter from=pcbox]{datastructboxfull}[2][]{
width=\textwidth, 
colback=datacolour!5!white,colframe=datacolour!75!black,title= Under the Hood Box~\thetcbcounter: #2,#1}

\newtcolorbox[auto counter]{conventionbox}[2][]{
floatplacement=t,float,
colback=conventioncolour!5!white,colframe=conventioncolour!85!black,title= Convention Box~\thetcbcounter: #2,#1}

\usepackage[normalem]{ulem}

\renewcommand{\arraystretch}{1.5}

\makeatletter
\renewcommand*\l@section{\@dottedtocline{1}{0em}{2.5em}}
\renewcommand*\l@subsection{\@dottedtocline{2}{2.5em}{2.5em}}
\renewcommand*\l@subsubsection{\@dottedtocline{3}{5.0em}{2.5em}}
\makeatother

\begin{document}

\title{Pauli Propagation: A Computational Framework for Simulating Quantum Systems}

\date{\today}

\author{Manuel S. Rudolph}
\email{manuel.rudolph@epfl.ch}
\affiliation{Institute of Physics, Ecole Polytechnique F\'{e}d\'{e}rale de Lausanne (EPFL),   Lausanne, Switzerland}
\affiliation{Centre for Quantum Science and Engineering, Ecole Polytechnique F\'{e}d\'{e}rale de Lausanne (EPFL),   Lausanne, Switzerland}

\author{Tyson Jones}
\affiliation{Institute of Physics, Ecole Polytechnique F\'{e}d\'{e}rale de Lausanne (EPFL),   Lausanne, Switzerland}
\affiliation{Centre for Quantum Science and Engineering, Ecole Polytechnique F\'{e}d\'{e}rale de Lausanne (EPFL), Lausanne, Switzerland}

\author{Yanting Teng}
\affiliation{Institute of Physics, Ecole Polytechnique F\'{e}d\'{e}rale de Lausanne (EPFL),   Lausanne, Switzerland}
\affiliation{Centre for Quantum Science and Engineering, Ecole Polytechnique F\'{e}d\'{e}rale de Lausanne (EPFL), Lausanne, Switzerland}

\author{Armando Angrisani}
\affiliation{Institute of Physics, Ecole Polytechnique F\'{e}d\'{e}rale de Lausanne (EPFL),   Lausanne, Switzerland}
\affiliation{Centre for Quantum Science and Engineering, Ecole Polytechnique F\'{e}d\'{e}rale de Lausanne (EPFL),   Lausanne, Switzerland}

\author{Zo\"{e} Holmes}
\affiliation{Institute of Physics, Ecole Polytechnique F\'{e}d\'{e}rale de Lausanne (EPFL),   Lausanne, Switzerland}
\affiliation{Centre for Quantum Science and Engineering, Ecole Polytechnique F\'{e}d\'{e}rale de Lausanne (EPFL),   Lausanne, Switzerland}
\affiliation{Algorithmiq Ltd, Kanavakatu 3 C, FI-00160 Helsinki, Finland}

\begin{abstract}
Classical methods to simulate quantum systems are not only a key element of the physicist's toolkit for studying many-body models but are also increasingly important for verifying and challenging upcoming quantum computers. Pauli propagation has recently emerged as a promising new family of classical algorithms for simulating digital quantum systems. Here we provide a comprehensive account of Pauli propagation, tracing its algorithmic structure from its bit-level implementation and formulation as a tree-search problem, all the way to its high-level user applications for simulating quantum circuits and dynamics. Utilising these observations, we present \ppjllink, a \texttt{Julia} software package that can perform rapid Pauli propagation simulation straight out-of-the-box and can be used more generally as a building block for novel simulation algorithms.
\end{abstract}

\maketitle

{
\hypersetup{linkcolor=black} 
\tableofcontents
}

\section{Introduction}

What makes a quantum system hard to simulate classically? This is a question that we cannot answer in this work. However, it is clear that not all quantum systems are intractable to simulate, and not all are simulable by the same computational frameworks. Tasks that are out of range of one approach are not precluded from study by another, perhaps bespoke, method. 
A ubiquitous and widely applicable framework for simulating quantum systems is the use of \textit{tensor network methods}~\cite{orus2014practical,verstraete2008matrix,shi2006classical,verstraete2004renormalization,biamonte2017tensor}. Such methods have been proven over again to be powerful and adaptable, and are still evolving, for example, for continually simulating quantum computations~\cite{tindall2023efficient,tindall2025dynamics,patra2023efficient}. Yet despite their broad domains and popularity, other frameworks such as \textit{neural quantum states}~\cite{carleo2017solving} can more effectively tackle some specific tasks of interest~\cite{wu2023variational}. 
Indeed no single method is optimal in all regimes, and many challenging problems require drastically different approaches.

Pauli propagation is a relative newcomer to the corpus of classical simulators and yet is already competitive with other state-of-the-art methods for certain tasks~\cite{beguvsic2023fast,rudolph2023classical,shao2023simulating,angrisani2024classically}. At their core, Pauli propagation methods approximate the evolution of a quantum operator (typically, an observable in the Heisenberg picture) via a truncated Pauli path integral. 
While first shown to be powerful in noisy~\cite{fontana2023classical,shao2023simulating} and random~\cite{angrisani2024classically} settings, the method has also been applied more broadly~\cite{beguvsic2024real,lerch2024efficient}. 
The approach looks particularly promising for quickly obtaining rough approximations of expectation values for quantum circuits on arbitrary topologies, and may find use in variational scenarios~\cite{bermejo2024improving,angrisani2024classically}. It is also well-suited for interfacing with quantum hardware, enabling the hybridization of classical and quantum compute power~\cite{cerezo2023does,lerch2024efficient,fuller2025improved}.

Readers may be familiar with alternative names for Pauli propagation algorithms that reflect particular implementations, approximations, or goals. 
To the best of our knowledge, Ref.~\cite{rall2019simulation} first introduced the term `Pauli propagation', contrasting it to stabilizer simulation, and highlighting both Schrödinger and Heisenberg picture approaches.
More recent implementations include \textit{observable back-propagation on Pauli paths} (OBPPP)~\cite{shao2023simulating}, the \textit{low-weight efficient simulation algorithm} (LOWESA)~\cite{fontana2023classical,rudolph2023classical}, the \textit{exclusive-or represented quantum algebra} (ORQA)~\cite{broers2024exclusive,broers2025scalable}, and \textit{Clifford pertubation theory} (CPT)~\cite{beguvsic2023simulating}, which owe their names to specific implementation details and truncation strategies. One more general name is \textit{sparse Pauli dynamics} (SPD)~\cite{beguvsic2023fast,beguvsic2024real}, which also falls under the framework of Pauli propagation with the specific goal of simulating real-time quantum dynamics. Very recently, a breakaway from the Pauli basis was proposed that works in the Majorana basis~\cite{miller2025simulation,d2025majorana,alam2025fermionic,alam2025programmable}, but which shares most algorithmic features with Pauli propagation. 

Most work so far on Pauli propagation has focused either on providing theoretical guarantees~\cite{aharonov2022polynomial,fontana2023classical,schuster2024polynomial,angrisani2024classically,angrisani2025simulating} or specific case studies~\cite{beguvsic2023fast,beguvsic2024real,rudolph2023classical,bermejo2024quantum}. Here we will take a step back and provide an end-to-end account of the algorithmic considerations underlying Pauli propagation methods. We start on the high-level with a discussion of the \textit{mathematical foundations} of the approach and its applications. Next we proceed to the mid-level, where we discuss \textit{algorithmic nuances} such as search and truncation strategies as well as methods to estimate simulation errors. Then we dive under the hood and discuss the \textit{implementation details}; the nitty gritty of how to get your money's worth from your CPU (or GPU). Finally, we swoop back to the bigger picture, looking out to where Pauli propagation might find its niche and avenues for future developments. 

To facilitate research \textit{on} and \textit{with} this framework, we provide an open-source library \ppjllink{}. This high-level \texttt{Julia} package can be used straight out of the box to simulate large-scale quantum circuits. Through careful consideration of \texttt{Julia}'s multiple dispatching facilities, the package remains adaptable and highly extensible. We hope that the package will prove useful to users simulating quantum systems, developing quantum algorithms, finding novel applications of Pauli propagation, and devising new propagation methods.

\section{A bird's-eye view}

\subsection{Defining Pauli propagation}\label{sec:framework}

Propagation methods in their essential form compute the evolution of an object $A$ under an operation $\EC$
expressed as a sequence of smaller, \textit{individually tractable} operations $\EC_l, \; l=1\dots m$,
\begin{equation}\label{eq:propagationbasic}
    \mathcal{E}[A] = \EC_m\left[\dots \left[\EC_1\left[A \right]\right]  \right]\,.
\end{equation}
Often, but not always, 
we then compute the overlap between $\mathcal{E}[A]$ and some other object $B$
to produce an output scalar,
\begin{equation}
    \langle B, \, \mathcal{E}[A] \rangle \; \in \mathbb{C}.
\end{equation}
In a typical statevector simulation, the object $A$ would usually be an initial state, $\mathcal{E}$ a series of gates and channels, and $B$ an observable operator of which the output expectation value is evaluated via a trace.
By contrast, in ``\textit{Pauli}'' propagation, this is generally reversed and $A$ is an observable expressed in the Pauli basis and evolved by the adjoint circuit $\mathcal{E}$, where the $\mathcal{E}_l$ are maps between Pauli strings.
The object $B$ is typically the initial quantum state, and the Hilbert-Schmidt inner product is the trace $\Tr[B\cdot\mathcal{E}[A]]$. More concretely, the following (non-exhaustive) list of properties can in principle be computed using the primitive of Pauli propagation:

\begin{figure*}
    \centering
    \includegraphics[width=0.98\linewidth]{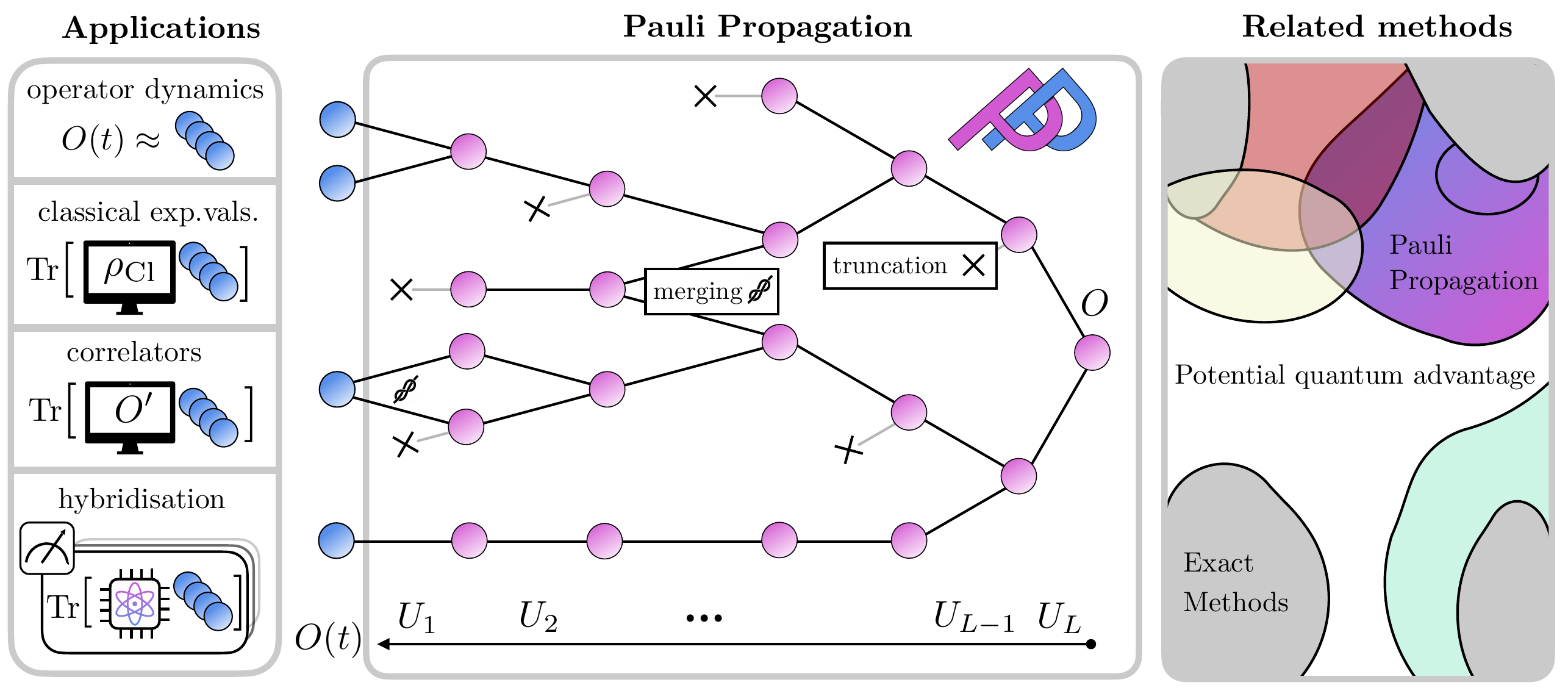}
    \caption{\textbf{Pauli propagation: Applications and Related methods}. Quantum systems can be tackled via various classical simulations methods, each for different applications and systems of interest. \textit{Pauli propagation} is a recent computational framework computing the evolution of Pauli strings under quantum dynamics or arbitrary quantum circuits. Often but not necessarily, Pauli propagation involves Heisenberg-picture backpropagation of an observable. This evolved observable can be utilized for applications ranging from transparently studying operator dynamics, expectation values with states encoded on classical or quantum hardware, or operator correlation functions. Its placement in the classical simulation toolbox with other, more established methods is not fully known, but within few years, this approach has already shown promise. Pauli propagation is most closely related to stabilizer simulation and its extensions for general gates, but it is fundamentally different to, e.g., tensor network approaches. Our library \ppjl\ provides state-of-the-art Pauli propagation techniques, and can be used for large-scale quantum simulation or as a building block in larger algorithmic frameworks.}
    \label{fig:pp_main}
\end{figure*}   

\begin{description}

    \item[The evolution of observables] We can compute the Pauli basis decomposition of an observable $O$ evolved under some channel $\mathcal{E}$, i.e., $  \mathcal{E}[A]  \equiv \mathcal{E}(O) = \sum_P c_P P$. Typically, we will work in the Heisenberg picture and so for unitary evolutions we compute $\mathcal{E}[O]  \equiv U^\dagger O U $, or for a noise channel $\mathcal{N}$  we would compute the action of its adjoint $\mathcal{N}^\dagger[O]$. Evolution of \textit{states} is also possible, i.e., we can also take $A = \rho$ and compute $U \rho U^\dagger$, but this is only efficient when the initial state is sparse in the Pauli basis, e.g., in the case of high-temperature states~\cite{rudolph2026thermal}. An extreme case is the maximally mixed state, which is representable by a single Pauli string, the all-identity $I/2^n$.
    
    \item[State expectation values] It is possible to compute $\Tr[ \EC[\rho] O ]$ where $\rho$ is a quantum state, $O$ is an observable and $\EC$ is a quantum channel as this is equivalent to applying the inverse of the channel on the observable. For example, in the case of unitary evolution, the evolved object will typically be the observable $O$ and the operation $U^\dagger  O U$ describes the unitary Heisenberg evolution of the system. We then compute the overlap of the propagated observable $U^\dagger  O U$ with the state $\rho$, which can be very hard in general, but is trivial for many common states like the all-zero state or other stabilizer states.
    
    \item[Correlation functions in time] We can compute correlation functions like $\Tr[O_1 U O_2 U^\dagger]$ using the same approach as above, except the final overlap is taken with another (or the same) observable rather than a state $\rho$. 

    \item[Variational optimization and machine learning] The code in PauliPropagation.jl is designed to be automatically differentiable by use of common \code{Julia} libraries. This unlocks variational optimization, for example, for training a quantum machine learning models, which has been used to simulate quantum convolutional neural networks at large scale~\cite{bermejo2024quantum}.
    \item[ Hybridization with other classical/quantum tools] Simulations of a single quantum circuit can be hybridized between different types of hardware (quantum and classical) or simulation methods (classical and classical). For example, $\Tr[\rho U_2^\dagger U_1^\dagger O U_1  U_2] = \Tr[\mathcal{E}_2(\rho) \mathcal{E}_1(O)]$, where $\mathcal{E}_1(\cdots) = U_1^\dagger (\cdots) U_1$ and $\mathcal{E}_2(\cdots) = U_2 (\cdots) U_2^\dagger$ denote the unitary evolution in the Schr\"{o}dinger and Heisenberg picture respectively. If $\mathcal{E}_2(O) = \sum_P c_P P$ is the half-backpropagated observable, the full expectation can be evaluated as $\sum_P c_P \Tr[\mathcal{E}_1(\rho) P]$. In this manner, one can trade shorter depth circuits for either simulation with the additional demands of computing the overlap of the two \textit{half-evolved} objects $\mathcal{E}_2(\rho)$ and $\mathcal{E}_1(O)$.
\end{description}

For concreteness, we will focus on detailing the Pauli propagation framework for computing the evolution of observables in the Heisenberg picture. However, the reader should bare in mind that the framework can be applied more generally. 

The ``Pauli'' propagation propagates \textit{Pauli strings}. Namely, tensor products of $n$ single qubit Pauli matrices $I, X, Y, Z$. These form a basis for all $2^n \times 2^n$ matrices.
We write an object $O$ in this Pauli basis as 
\begin{equation}\label{eq:initialopsum}
    O = \sum_{i=1}^{N_O} c_i P_i\, ,
\end{equation}
where $c_i$ are coefficients for the Pauli strings $P_i$, and $N_O$ is the number of non-zero-coefficient Pauli strings in the Pauli decomposition of $O$.

The action of an operation $\EC_l$ onto a Pauli string $P$ can generally be written as
\begin{equation}\label{eq:operation_action}
    \EC_l[P] = \sum_{j=1}^{N_{\EC_l,P}} c_j^{\EC_l} P_j\, ,
\end{equation}
where $c_j^{\EC_l}$ are the evolved coefficients for the Pauli strings $P_j$, and $N_{\EC_l,P}$ is the number of non-zero coefficient Pauli strings in the sum created from acting $\EC_l$ onto Pauli $P$.
We also call this number the \textit{branching factor}, and we will say $\EC_l$ is 1-branching if it maps one Pauli string to one Pauli string (i.e., it does not branch), 2-branching if it creates two Pauli strings from one, etc.
In Section~\ref{sec:gates&ops} we will give examples of how an operator's action can be written in the Pauli basis for some commonly encountered operators. In Box~\ref{box:PTM} we further discuss how the action of an operator on Pauli strings can be conveniently captured via the \emph{Pauli transfer matrix} (PTM) formalism.

\begin{theorybox}[label={box:PTM}]{Pauli Transfer Matrices}
In the Pauli Transfer Matrix (PTM) formalism~\cite{chow2012universal} an $n$-qubit operator can be represented as a $4^n$-dimensional vector, where the $i$'th entry is the  coefficient of the corresponding \emph{normalized} Pauli string $\mathsf{P}_i = 2^{-\frac{n}{2}} P_i$. That is, an operator $O$ is represented as the vector $|O \rrangle_i = \text{Tr}[ O \mathsf{P}_i]$, such that the Hilbert-Schmidt inner product can be expressed as $\Tr[A^\dag B] = \llangle A | B\rrangle $.
Any operation $\mathcal{E}$ is then represented in the PTM formalism as $\bm{\mathcal{E}}$ where
\begin{equation}
    \label{eq:definition_of_ptm}
    [\bm{\mathcal{E}}]_{ij}  = \text{Tr}[\mathsf{P}_i \mathcal{E}(\mathsf{P}_j)] \, .
\end{equation}
Equivalently, the PTM element $[\bm{\mathcal{E}}]_{ij}$ is (up to normalization) the coefficient $c_i$ in Eq~\eqref{eq:operation_action} for the operation $\mathcal{E}$ applied to the Pauli string $P_j$. Thus the PTM framework formalises the transformations an operation applies in the Pauli basis. 
The expectation value of a state $\rho$ evolved under an operation ${\mathcal{E}}$ with an operator $O$ can be written as 
\begin{equation}\label{eq:ptmexp}
      \Tr[O\EC[\rho]]  =  \llangle O | \bm{\mathcal{E}} | \rho \rrangle  \,.
\end{equation}
Equivalently, we can consider the dual map $\ECdag$ evolving the operator $O$, such that 
$\Tr[\ECdag[O]\rho] = \Tr[O \EC[\rho]]$:
\begin{equation}
      \Tr[\ECdag[O]\rho]  =  \llangle \rho | \bm{\ECdag} | O \rrangle  \,.
\end{equation}
The matrix $ \bm{\ECdag}$ is the \emph{transpose} of the matrix $\bm{\mathcal{E}}$, i.e. $[\bm{\ECdag}]_{ij} = [\bm{\mathcal{E}}]_{ji}$.
\end{theorybox}

Algorithmically speaking, if the action in Eq.~\eqref{eq:operation_action} can be defined for each $\EC_l$, i.e., if it is known (or can be efficiently computed) which Pauli strings arise and with which coefficients, then Pauli propagation can be used to estimate $\EC[O]$ via the sequential application of the operations to all arising Pauli strings. Concretely, we can write 
\begin{align}
    \EC[O] &= \EC_1\left[\dots \left[\EC_m\left[O\right]\right]  \right]\\
    &= \EC_1\left[\dots \left[\EC_m\left[\sum_{i=1}^{N_O} c_i P_i\right]\right] \right]\\
    &= \sum_{\alpha=1}^{N_{\EC,O}} c_\alpha^{\EC} P_\alpha    \,.\label{eq:evolved_operator}
\end{align}
where the final line captures the fully evolved observable in the Pauli basis. 
One might wonder whether writing down this action is as hard as solving the entire simulation task. Fortunately, the answer is often `No'. Typical gates are \textit{local} acting only on $k$ sites, for some small $k$, which means that the $k$-local gate can map one Pauli string only to at most $4^k$ Pauli strings. The mapping can be defined on those active sites, with all other Paulis in the Pauli string remaining unaffected.

We can then use the propagated operator in Eq.~\eqref{eq:evolved_operator} to compute properties of interest. For example, to estimate observable expectation values one simply computes the overlap between each of the Pauli strings in Eq.~\eqref{eq:evolved_operator} with the initial state of interest. That is, 
\begin{equation}\label{eq:expectation_value}
    \Tr[\rho \EC[O]] = \sum_{\alpha=1}^{N_{\EC,O}} c_\alpha^{\EC} \Tr[\rho P_\alpha] \, .
\end{equation}
In practice, estimating expectation values (i.e., inner products) with states is often drastically simpler than full state or operator evolutions. In Sec.~\ref{sec:truncations}, we outline several truncation strategies that can be employed to approximate the computation of Eq.~\eqref{eq:expectation_value} while not always faithfully approximating the object in Eq.~\eqref{eq:evolved_operator}.

\begin{softwarebox}[label={box:initial-state}]{Basics} The \ppjllink{} package can be used to compute or approximate objects represented and evolved in the Pauli basis, as well as to estimate expectation values with, e.g, quantum states.
A circuit is a vector of gates, and the parameters of the parametrized gates (below \code{thetas}) are typically vectors of numbers.
Importantly, both are defined in the order that they would act in the Schrödinger picture. This should be intuitive for users who are used to simulating quantum states. Our high-level type for Pauli strings is \code{PauliString}, which can be defined via its non-identity components on \code{nq} qubits. Here is a simple example code:

\medskip
\code{nq = 3\\
circuit = [\\
\text{\hspace{4mm}}PauliRotation(:X,~1), \\
\text{\hspace{4mm}}PauliRotation(:X,~2), \\
\text{\hspace{4mm}}CliffordGate(:CNOT, [2, 3])\\
]\\
thetas = [0.5, -1.2]\\
pstr = PauliString(nq, [:Z, :Z], [1, 3])\\
psum = propagate(circuit, pstr, thetas)
} 
\medskip

The output \code{psum} is a \code{PauliSum} object, our container for the propagating Pauli strings. As hinted at above, the circuit definition is in the Schrödinger picture, but \code{propagate()} applies the gates with its parameters in the \textbf{reverse order}, i.e., the Heisenberg picure. Finally, an expectation value with the all-zero initial state can be computed via
\code{overlapwithzero(psum)}. \\
\textit{Note, all code examples in this manuscript are for the current version of the package (v0.7), and future changes may be introduced.}

\end{softwarebox}

\subsection{Common gates and operations}\label{sec:gates&ops}

In this section we detail the action of commonly encountered gates and noise channels in the Pauli basis by specifying the form Eq.~\eqref{eq:operation_action} takes in each case. \\

\emph{Clifford Gates:} These gates form a group defined by their action on Pauli strings. Each Clifford gate $C$ transforms one Pauli string to another Pauli string, potentially itself, and potentially with a change in sign. That is, we have
\begin{equation}\label{eq:CliffordGate}
    C[P] = s_{C,P} \cdot P^\prime,
\end{equation}
where $s_{C,P} \in \{\pm1 \}$ is a sign depending on the definition of $C$ and its action on $P$, and $P^\prime$ is a Pauli string. This corresponds to conjugation by the corresponding unitary $U_C$ of the Clifford gate in the computational basis, i.e., $U_C^\dagger P U_C$.
Due to their 1-branching (i.e., non-branching) nature, any circuit consisting of solely Clifford gates can be simulated in a time scaling polynomially in the number of gates. 

Here are some examples of the H and CNOT Clifford gates acting upon Pauli strings in the Heisenberg picture:
\begin{align}
    \nonumber&\text{H}[I] &&= 1 \cdot I\\
    \nonumber&\text{H}[X] &&= 1 \cdot Z\\
    \nonumber&\text{H}[Y] &&= -1 \cdot Y\\
    \nonumber&\text{H}[Z] &&= 1 \cdot X\\
    \nonumber&\text{CNOT}[ZX] &&= 1 \cdot ZX\\
    \nonumber&\text{CNOT}[IZ] &&= 1 \cdot ZZ\\
    \nonumber&\text{CNOT}[YX] &&= 1 \cdot YI\\
    \nonumber&\text{CNOT}[XZ] &&= -1\cdot YY
\end{align}
Note, we here assume that the CNOT is controlled on the first (leftmost) qubit and we do not explicitly write the tensor product between the Pauli operators. That is, $Z X \equiv Z \otimes X \equiv Z_1 X_2$.

A few observations from these examples. The Hadamard H gate is its own inverse (applying it twice returns the original Pauli string) and transforms $Z$ to $X$ and vice versa. Even though the CNOT gate induces entanglement in states, it does not directly increase computational complexity. However, because it can increase (or decrease) the number of non-identity Paulis, it may make \textit{subsequent} gate applications branch more.

\medskip

\paragraph*{Pauli Rotations:} Gates of the form $R_G(\theta) = e^{-i\theta G/2}$, where the generator $G$ is an $n$-qubit Pauli string and $\theta$ is the angle of the rotation, are ubiquitous in quantum computing.
They appear in many canonical circuits, including the Trotterisation of Hamiltonian simulation.
A Pauli rotation is always either 1-branching (i.e., non-branching) or 2-branching, irrespective of how many qubits the gate acts non-trivially upon.
In particular, in the Heisenberg picture we have 
\begin{equation}
    R_G(\theta)
    [P] \coloneq e^{i\theta G/2} P e^{-i\theta G/2}
\end{equation}
such that for $\theta \in \mathbb{R}$ 
\begin{equation}\label{eq:PauliRotationPtmAsCommutator}
    R_G(\theta)
    [P] = \begin{cases}

    P & \text{if } [P,G] = 0\\
    \cos(\theta)P + \sin(\theta)P^\prime & \text{else}
    \end{cases}
\end{equation}
where $P^\prime = i[G, P]/2$ is a new Pauli string.

Given the 2-branching action of Pauli rotations, a circuit composed of $m$ Pauli rotations can generate up to $2^m$ Pauli strings. However, in practice, substantially fewer terms will typically need to be computed for two reasons. Firstly, a Pauli rotation causes branching only when the generator does not commute with the Pauli string operated on (as so for \textit{half} of all strings) and commutations tend to dominate if $O$ consists of \textit{local} Pauli strings, i.e., ones that have few non-identity Paulis. Secondly, duplicate copies of the same Pauli string can be generated during propagation which can then be \textit{merged} to reduce memory costs (see Section~\ref{sec:tree-traversal} for more details). It is hard to overstate how much these two effects simplify the computation compared to the worst-case scaling. For an example of this effect see Theory Box~\ref{box:MergeExample}. 

\begin{theorybox}[label={box:MergeExample}]{Example of the importance of merging and commutativity}
Consider the practical example of a 2-qubit circuit $ U =  RX1(0.3)RX2(\pi/3) RZZ(-0.8)$ where the right-most unitary is applied first to a hypothetical state in the Schrödinger picture.  We take our final observable to be the 2-qubit Pauli string $O = ZI$. Working in the Heisenberg picture, we first obtain
\begin{align}
    RX1(0.3)[O] &= 0.96 \cdot ZI + 0.30 \cdot YI,  \\ 
    &= O_3\,,
\end{align}
with coefficients shortened for display. Then the next rotation
\begin{align}
    RX2(\pi/3)[O_3] &= 0.96 \cdot ZI + 0.30 \cdot YI,\\
    &= O_2
\end{align}
leaves the Pauli sum unaffected because the $RX2$ gate commutes with both Pauli strings that carry the identity Pauli on the second site.
And finally, 
\begin{align}
    RZZ(-0.8)[O_2] &=  0.96 \cdot ZI +  0.21 \cdot YI - 0.21 \cdot XZ, \\
    &= O_1\,,
\end{align}
where the $ZI$ Pauli string commutes with the $RZZ$ gate, $0.21 = 0.30 \cdot \cos(-0.8) $, and $0.21 = 0.30 \cdot \sin(-0.8)$. This example resulting in three Pauli strings in $O_1$ highlights that the expected $2^m = 8$ scaling is not representative of the true computational complexity of simulating circuits with $m$ Pauli rotations.
\end{theorybox}

\medskip

\emph{T gates:} T gates are special cases of Pauli-Z rotations with $\theta=\pi/4$. Thus it follows from Eq.~\eqref{eq:PauliRotationPtmAsCommutator} that if there is an $X$ or $Y$ at the location in the Pauli string on which T gates acts then the operation is 2-branching (and the new coefficients have equal magnitude), otherwise the gate leaves the Pauli string unchanged. Famously the worst-case simulation of a circuit composed only of Clifford and T gates is known to be exponential in the number of $T$ gates, but not in the number of Clifford gates.

\medskip

\emph{Pauli Noise:}
Pauli noise is an umbrella term for \textit{unital} quantum channels that are diagonal in the Pauli basis. This encompasses both homogeneous and inhomogeneous dephasing and depolarising channels.
Such noise multiplies the coefficients of some or all non-identity Paulis by a damping factor that is smaller than 1. This does not change the Pauli strings themselves and so a Pauli noise channel is non-branching. 
It is often sufficient to consider $1$ and $2$-local Pauli channels to capture passive and active (following $2$-qubit operations) error processes in quantum devices~\cite{emerson2007symmetrized,angrisani2025simulating}. 
For example, the one-qubit inhomogeneous depolarising channel, parameterised by the individual probabilities $p_i$ of $X$, $Y$ and $Z$ errors upon the decohering qubit, modifies a density matrix $\rho$ as
\begin{align}
    \mathcal{E}(\rho) = p_I \rho + p_x X \rho X + p_y Y \rho Y + p_z Z \rho Z,
\end{align}
where $p_I = 1-(p_x+p_y+p_z)$. Its corresponding PTM, agnostic to whether working in the Schr\"odinger or Heisenberg pictures, maps Pauli strings to
\begin{align*}
    \mathcal{E}[I] &= 1 \cdot I, 
        & \mathcal{E}[Y] &= (1-2p_x-2p_z) \cdot Y,  \\
    \mathcal{E}[X] &= (1-2p_y-2p_z) \cdot X,
        & \mathcal{E}[Z] &= (1-2p_x-2p_y) \cdot Z.
\end{align*}
Two-qubit depolarising noise with uniform probability $p$
scales all strings by $1 - 16p/15$, \textit{except} for $II$ which is unchanged. One (or two)-qubit dephasing (driving the qubits towards the $z$-axis of the Bloch sphere) scales all strings containing an $X$ or $Y$ by $1-2p$ (or $1-4p/3$).
\medskip

\emph{Amplitude damping and arbitrary local noise}:
Moving beyond Pauli noise, any single-qubit channel $\mathcal{N}$ can be written in the Schr\"{o}dinger picture as~\cite{king2001minimal,ben2013quantum, mele2024noise}
\begin{align}\label{eq:normalform}
    \mathcal{N}(\cdot) = U   \mathcal{N}'(V(\cdot)V^\dag)U^\dag,
\end{align}
where $U,V$ are arbitrary single-qubit unitaries and the action of $\mathcal{N}'$ on the single-qubit Pauli matrices is given by
\begin{align}
    &\mathcal{N}'(I) = I + t_X X + t_Y Y + t_Z Z,
    \\&\mathcal{N}'(X) = D_X X,
    \\&\mathcal{N}'(Y) = D_Y Y,
    \\&\mathcal{N}'(Z) = D_Z Z,
\end{align}
where $\bold{D} = (D_X, D_Y, D_Z) \in [-1,1]^3$ and $ \bold{t} = (t_X, t_Y, t_Z)\in [-1,1]^3$ are two 
vectors. If $\bold{t} = 0$ such that $\mathcal{N}(I) = I$, then the noise channel is said to be \emph{unital} and reduces to Pauli noise (up to a change of basis). Otherwise, the channel is non-unital, i.e. $\mathcal{N}(I)\neq I$.
A common example of non-unital noise is the amplitude-damping channel, characterized by $\bold{D} = (\sqrt{1-\gamma}, \sqrt{1-\gamma}, 1- \gamma )$ and $\bold{t} = (0,0, \gamma)$ for some $\gamma \in (0,1]$. 
Repeatedly applying an amplitude damping channel pushes a quantum state
to the computational zero state.

\smallskip

In most applications of Pauli Propagation, we are concerned with the dual representation $\ECdag$ of a channel $\EC$, since we typically operate in the Heisenberg picture.
Notably, $\ECdag$ is always unital, i.e., $\ECdag(I) = I$, but if $\EC$ is non-unital, then $\ECdag$ is not trace-preserving. Thus, in the Heisenberg picture, noise channels do not affect or branch on $I$, but they may branch on $X,Y,Z$ Paulis.

\medskip

\emph{Controlled gates and measurements}: 
Pure states like $\ket{0}\bra{0}^{\otimes n} = (I+Z)^{\otimes n}/2^n$ contain exponentially many terms in the Pauli basis. This means the introduction of control qubits to few-branching unitaries (or even most Cliffords) can make them prohibitvely many-branching. As such, Pauli propagation techniques are typically ill-suited for the simulation of many-controlled unitaries. 
Despite this, \textit{projectors} can often be efficiently implemented, even when global and with exponentially large PTMs. This is because inner products with stabilizer states like the all-zero state are efficiently computable.
For example, the one-qubit zero projector $\Pi = \ket{0}\bra{0}$ is either two-branching, else \textit{removes} a Pauli string.
\begin{align}
    \Pi[I] = \Pi[Z] =  (I + Z)/2,
    \hspace{1cm}
    \Pi[X] = \Pi[Y] = 0.
\end{align}
The $n$-qubit zero projector factorises into a product of $n$ one-qubit projectors and is ergo principally $2^n$-branching, but only when the targeted substring contains no $X$ or $Y$. And strings differing only by $I \leftrightarrow Z$ in the targeted substring (of which there are $\le 2^n$) are mapped to the \textit{same} $2^n$-term sum.
Mid-circuit \textit{measurements} which normalise the projector are also possible to efficiently simulate, though with dissimilar logic, namely, via ancilla qubits and controlled unitary operations~\cite{deshpande2024dynamic}.

The \texttt{QuESTlink} Mathematica library~\cite{jones2020questlink} is useful for quickly obtaining and visualising Pauli transfer matrices of symbolic operations.
For example, \texttt{DrawPauliTransferEval} can visualise the Heisenberg evolution of the string $YX$under the circuit $CR_X(2\phi) \cdot (\text{H} \otimes I) \cdot (R_Y(\theta) \otimes \ket{0}\bra{0})$ as:
\begin{center}
    \includegraphics[width=.6\columnwidth]{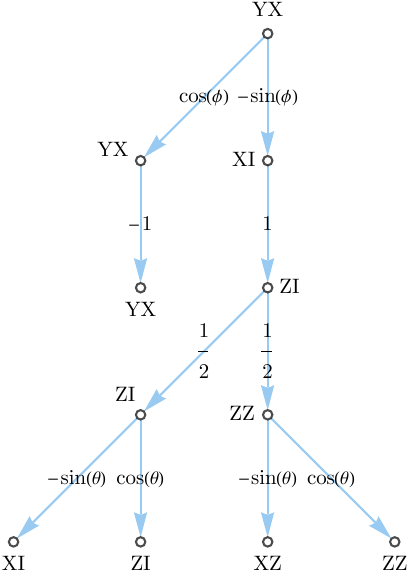}
\end{center}

\begin{softwarebox}[label={box:topologies}]{Topologies \& Circuits}
Multi-qubit gates can be applied between any pair (or more) of qubits, which is an advantage over more topology-tailored simulation methods. How qubits are connected does not directly determine the difficulty of the simulation, but it will likely affect how Pauli strings evolve and consequently how much they branch throughout the circuit. 
\ppjllink{} provides a suit of pre-defined entangling topologies in 1D and 2D,
for example 
\code{bricklayertopology()} and \code{rectangletopology()}
which can be passed to common circuit builders, for example, \code{tfitrottercircuit()} for simulating the dynamics of the transverse-field Ising model. Topologies are vectors of tuples like \code{topology = [(1, 2), (2, 3), (1, 3)]}, 
indicating which sites are connected via multi-qubit gates, and circuits are vectors of gates.
\end{softwarebox}

\begin{softwarebox}[label={box:custom-gates}]{Custom gates}
The Pauli propagation framework can of course be applied to a broader class of operations than the ones listed in Section~\ref{sec:gates&ops}. There are a number of ways in which a new gate/operation can be defined in \ppjllink{}. The simplest is via the Pauli Transfer Matrix formalism (see Theory Box~\ref{box:PTM}) and our \code{TransferMapGate} type. Alternatively, gates can be defined by directly specifying their action on Pauli strings in a custom \code{apply()} function, which may be more efficient but more complicated to implement. Depending on the gate type, it may be more efficient to manually control the branching pattern and implement more specialized memory management by overloading the \code{applytoall!()} function. We explain all approaches in our example Jupyter notebooks.
\end{softwarebox}

\subsection{Pauli propagation surrogates}\label{sec:surrogate}
\begin{figure}
    \centering
    \includegraphics[width=1\linewidth]{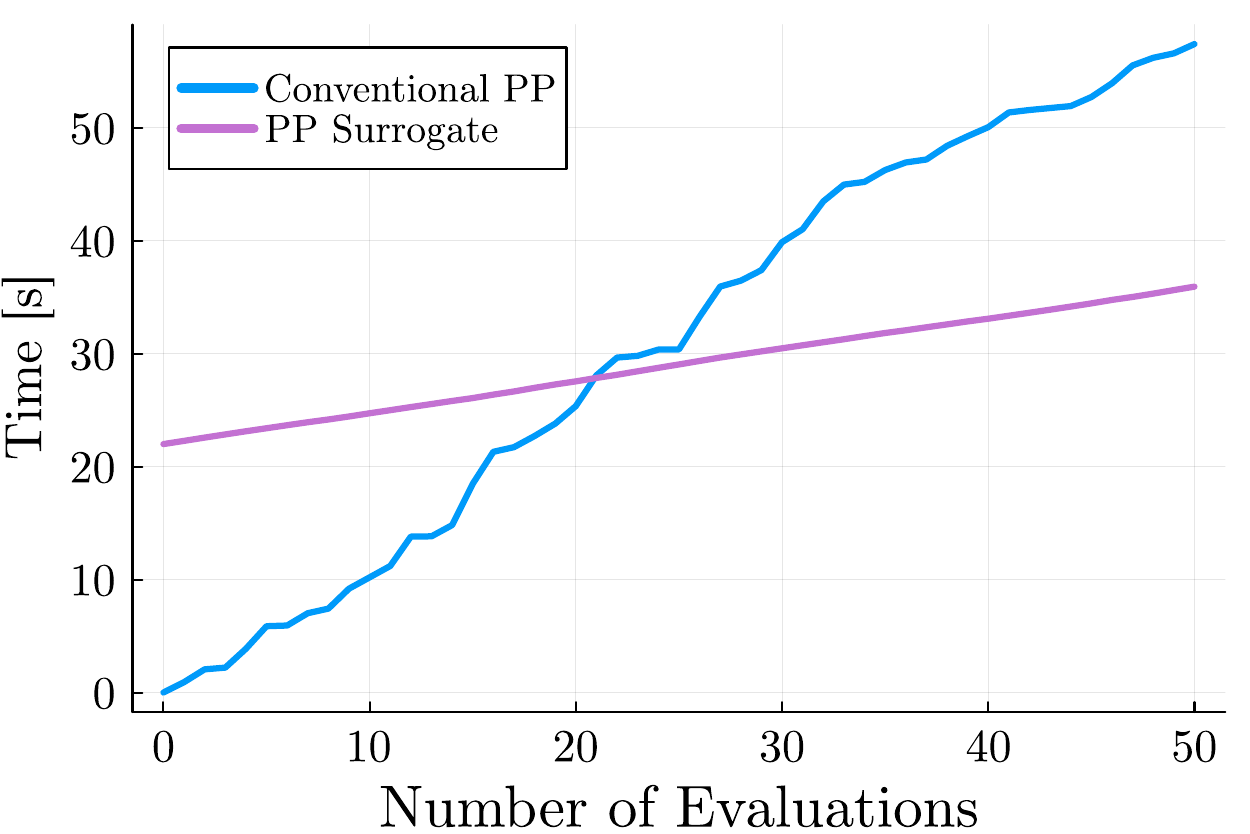}
    \caption{\textbf{Potential advantage of a Pauli propagation surrogate}. A Pauli propagation surrogate invests time and especially memory in an initial surrogation stage, then repeated evaluations can be significantly faster than running conventional Pauli propagation over and over for different parameters.
    This is demonstrated on a 20-qubit example with 10 layers and random correlated parameters within each brickwork layer. We estimate the operator correlation value of Pauli Z in the middle of the qubit chain, which represents one of the strongest sparsifications of the surrogate relative to conventional Pauli propagation. The truncation parameters are chosen such that both models achieve approximately $0.005 - 0.010$ average absolute error compared to statevector simulation, and both simulations use a single CPU thread for a fairer comparison. We emphasize that this setup is chosen to demonstrate the potential advantages of surrogates, but our current surrogate implementation may not scale to the most interesting cases.}
    \label{fig:surrogate-time}
\end{figure}

The word ``surrogate" has been used within the fields of machine learning and quantum computing and has various meanings. What we refer to as a \textit{Pauli propagation surrogate} is a compiled computational model for expectation values in the form of Eq.~\eqref{eq:expectation_value} that trades an initial overhead in time and especially memory against significantly faster re-evaluation of expectation values for different circuit parametrizations. One of the few Pauli propagation surrogates that we are aware of is known as LOWESA (low-weight efficient simulation algorithm). LOWESA was proposed in Ref.~\cite{fontana2023classical} for the classical simulation of noisy variational quantum algorithms, but the later works in Ref.~\cite{rudolph2023classical,bermejo2024quantum,lerch2024efficient, miller2025simulation} fully embraced the surrogation aspect of the initial proposal.

We would like to draw attention to two features in a simulation that can give rise to powerful surrogate models in the spirit of this work. First, if there is a lot of high level logic in the algorithm, e.g., many if-else statements potentially with truncations. When this logic dictates the flow of the computation and only depends on the structure of the problem not its parameters, then the problem can be compiled once and subsequently evaluated deterministically without this logic. This is reminiscent of how software libraries for \textit{automatic differentiation} work. A computational graph for the derivative of a function is built once and can then be evaluated again and again during common gradient descent routines. The graph does not change because it is fixed from the time that the function is defined, independent of the input parameters. 

A second feature of quantum expectation value simulations, that can further drastically reduce re-evaluation time, is the fact that many Pauli terms annihilate when overlapped with an initial state (or operator). Consider the decomposition of an observable expectation function in the Pauli basis,
\begin{align}
    \langle O \rangle &= \Tr[\rho U^\dagger(\thv)OU(\thv)]\\
    &= \sum_\alpha c_\alpha(\thv) \Tr[\rho P_\alpha]
\end{align}
where $U^\dagger(\thv)OU(\thv) = \sum_\alpha c_\alpha(\thv) P_\alpha$ is a unitarily evolved observable written in Pauli basis and  $\rho$ is the initial state.
Usually, the sum over Pauli strings with their coefficients can instead be replaced or approximated with a significantly smaller sum over fewer terms where $\Tr[\rho P_\alpha]$ is large. Pauli paths leading to small or zero contributions do not need to be revisited when changing the parameters $\thv$. See Fig.~\ref{fig:surrogate-time} for a proof-of-principle demonstration of the potential speed-ups.

Pauli propagation surrogates can be used naturally in the context of classically simulating variational quantum algorithms. Namely, one could optimize the parametrized quantum circuit fully classically on the approximate surrogate landscape. In some cases, one may need to be careful that the truncations used do not artificially change the depth or the positions of minima. For example, if aiming to find a ground state of a given Hamiltonian, a truncated Pauli propagation surrogate does not strictly simulate a normalized quantum state and thus could lead to energies that are lower than classically possible for a genuine quantum state. However, in other contexts, for example when classically simulating a quantum convolutional neural network for solving phase classification problems, as in Ref.~\cite{bermejo2024quantum}, this is not an issue. The surrogated quantum convolutional neural network can simply act as a new quantum-inspired classical model for the task at hand. 

Beyond variational problems, the prospects of Pauli propagation surrogates in the context of dynamical simulations are particularly intriguing. For example, when simulating the dynamics of a Hamiltonian $H(\vec{\lambda})$ the parameters of that system, $\vec{\lambda}$, appear as parameters in the gates of the circuit. A Pauli propagation surrogate of this circuit therefore has the potential to simulate a family of Hamiltonians. Similarly, if the initial state preparation circuit is included in the circuit description, one could include this circuit in the surrogate and explore the evolution of the system for a whole family of different initial states. We expect these capabilities to be most useful in the context of averaging over the effect of disorder or for meta-learning optimal Hamiltonians or initial states for various tasks.

We end by stressing that the potential of Pauli propagation surrogates have not yet been fully realized. Writing efficient code that builds a computational graph on the fly and compiles a function that can be evaluated very quickly is not trivial. \ppjl{} currently contains a surrogate version close to the one used in Refs.~\cite{bermejo2024quantum,lerch2024efficient}, but it does not leverage sophisticated code compilation techniques that could drastically accelerate evaluation. 
As of now, all non-surrogate code in \ppjl{} is differentiable for learning and optimization tasks, and currently is likely the better choice for almost all concrete applications.

\subsection{Prior work and related methods}
Pauli path integrals are a relatively recent theoretical tool for quantum information which gained traction after the seminal work of Ref.~\cite{aharonov2022polynomial}, providing polynomial-time classical algorithms for sampling from the output distribution of certain noisy circuits.
Pauli-path based techniques also play a central role in spoofing the Linear Cross-Entropy benchmark used in quantum supremacy experiments\ \cite{barak2020spoofing, tanggara2024classically}. There have further been numerous papers providing guarantees for noisy~\cite{aharonov2022polynomial, fontana2023classical, shao2024simulating, gonzalez2024pauli, schuster2024polynomial, angrisani2025simulating} and noiseless~\cite{angrisani2024classically,lerch2024efficient} circuits as well as efforts to understand such methods through Fourier analysis~\cite{nemkov2023fourier,fontana2023classical,cirstoiu2024fourier}. 
However, in this section we focus on the even more recent developments in  practical implementations of algorithms related to Pauli propagation.

Pauli propagation is related to so-called \textit{stabilizer simulation}~\cite{aaronson2004improved} for simulating Clifford circuits and the \textit{extended stabilizer simulation}~\cite{bravyi2016improved} for Clifford and few non-Clifford circuits. Those frameworks are concerned with how states can be compactly represented in the Pauli basis, and how that efficient representation can be used to estimate expectation values or sample. If an $n$-qubit quantum state is a Pauli stabilizer state~\cite{gottesman1997stabilizer,aaronson2004improved}, i.e. if the state can be fully characterized by $n$ Pauli strings, how those Pauli strings are updated under the action of common gates is in the same spirit (and sometimes exactly the same) as in Pauli propagation. One software package for stabilizer simulation is STIM~\cite{gidney2021stim}, which is vastly faster for simulating purely Clifford circuits, but does not support non-Clifford gates. For circuits consisting of Clifford gates and Pauli rotations (including T gates), for example, the resulting state will be a weighted sum of stabilizer states, each of them represented via $n$ Pauli stabilizers. Due to the flexibility in our code design, it is possible to implement extended stabilizer simulation in the Schrödinger picture without redefining the propagation -- only the action of gates on the so-called stabilizer \textit{tableaus}~\cite{aaronson2004improved}. More concrete implementations can be found via, e.g., Refs.~\cite{fast2022pashayan,bravyi2019simulation}.

\begin{softwarebox}[label={box:literature}]{Works using our code}
Very early versions of our Pauli propagation surrogate has been used to simulate the IBM ``utility'' experiment on 127 qubits~\cite{rudolph2023classical}. More refined versions of this code were used to demonstrate that quantum convolutional neural networks (QCNNs) may be more broadly classically simulable than previously thought~\cite{bermejo2024quantum}. The current version of the Pauli propagation surrogate in \ppjllink{} was demonstrated in Ref.~\cite{lerch2024efficient} in conjunction with theoretical guarantees for hyrid simulation schemes involving quantum computers. Modern versions of our more flexible non-surrogate code as been used in Ref.~\cite{angrisani2024classically}, demonstrating that Pauli propagation with Pauli weight truncation leads to efficient average-case simulation of circuits that were previously believed to be intractable. In Ref.~\cite{angrisani2025simulating} it was used to demonstrate efficient large-scale simulation of quantum circuits with arbitrary local noise. Ref.~\cite{danna2025circuit} compresses 2D quantum dynamics circuits up to $30\times30$ grid size via automatic differentiation. Ref.~\cite{teng2025leveraging} demonstrates how spatial symmetries in quantum systems can be leveraged to reduce memory via symmetry merging strategies now available in our library. Ref.~\cite{rudolph2026thermal} prepare thermal states via imaginary time evolution-- a feature which has since been added to our library. Finally, \code{MajoranaPropagation.jl}~\cite{d2025majorana} is built on top of \ppjl{} to flexibly simulate fermionic systems in the related Majorana basis.
\end{softwarebox}

One open-source package embracing Heisenberg evolution in the Pauli basis is \code{PauliStrings.jl}. Its focus is on simulating real-time quantum dynamics by solving the von Neumann equation directly via calculation of the nested commutator of the evolving observable $O(t)$ with the Hamiltonian $H$. The time derivative at time $t$ is estimated like 
\begin{equation}
    \frac{\partial O(t)}{\partial t} = i[H, O(t)],
\end{equation}

\noindent which can then be plugged into a partial differential equations solver of choice, for example, 4th-order Runge Kutta. For continuous-time dynamics, this approach could potentially be compared to \ppjl{} simulating a finely Trotterized quantum circuit, but a fair comparison would be fiddly given the different focuses of the two packages (e.g., real-time dynamics versus circuit simulation / Floquet dynamics). Alternatively, \code{PauliStrings.jl} can be used for Lanczos or Krylov methods. 
For more information we refer to the accompanying manuscript in Ref.~\cite{loizeau2025quantum}. Under the broad umbrella of Lanczos Pauli basis simulation, a number of themes have been explored that have been, or could be, applied to vanilla Pauli Propagation. These include: surrogation~\cite{uskov2024quantum}, weight truncation for random circuits~\cite{ermakov2024unified}, extrapolation methods~\cite{teretenkov2025pseudomode}, generalizations beyond the Pauli basis~\cite{ermakov2025operator} and limitations due to dynamical phase transitions~\cite{shirokov2025quench}.

Mirroring the progress in the field of Pauli propagation in general, the works in Refs.~\cite{beguvsic2023simulating,beguvsic2023fast,beguvsic2024real} demonstrate large-scale simulations with various truncation strategies and design choices. The source code for the corresponding publications is partially available therein, and the results can be replicated out-of-the-box via \ppjl{}. 

One tensor network approach from the physics community that mirrors some motivating principles of Pauli propagation is known as \textit{dissipation-assisted operator evolution} (DAOE)~\cite{rakovszky2022dissipation}. It uses matrix product states (MPS) in the Pauli basis with the specific goal of estimating expectation values after dynamical evolution. DAOE shares several conceptual similarities with Pauli propagation. Firstly, in the language of Pauli propagation, DAOE can be viewed as encoding the Pauli coefficients of a Heisenberg evolved observable in a tensor network (and thereby allowing access to a wide repertoire of tensor network tools and compression). Secondly,  it employs a strategy similar to Pauli weight truncation~\cite{angrisani2024classically} by imposing exponentially decaying coefficients for higher-weight Pauli operators.DAOE has been successfully employed to extract transport quantities from highly damped Paulis by extrapolating to the zero-damping limit, thus showing that the fields of tensor networks and Pauli propagation may continually and jointly progress for targeted applications.

There are, however, both conceptual and practical differences between Pauli propagation and DAOE-- or rather, tensor networks in general. One difference is that Pauli propagation is limited precisely by the number of Pauli strings $N$, whereas an MPS bond dimension $\chi$ is only (very loosely) upper-bounded by it. At the same time, the memory and runtime of Pauli propagation are $\OC(N)$, while the memory of MPS is $\OC(\chi^2)$ and the runtime is $\OC(\chi^3)$. In practice, the bond dimension $\chi$ is determined by how spatially factorizable the underlying Pauli sum is into the tensor network graph, and almost always we find that significant compression is possible, i.e., that $\chi \ll N$. An extreme example is the uniform superposition of all $N=4^n$ Pauli strings over $n$ sites, which is a propagation algorithm's nightmare, but it remains $\chi=1$ as an MPS. Removing one Pauli string from that sum actually \textit{increases} the bond dimension to 2. Similarly, the MPS bond dimension in DAOE used for encoding operator entanglement can grow rapidly when Pauli strings are fully truncated or with the size of the window of undamped Pauli strings. The other extreme is a Pauli sum consisting of a polynomial number of random Pauli strings. Its factorization likely remains full rank, which results in a polynomial memory and runtime disadvantage over Pauli propagation.  That being said, when simulating physical systems, the required computational resources for both methods tend to correlate quite strongly~\cite{dowling2025bridging}, making the comparison a case-by-case study.

Above we have detailed some classical simulation algorithms that are most closely related to Pauli propagation. However, a complete survey would be impossible. Let us end this section by simply highlighting that tensor network methods for evolving operators have substantially outgrown their origins to a highly flexible approach. In particular, state of the art tensor network methods can work in the Heisenberg picture~\cite{verstraete2004matrix,hartmann2009density} and tackle beyond 1D topologies~\cite{beguvsic2023fast,liao2023simulation}. The relative merits of Pauli propagation compared to tensor networks, and other approximate simulation methods, remains an open question.

\medskip

\section{Mid level}

\begin{figure}
    \centering
    \includegraphics[width=\linewidth]{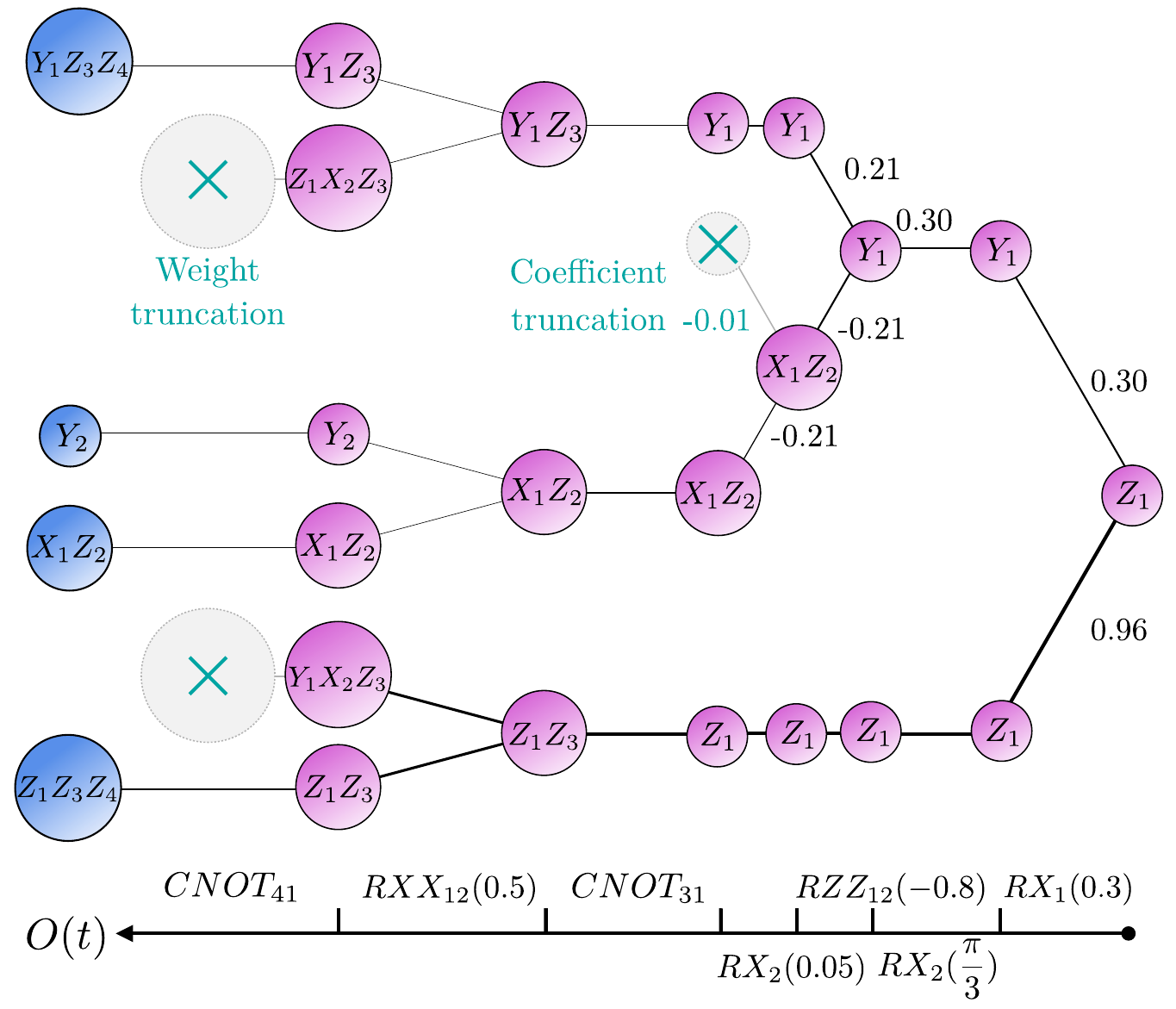}
    \caption{\textbf{Truncations during Pauli propagation}. A tree graph illustrates the evolution of Pauli strings (nodes) and their associated coefficients (edges) under a unitary circuit, applied in sequence from right to left along the bottom axis. The shade of the edges indicates the magnitude of the coefficient size and the size of the node the weight of the corresponding Pauli string. The first three gates are illustrated in Theory Box~\ref{box:MergeExample}. In the worst case, the leaves of the tree will grow exponentially with the number of gates, leading to an intractable tree. However, for many simulation tasks, this growth can be controlled using systematic truncation strategies described in Section~\ref{sec:truncations}.  Two types of truncations are depicted: weight truncation, which discards Pauli strings with weight above a chosen threshold (e.g., weight $4$), and coefficient truncation, which prunes strings with small coefficients (e.g., $-0.01$). Custom truncations are also supported by PauliPropagation.jl.
    }
    \label{fig:pp_truncations}
\end{figure}

\subsection{Truncation strategies}\label{sec:truncations}
Truncations are targeted approximations in the simulation of quantum systems that aim to achieve drastic reductions in computational resources while maintaining high accuracy. The degree of required accuracy strongly depends on each task. Some tasks require computations to be precise up to an exponentially small scale, while others only need to be visually close at unit scale.

Looking at Eq.~\eqref{eq:expectation_value} there are currently two main families of truncations: 1) ones that truncate small (or likely small) coefficients $c_\alpha$ and 2) ones that truncate Paulis $P_\alpha$ where $\Tr[\rho P_\alpha]$ is small (or likely small). Expectation values can be significantly easier to simulate than full operators as the second truncation scheme only works in the latter case. In Fig.~\ref{fig:pp_truncations}, we illustrate these two truncation strategies, using an example of a sequence of unitary gates evolving an initially local Pauli string. The resulting operator contains fewer Pauli strings when truncations are applied, compared to the full evolution.

\medskip 

\emph{Small coefficient truncation.}
The most straightforward truncation strategy is to discard Pauli strings whose current absolute coefficients fall below some threshold. While it is intuitive that truncating the smallest coefficients should result in the lowest approximation errors, and it practically often does, those errors may still accumulate to result in large errors. 
This is because complex quantum dynamics may generate a considerable number of new Pauli strings -- exponential in the number of non-Clifford gates --  each with a small coefficient. As a result, approximating the evolved observable with few Pauli terms can become extremely challenging, if not impossible.

This truncation strategy is tightly connected to the concept of magic (see box~\ref{box:magic} for more details). 
Nevertheless, truncating small absolute coefficients is the current go-to truncation strategy for Pauli propagation, without which practical simulations are unimaginable. It has, for example, been used in Refs.~\cite{beguvsic2023fast,beguvsic2024real} under the name of \textit{sparse Pauli dynamics} (SPD), and is currently the only truncation in PauliPropagation.jl that is active by default. We note that some works report the coefficient truncation value \textit{before} merging, while we currently truncate \textit{after} merging. Truncating after merging is safer for general gates (our focus), and it is somewhat more accurate and resource-intensive. This means that, when using truncation thresholds reported in other works, \ppjl{} simulation may be more accurate as a consequence of maintaining more Pauli strings (and, correspondingly, potentially slower). Thus slightly modified truncations may yield the most comparable results.

\begin{softwarebox}[label={box:truncations}]{Truncations}
\ppjllink{} provides most truncations mentioned in Sec.~\ref{sec:truncations} by default. The most important keyword arguments to the \code{propagate()} function will be the minimum coefficient truncation threshold \code{min\_abs\_coeff}, and the weight truncation threshold \code{max\_weight}. Custom truncation functions can be passed via the \code{customtruncfunc} keyword argument. 
The only truncation active by default is \code{min\_abs\_coeff=1e-10}, leaving typical calculations practically exact. 
We recommend approaching simulations \textit{from fast to faithful}, i.e., starting with truncation values that clearly truncate too much and slowly making them more faithful. This is because the hardness of simulating quantum systems/circuits can vary drastically, and one will want to observe \textit{convergence} similar to Fig.~\ref{fig:convergence}. For typical expectation value calculations, we might start off with \code{min\_abs\_coeff=1e-3} and \code{max\_weight=5}. Often, such simulations will finish in (milli)seconds but provide imprecise results. This will be highly case-dependent.
\end{softwarebox}

\emph{Pauli weight truncation.}
This truncation scheme drops Pauli strings with large weight, i.e., many non-identity Paulis, which are often likely to have a small overlap with the initial state. An intuitive example is that of the all-zero initial state $\rho = |0\rangle\langle0|$, which can be written as a uniform sum of all Pauli strings consisting of I and Z Paulis. Thus, if we consider a Pauli string $P$ with $W$ non-identity Paulis and assume that those Paulis are drawn randomly from the set $\{X, Y, Z\}$, then the ``probability'' that $\Tr[\rho P] = 0$ is $1- (1/3)^W$ because a Pauli string containing \textit{any} $X$ or $Y$ Pauli is orthogonal to $|0\rangle\langle0|$. This argument in fact holds not just for the  all-zero initial state but also for all stabilizer states and indeed random quantum states~\cite{angrisani2024classically}.

Ref.~\cite{angrisani2024classically} goes further and proves that this truncation strategy yields small average-case simulation errors for \textit{locally-scrambling} circuits, i.e., for circuits that are least locally random for any observable and initial state. However, this provable guarantee is on average and has high error for specific domains in parameter space. Quantum circuits describing real-time quantum dynamics, for example, do not satisfy the theoretical assumptions and practically lead to larger errors. One may expect that this is due to Pauli strings $P$ with large weight having larger overlaps with the initial state, i.e., larger $\Tr[\rho P]$ than in unstructured simulations, but this does not seem to be the only reason. While it has not yet been rigorously explored in the context of quantum circuit simulation, there exists indication that structured circuits with conserved quantities exhibit systematic operator \textit{backflow} from high to low weight~\cite{rakovszky2022dissipation}. Thus, truncating high-weight Pauli strings may result in this backflow being interrupted, leading to strongly accumulating errors over time.

A variant on weight truncation whereby $Z$ terms are not counted for Pauli weight (i.e., only Pauli strings with large numbers of $X$ or $Y$ terms are truncated) was employed in Ref.~\cite{beguvsic2024real} for quantum dynamics where the initial state is the all-zero state. While this heuristic appears sensible, it is unclear whether it provides practical improvements over Pauli weight truncation, which is less accurate per truncation threshold but substantially cheaper. \\

There are a number of truncation strategies that can naturally be applied in the context of Pauli propagation surrogates. In the context of fixed angle simulations these truncation schemes are not strictly necessary/advantageous but in certain cases can be used to provide average case theoretical guarantees in some regimes. We detail these below.  \\

\emph{Frequency truncation.}
The \textit{frequency} of a path is the number of times a Pauli path picks up either a cosine or a sine coefficient -- i.e., how often it branches at Pauli rotations -- in a circuit consisting of Clifford gates and Pauli rotations. Truncating based on this quantity is effectively truncating according to the \textit{average} coefficient over the entire parameter range. Consider that $\frac{1}{2\pi}\int_{0}^{2\pi}\sin^2(\theta)d\theta = \frac{1}{2\pi}\int_{0}^{2\pi}\cos^2(\theta)d\theta = \frac12$. It becomes clear that if a path has picked up $\ell$ cosine or sine coefficients, then the average coefficient will be $\left( \frac12\right)^\ell$. 
This has been employed for the Pauli propagation surrogate algorithm LOWESA for simulating the so-called IBM quantum utility experiment~\cite{rudolph2023classical}.
We further note that in noisy quantum circuits and particular ansatz structures, frequency truncation provides provable average-case polynomial runtimes for inversely polynomially small errors~\cite{fontana2023classical,martinez2025efficient}.

\medskip

\emph{Sine truncation.} 
A more targeted version of frequency truncation, that we sometimes refer to as \textit{small-angle truncation}, is the \textit{sine truncation} dropping paths if the number of sine factors picked up in a circuit consisting of Clifford gates and Pauli rotations is above some threshold. This is the truncation at heart of the so-called \textit{Clifford perturbation theory} (CPT) approach~\cite{beguvsic2023simulating} because this truncation can be understood as an expansion of the evolved observables in the power of sines. If the angles parametrizing Pauli rotations are small, this truncation tends to truncate the smallest coefficients. In fact, Ref~\cite{lerch2024efficient} proves that sine truncation can be applied to faithfully simulate small-angle regions-- so-called \textit{near-Clifford patches} -- in polynomial time.

\medskip

\emph{Path-weight truncation.}
What is commonly referred to as path weight is the \textit{cumulative} Pauli weight of a Pauli as it propagates. Where precisely in the circuit the path weight is updated by adding the current Pauli weight is arbitrary and could be every gate or every layer. The practical benefits of this truncation strategy are so far largely unexplored. However, it has been used to great success in formalizing the classical simulability of noisy quantum circuits~\cite{aharonov2022polynomial,schuster2024polynomial,angrisani2025simulating}. In those theoretical results, path weight is a tight proxy for the magnitude of Pauli path coefficients due to the damping nature of common noise models. However, it is not clear whether this truncation scheme has advantages over direct small coefficient truncation in noisy circuits or Pauli weight truncation for noise-free circuits.

\medskip

\begin{figure*}
    \centering
    \includegraphics[width=1\linewidth]{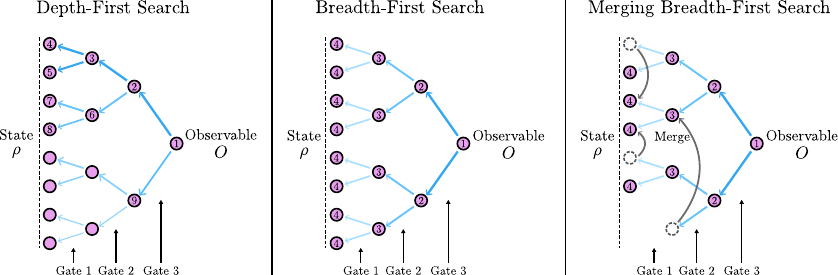}
    \caption{\textbf{Traversing the Pauli propagation tree}. Repeated application of a gate to an observable represented as a sum of Pauli strings can cause repeated \textit{branching} of Pauli strings into other Pauli strings with new coefficients. The exact branching pattern depends on the type of gate, but here we explore 2-branching gates (e.g., Pauli rotations) for simplicity. The nodes in the trees represent Pauli strings at different positions in the circuit, and the strength of the arrows and the numbers on the nodes denote the order in which the algorithm visits the nodes. \textit{Depth-first search} (DFS) is a strategy that fully propagates a Pauli string through the entire circuit before returning to propagating the next. This is very light on memory. \textit{Breadth-first search} (BFS) instead evolves Pauli strings gate-by-gate, consequently being heavier in memory. However, knowledge about all Pauli strings after each gate allows for \textit{merging} of identical Pauli strings. This \textit{merging BFS} approach is asymptotically faster than DFS or BFS, and its memory requirements are in between the other two.}
    \label{fig:traversal}
\end{figure*}

\subsection{Pauli propagation is a tree search problem}\label{sec:tree-traversal}

We now face the question of how to compute all the contributing terms to Eq.~\eqref{eq:evolved_operator} or at least the most important ones. Due to the branching nature of the simulation problem, this problem can be framed as a \textit{tree search}. This is a widely-studied topic in computer science, with two general search approaches being the \textit{depth-first search} (DFS) and \textit{breadth-first search} (BFS). In our case, DFS amounts to simulating in a \textit{path by path} fashion, and BFS as in a \textit{gate by gate} fashion. This is sketched in Fig.~\ref{fig:traversal}.

The main advantage of DFS is its low memory usage. Whenever a Pauli string branches into more than one Pauli string, one always follows through with the simulation of one of those Pauli paths until the end of the circuit. The other is saved (e.g., via \textit{stack} data structures or recursion) until it is their turn to be propagated. At the end of the circuit, one computes the overlap of the final Pauli string with the initial state/observable and its contribution can be added to a running sum. Then the simulation can proceed to simulating the next path without storing in memory anything about the previous one. Thus DFS always has polynomial memory (likely linear) requirements in the number of gates and qubits, which is exemplified by the credibility of the claims to have simulated the so-called ``IBM quantum utility experiment'' on a Commodore 64 computer with 64kB of RAM~\cite{anonymous2024quantum}. The approach can also be efficiently parallelized as was demonstrated in Ref.~\cite{rudolph2023classical} simulating that same quantum experiment. The major disadvantage of DFS is its runtime which is substantially larger than BFS due to the impossibility of merging paths. 

BFS computes the Pauli sum at each point in the circuit. Commonly this causes an exponential increase in the number of stored Pauli strings with the number of branching gates, and thus a strictly higher peak memory requirement than DFS. However, BFS allows for the \textit{merging} of paths where the current Pauli strings are identical by simply adding their coefficients together and then continues propagating them as one. This not only reduces the self-inflicted memory requirements relative to naive BFS, but crucially also the runtime relative to DFS. This is because numerous Pauli paths have collapsed into one, with every subsequent gate application effectively being parallelized among all merged paths. We call this approach \textit{merging-BFS}.

\begin{figure}
    \centering
    \includegraphics[width=1\linewidth]{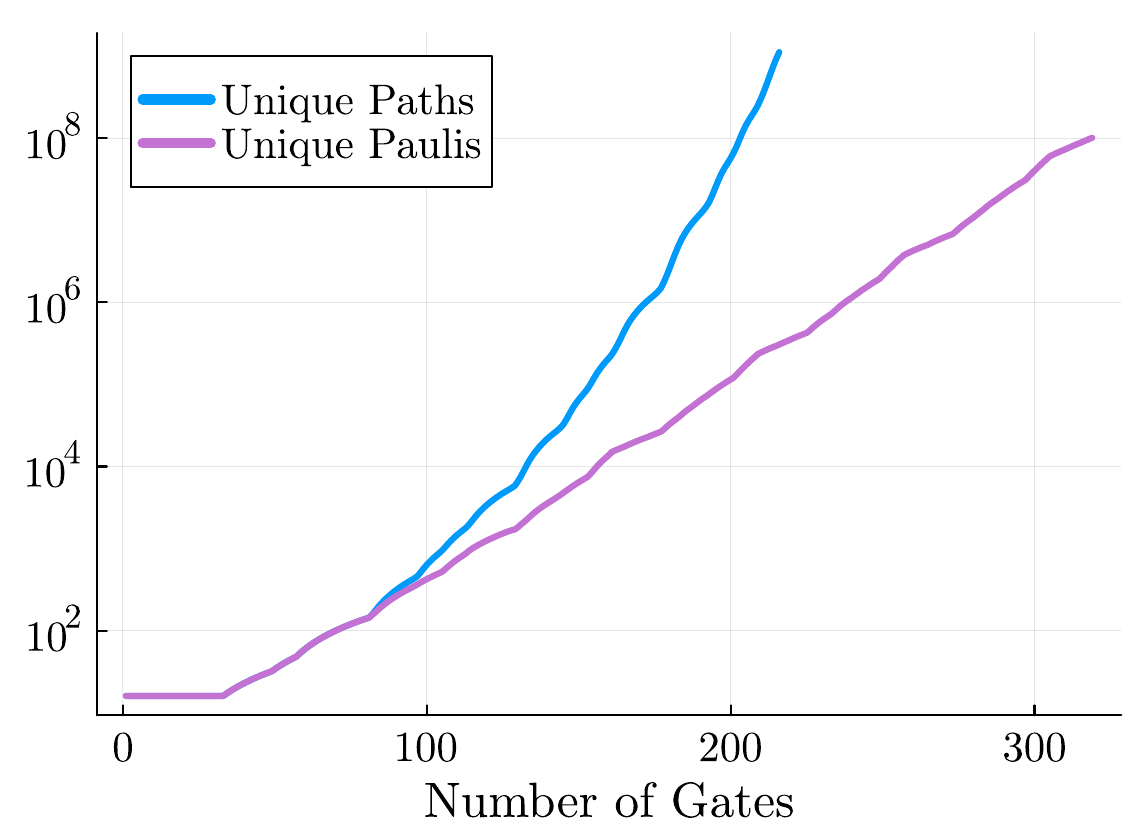}
    \caption{\textbf{The importance of merging Pauli paths}. We consider a 16-qubit quantum circuit consisting of layers of RX, RZ and RZZ Pauli rotations on a periodic bricklayer topology and compare the number of Pauli paths to the number of unique Pauli strings in the simulation. This can be understood as the comparison between DFS or BFS, and the merging-BFS approach. The initial observable is the sum of single-qubit Z Pauli strings, which initially gives rise to operator lightcone effects. After the initial stages, the number of paths scales with a stronger exponential in the number of Pauli rotations, which translates into a relative slowdown of pure DFS/BFS approaches without merging. No truncations are employed here.}
    \label{fig:merging}
\end{figure}

Merging-BFS trades a likely exponential memory disadvantage for likely strong polynomial speed-up over DFS. Considering the common case of $m$ 2-branching Pauli rotations, the runtime measured by the number of paths can be up to $2^m$. In practice, as discussed in Section~\ref{sec:gates&ops}, the exponent will not be exactly $m$, given that gates can commute with Pauli strings, especially for local Pauli strings consisting of mostly identity Paulis. Even so, the number of unique Pauli strings that merging-BFS would compute is upper-bounded by that same number, but for non-pathological cases, the rate of unique Pauli strings will be drastically lower than the number of paths. This is exemplified in Fig.~\ref{fig:merging}. In the limit of deep circuits, merging-BFS is limited by the full Pauli space, i.e., $4^n$ unique Paulis, wheras the number of paths $\OC(2^m)$ can keep growing. 
Thus, even simulating circuits with a few qubits, e.g., $n \lessapprox 12$, could run eternally using DFS if there are enough non-Clifford gates. For a schematic representation of the tree traversal algorithms we point to Fig.~\ref{fig:traversal}.

In the PauliPropagation.jl package we currently optimize for runtime over memory and focus on a BFS strategy. Indeed, a purely DFS strategy will require impossibly demanding runtimes except for simple case studies where truncations can be tailored to only explore relatively few contributing paths~\cite{rudolph2023classical}.
Eventually, we may need to develop hybrid DFS and (merging-)BFS approaches, potentially involving randomized techniques such as \textit{Monte Carlo} sampling of paths. While currently not explored, such approaches may become necessary when memory becomes the limiting factor. Merging-BFS Pauli propagation can fill 1TB of memory in a few hours using a single CPU thread when simulating physical problems of interest. Future versions of this code may be parallelizable over plenty CPU threads to the point where typical computational cluster nodes can run out of memory in little time. Given this context, departing from purely merging-BFS for hard simulations seems to be a certainty, but how to combine the best of all worlds is currently not clear.

\subsection{Error estimates}

When pushing the boundaries of classical simulation, there remains the question of how accurate the results are. It is often not clear how much total error has been incurred via systematic approximations and truncations, and how \textit{converged} the simulations are in the relevant truncation parameters. This is a challenge for all large-scale classical simulation methods, including Pauli propagation and tensor networks, as well as non-error-corrected quantum devices. 

Two approaches for estimating errors in quantum circuit simulations are: i) using the known error per gate to upper-bound the total error via the triangle inequality, and ii) monitoring convergence by observing that quantities of interest stabilize as truncation thresholds are relaxed. 
The first approach of bounding errors per gate is, for example, used widely in tensor network simulations. For instance, in time-dependent variational principle (TDVP) algorithms, a gate is absorbed into the state, followed by a singular value decomposition (SVD). Small singular values are discarded to truncate the bond dimension, and the sum of the squared discarded singular values directly quantifies the infidelity of the gate application. This infidelity can then be used to bound deviations in expectation values. A similar strategy can be applied to Pauli propagation, which we discuss next.

\medskip

\begin{theorybox}[label={box:worst}]{Bounding the error per truncation}
Assume we want to approximate $\Tr[O \mathcal{C}(\rho_0)]=\Tr[\mathcal{C}^\dag(O)\rho_0]$, where $\mathcal{C} = \mathcal{C}_L \circ \mathcal{C}_{L-1}\circ \dots \circ \mathcal{C}_1$ is an $L$-layered quantum circuit.
Using Pauli Propagation, we can approximate the observable $\mathcal{C}^\dag(O)$ by iteratively back-propagating and truncating $O$ layer by layer. Given $O = O_{L}$, we start by computing the observable $\mathcal{C}_L^\dag(O_{L}) = \sum_{\alpha} c_\alpha P_\alpha$ and we truncate some of its Pauli terms, obtaining the observable $O_{L-1}$.

We can upper bound the truncation error via the triangle inequality:
\begin{align}
    &\left\lvert{\Tr[(O_{L-1} - \mathcal{C}_L^\dag(O_{L}) )\rho_{L-1}]} \right\rvert \\\leq &\sum_{\substack{\text{truncated} \\ \text{Paulis}}}\abs{c_\alpha} \coloneqq \Delta_L,
\end{align}
where $\rho_{L-1} = \mathcal{C}_{L-1}\circ \dots \circ \mathcal{C}_1(\rho_0)$. Iterating the same argument over $L$ layers and applying again the triangle inequality, we find that the total error is upper bounded by
\begin{align}
    \Delta \coloneqq \Delta_{L} + \Delta_{L-1} +\dots +\Delta_1,
\end{align}
where $\Delta_j$ is the $\ell_1$-norm  
discarded during the $(L-j + 1)$-th truncation.
We stress that this is a \emph{worst-case} bound, as we didn't make any assumption on the initial state $\rho_0$ or the circuit layers $\mathcal{C}_j$.

Under certain assumptions, this bound can be further tightened: if we want to approximate a two-point correlation function $2^{-n}{\Tr[O \mathcal{C}(P_0)]}$, the error per truncation can be bounded by the $\ell_2$ norm 
\begin{align}
    &\frac{1}{2^n}\left\lvert{\Tr[(O_{L-1} - \mathcal{C}_L^\dag(O_{L}) )P_{L-1}]} \right\rvert 
     \\\leq  
   &\sqrt{\sum_{\substack{\text{truncated} \\ \text{Paulis}}}c_\alpha^2 },
\end{align}
where $P_{L-1} = \mathcal{C}_{L-1} \circ \dots \circ \mathcal{C}_1(P_0)$ and we applied Cauchy-Schwarz's inequality.
A similar bound based on the $\ell_2$-norm also holds \emph{on average} when the initial state
$\rho_0$ is a random computational basis state\ \cite{schuster2024polynomial}.
\end{theorybox}

\emph{Error per gate.} 
During unitary evolutions in Pauli propagation, the 2-norm of the observable $O = \sum_\alpha c_\alpha P$ --  i.e. $2^{-n}\Tr[O^2] = \sum_\alpha c_\alpha^2$ --  is conserved and any loss in that can be directly measured. Unfortunately, however, it is the $\ell_1$-norm of the coefficients $c_\alpha$ 
that directly bounds the additive error on expectation values, as discussed in Theory Box~\ref{box:worst}.

While it is encouraging that Pauli propagation appears to perform significantly better currently known error estimates, it also means that having provable trust in Pauli propagation results is difficult to obtain.
Remarkably, although non-trivial worst-case error bounds have been established for restricted circuit classes that generate low magic or operate in high-noise regimes~\cite{rall2019simulation, gonzalez2024pauli, lerch2024efficient}, a series of works has shown that average-case analyses can capture a broader ensemble of circuits under both noiseless and noisy conditions~\cite{aharonov2022polynomial, schuster2024polynomial, angrisani2024classically, angrisani2025simulating}.  However, it is crucial to interpret these error guarantees in context. As we illustrate with a simple toy example in Box~\ref{box:worst-vs-avg},  worst-case and average-case errors can be within a multiplicative factor which is exponential in system size.

\begin{theorybox}[label={box:worst-vs-avg}]{Worst-case vs Average-case errors}
We provide two simple toy examples showing that worst-case and average-case errors can be substantially different.

\smallskip
\noindent \textbf{Weight truncation.} Let $O = Z^{\otimes n} + Z\otimes I^{\otimes (n-1)}$. Approximating $O$ via the \emph{weight-truncation} rule with some cutoff $k\geq 2$, we obtain $O' = Z\otimes I^{\otimes (n-1)}$. The truncation error is $1$ if we take a computational basis state as input, but it is exponentially small on average over tensor products of \emph{random} single-qubit stabilizer states:
\begin{align}
    &\abs{\Tr[(O-O')\ketbra{0^n}{0^n}]} = 1,
    \\& \mathbb{E}_{\ket{\psi} \sim \mathrm{Stab}_1^{\otimes n}} \abs{\Tr[(O-O') \ketbra{\psi}{\psi}]} = 3^{-n},
\end{align}
where $\mathrm{Stab}_1 =\left\{\ket{0}, \ket{1}, \ket{+}, \ket{-}, \ket{+}_y, \ket{-}_y\right\}$.

\smallskip

\noindent \textbf{Small-coefficient truncation.} Let $O =  Z_1 + \frac{1}{n-1} \sum_{i=2}^n Z_i$, where $Z_i$ denotes the Pauli string acting as $Z$ on the $i$-th qubit and as identity elsewhere. Approximating $O$ via the \emph{small-coefficient truncation} rule with some cutoff $\tau > \frac{1}{n-1}$, we obtain $O' =  Z_1  = Z\otimes I^{\otimes (n-1)}$. The truncation error is $1$ if we take the computational zero state as input, but it scales inverse-polynomially if instead we consider \emph{random} computational basis states:
\begin{align}
    &\abs{\Tr[(O-O')\ketbra{0^n}{0^n}]} = 1,
    \\&\mathbb{E}_{\boldsymbol{x} \sim \{0,1\}^{n}} \Tr[(O-O')\ketbra{\boldsymbol{x}}{\boldsymbol{x}}]^2 
    \\ \leq &\Tr[(O-O')^2 \left(\frac{\mathbb{I}}{2^n}\right)] = \frac{1}{n-1}.
\end{align}
\smallskip

These examples highlight the gap between worst-case and average-case error bounds. While both offer guidance for choosing a simulation strategy, they must be interpreted with care and in context.
\end{theorybox}

\medskip

\emph{Monte Carlo error estimates.} One rather remarkable capability of Pauli path techniques is to provide efficient protocols for estimating errors on average over an ensemble of circuit parameterizations. Consider the case of Pauli rotations multiplying sine and cosine coefficients onto Pauli paths. We remember that $ 1 \leq |\sin(\theta)| + |\cos(\theta)| \leq \sqrt{2}$ but $\sin(\theta)^2 + \cos(\theta)^2 = 1$. This implies that while expectation values cannot in general be estimated efficiently with Monte Carlo sampling of Pauli paths due to the exponentially increasing 1-norm of the coefficients, second order quantities such as variances and mean square errors can be efficiently estimated, as discussed in more detail in Box~\ref{box:MC}. However, it is important to stress that these error estimates are, as mentioned above, \textit{on average} over an ensemble of circuit parameters under the condition that Pauli rotations are \textit{independently parametrized}. While this can then be used to demonstrate efficient simulability on average over certain families of circuits, for any particular set of parameters, especially those not randomly drawn from a distribution, the Monte Carlo estimates are likely to drastically under-estimate the error.

\begin{theorybox}[label={box:MC}]{Monte Carlo error estimate}
Here we convey the general mathematical intuition behind the Monte Carlo error estimate, which is described in detail in  App.~H of Ref.~\cite{angrisani2024classically}.
Given a circuit $U(\boldsymbol{\theta})$ parametrized by $\boldsymbol{\theta} = (\theta_1, \theta_2,\dots, \theta_m)$, an observable $O$ and a state $\rho$, we aim at approximating the expectation value $
\langle O(\boldsymbol{\theta}) \rangle \coloneqq\Tr[OU(\boldsymbol{\theta}) \rho U(\boldsymbol{\theta})^\dag]$, which can be expressed as a sum over Pauli paths as follows
\begin{align}
   \langle O(\boldsymbol{\theta}) \rangle  =  \sum_{\text{Pauli paths }\gamma} \Phi_\gamma(\boldsymbol{\theta}) \Tr[P_\gamma \rho]\,.
\end{align}
$\Phi(\gamma)$ are the coefficients associated to the paths $\gamma$ and $P_\gamma$ is the last Pauli string resulting from the path $\gamma$. These paths can be understood as branches of the propagation tree without merging.

When approximating the true expectation value via $\langle O^{(\mathcal{S})}(\boldsymbol{\theta})\rangle$ with a truncated set of paths $\mathcal{S}$, we can efficiently estimate the Mean Square Error (MSE) $ \mathbb{E}_{\boldsymbol{\theta}} \left({\langle {O}(\boldsymbol{\theta}) \rangle} - {\langle {O^{\mathcal{(S)}}}(\boldsymbol{\theta}) \rangle} \right)^2$ on average over the parameters $\thv$. Assuming that the angles $\theta_i$ are \emph{independently} sampled, e.g., $\boldsymbol{\theta}$ is sampled uniformly from $(-\pi,\pi)^m$, the MSE can be rewritten in a strikingly simple form:
\begin{align}
    \mathrm{MSE} =
    &\sum_{\text{Pauli paths }\gamma} \underbrace{\mathbb{E}_{\boldsymbol{\theta}}\Phi_\gamma^2(\boldsymbol{\theta}) }_{\Pr[\gamma]}\underbrace{\mathbf{1}_{\gamma \not \in \mathcal{S}}  \Tr[P_\gamma \rho  ]^2}_{g(\gamma)}\\ = &\mathbb{E}_\gamma g(\gamma) \label{eq:mse},
\end{align}
where $\mathbf{1}_{\gamma \not \in \mathcal{S}}$ denotes the indicator function which equals 0 if $\gamma$ is in  $\mathcal{S}$ and 1 otherwise.
This characterization of the MSE ensures that it can be estimated with high probability and arbitrarily small error $\epsilon$ by sampling $\mathcal{O}(\epsilon^{-2})$ Pauli paths with probability $\Pr[\gamma]\coloneqq \mathbb{E}_{\boldsymbol{\theta}}\Phi_\gamma^2(\boldsymbol{\theta})$ and averaging the associated values of $g(\gamma)$. In practice, the average error may be smaller due to the merging of Pauli paths, but also a lot larger for particular choices of $\thv$, i.e., this value is not representative of the worst-case error.
\end{theorybox}

\medskip

\emph{Apparent convergence.} A more concrete approach for estimating errors in approximate simulations is testing for convergence in the truncation values. It is reasonable to assume that results, e.g., expectation values, stabilize on any fixed scale when more and more computational resources are spent. This is strictly true for unbiased estimators, but it is still a working heuristic in the domain of tensor networks where truncation via bond dimension biases calculations. Many state-of-the-art results are attained by \textit{extrapolating} from finite bond dimension simulations to the infinite bond dimension regime (e.g., Ref~\cite{rams2018precise}, but it is common practice), yet there is no theoretical guarantee that this should work as well as it does. Extrapolating in Pauli propagation truncations appears more challenging in the Pauli picture where we commonly see oscillating expectation values rather than monotonically converging ones.

\begin{figure}
    \centering
    \includegraphics[width=1\linewidth]{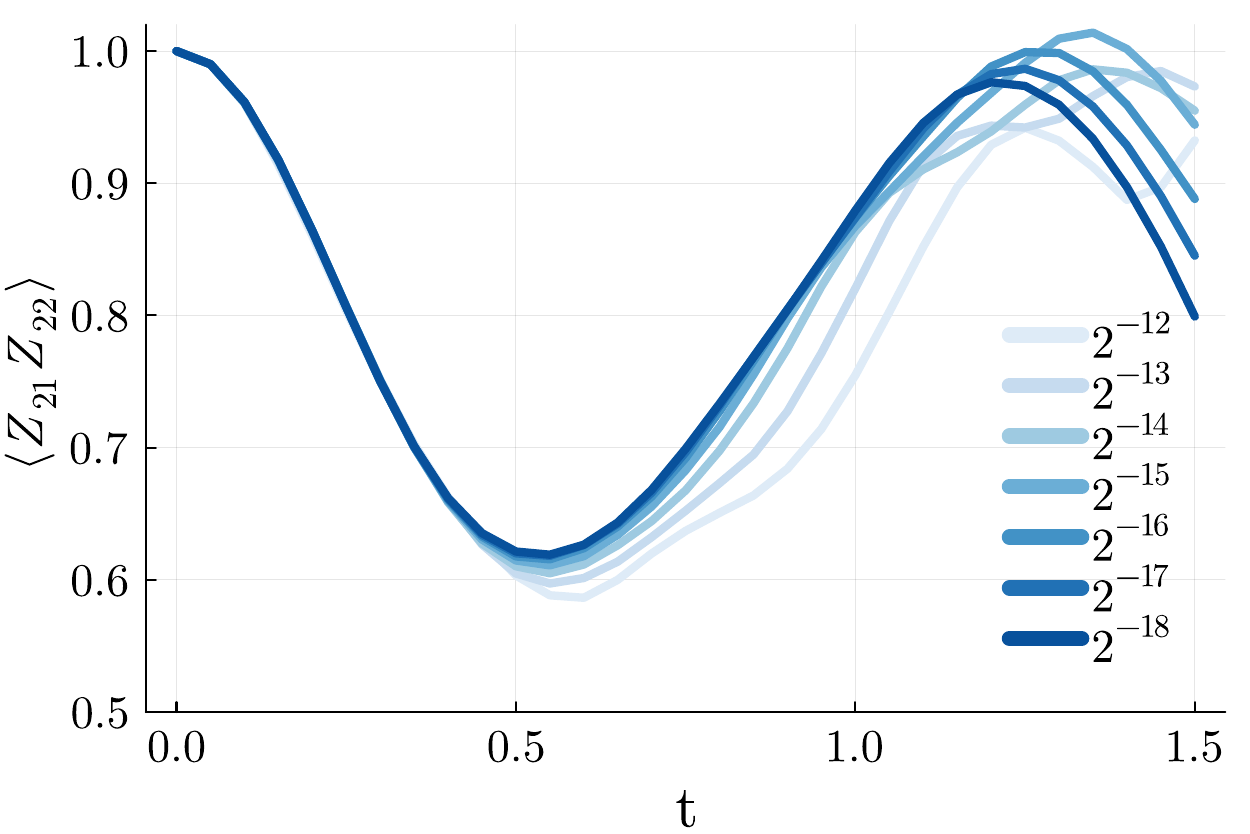}
    \caption{\textbf{Testing convergence in Pauli propagation}. This is a 6x6 qubit system evolving under the so-called tilted-field Ising Hamiltonian consisting of Z, X and ZZ terms. We choose a time step of $dt=0.05$ and simulate up to 30 Trotter layers. The different lines are different choices of the minimum absolute coefficient threshold. An additional Pauli weight truncation of 9 is imposed, which we have not witnessed visually affecting the results on this time scale. We observe apparent convergence in the sense that the different simulations agree until longer and longer times, which increases trust in certain features in time. }
    \label{fig:convergence}
\end{figure}

What remains is to visually observe convergence, as exemplified in Fig.~\ref{fig:convergence}. Choosing lower and lower coefficient truncation cutoffs, requiring additional computational resources, we see the expectation values in the real-time dynamics of a $6 \times 6$ system under the tilted-field Ising Hamiltonian coincide for longer and longer times.
That being said, it is not a priori clear what a \textit{good} accuracy is. The general features of the dynamics on this timescale can arguably be recovered with a threshold in the range $2^{-14} - 2^{-16}$, which for us took between 35 seconds and 5 minutes with off-the-shelf PauliPropagation.jl code and a single CPU thread. These truncations produced between 1.2 million and 9.3 million unique Pauli strings. The precise expectation values between $t=1.0$ and $t=1.5$ seemingly require a cutoff of $2^{-18}$ or below, which took 26 minutes with a single CPU thread and produced just under 50 million Pauli strings.

\section{Under the hood}

This section considers the implementation nuances of Pauli propagation such as choices of data structures and high-performance optimisations. We will both motivate \ppjllink's implementation choices and propose further generalisations relevant to the framework.

The goal of a high-level library like \ppjl{} is to abstract many of the technical complications away, freeing users from worrying about low-level optimizations. This section will lift the veil to some of the challenges under the hood: the choice of suitable data structures, memory management, control flow, parallelisation and compilation. Such considerations are essential in the development of high-performance simulators like \ppjl{}.

\begin{glossarybox}[label={box:glossary}]{CS jargon}
Explaining the internal engine of Pauli propagation requires using language from the computer science (CS) community. This box clarifies several such terms used below.

\medskip

\textbf{\emph{Amortised complexity}} averages the computational complexity of many \textit{sequential} operations, as an alternative to the often overly pessimistic worst-case complexity. This is useful for capturing the practical performance of growing data structures, like when appending elements in a list (amoritsed $\OC(1)$ time), for which the worst-case complexity corresponds to an non-representative, rare event ($\OC(N)$ resizing).

\medskip

\textbf{\emph{Branching}}, in the context of a program's control flow, occurs when executing conditional instructions like \code{if} statements. This is distinct from when we say an operator induces ``branching" of a Pauli string.

\medskip

\textbf{\emph{Endianness}} indicates the ordering of bits in a binary number; whether the rightmost bit is the least (little-endian) or most (big-endian) significant. So too it defines the ordering of qubits in a quantum register and in our notation.

\medskip

\textbf{\emph{Enumeration}} involves iterating over the elements within a well-ordered container. In our context, ``enumerating a Pauli sum" refers to sequentially processing each constituent Pauli string.

\medskip

\textbf{\emph{Hot loops}} are control flow loops (like induced by \code{for}) with many iterations and which dominate runtime, creating a ``hotspot" in performance analysis. Accelerating the body of such loops is typically a priority during optimisation.
Like many quantum simulations, Pauli propagation involves exponentially-many repetitions of a numerical subroutine, forming a hot loop.

\end{glossarybox}

\begin{softwarebox}[label={box:paulistrings}]{Pauli String Example}
As an example, the high-level object representing the Pauli string $XIZY$ can be defined in \ppjllink{} via

\vspace{1mm}
\code{PauliString(4, [:X, :Z, :Y], [1, 3, 4])}
\vspace{1mm}

A Pauli sum object representing $0.5 \cdot XIZY - 1.3 \cdot IZII$ can be defined by 

\vspace{1mm}
\code{psum = PauliSum(4)} \\
\code{add!(psum, [:X, :Z, :Y], [1, 3, 4], 0.5)} \\
\code{add!(psum, :Z, 2, 1.3)} 
\vspace{1mm}

\end{softwarebox}

\subsection{Representing a Pauli string}\label{ssec:Paulis}

\begin{conventionbox}[label={box:BitEndianness}]{Bit endianness}
We notate binary numbers with the least significant bit \textit{on the right} (so called ``little-endian"). Where relevant, we use a subscript suffix to notate when a number is a base-$n$ numeral, for example 
$$6_{\color{gray}10} = 110_{\color{gray}2}.$$
Note this is the reverse of the big-endian convention so far adopted for notating Pauli strings, whereby the \textit{leftmost} qubit is the lowest index, e.g.
$$
Z X \equiv Z \otimes X \equiv Z_1 X_2 \, .
$$
Note too we index from \textbf{\textit{one}} (not zero) for both Pauli strings \textit{and} bits, such that the bit at index $1$ of $110_{\color{gray}2}$ is $0$. This further aligns with \texttt{Julia}'s indexing convention which  starts from one (not zero).

\end{conventionbox}

This manuscript has so far taken for granted the ability to symbolically denote a Pauli string, such as $X_1Y_3Z_4$, and perform operations on them.
We here describe how to perform this in a program, and the numerous considerations for memory and runtime efficiency. 

Let us first summarise the relevant properties and demands of the Pauli string structure in the Pauli Propagation framework.
\begin{enumerate}[i]
    \item \textit{Limited alphabet:} A Pauli string has an ``alphabet" of \textit{four}: $I$, $X$, $Y$ and $Z$. Since $n$ bits can enumerate $2^n$ symbols, a single Pauli can be optimally represented by two bits (a ``dibit" or ``crumb"). In contrast, an intuitive but wasteful representation as an ASCII character in a string like \code{"I", "X", "Y", "Z"} would consume eight bits. We seek a representation which takes advantage of the limited alphabet to save memory.

    \item \textit{Bounded length:} Our Pauli strings contain only as many operators as there are qubits in the simulated circuits. This empowers us to bound the number of represented operators in the string, for example to $128$ qubits. Such a ``compile-time" bound enables the use of a static structure (rather than some dynamically growing structure) which asymptotically improves some operations and invites significant compiler optimisations. 
    
    \item \textit{Primitive:} During Pauli propagation, we will instantiate extremely many Pauli strings, motivating the use of a primitive type (like an integer) over an object (like a string or a custom class). This shrinks memory costs and avoids the substantial runtime overheads of object construction and garbage collection. It also permits a Pauli string to fit into a single CPU register for faster processing, and may enable a clever compiler (like \texttt{Julia}'s) to autovectorise looped processing, such as when we process all strings within a large Pauli sum.

    \item \textit{Fast operations:} Operations between Pauli strings, like products and commutators, must be \textit{as fast as possible} - ideally in effectively $\OC(1)$ time - since they are repeatedly performed in ``hot loops", i.e., performance-critical loops over many fast operations. This necessitates the use of a type compatible with single-instruction-multiple-data (SIMD) processing, such that multiple Paulis in a string are processed simultaneously.

    \item \textit{Individually addressable:} The Paulis in our string must be \textit{individually addressable} so that we can quickly obtain a Pauli at a particular index, ideally in $\OC(1)$ time. This is necessary to implement an operation on a particular qubit/site of a string.

    \item \textit{Quickly comparable:} Pauli strings must be \textit{quickly comparable} to keep merging-BFS efficient. That is, we should be able to in $\mathcal{O}(1)$ time decide whether two strings are identical. In Sec.~\ref{sec:representing_pauli_sum}, we will additionally motivate that our Pauli string type is quickly ``{hashable}" for integration with containers like dictionaries, or can be \textit{sorted} for containers based on arrays.
    
\end{enumerate}

For these reasons, a natural choice of type for a Pauli string is one single \textit{unsigned integer}; the representation of the string as a base-four numeral. In this picture, each bit pair denotes a Pauli operator, and half as many Paulis can be represented as the integer type has bits. For example,
\begin{gather*}
    \text{ZYX} 
        \;=\;
        27_{\color{gray} 10}
        \;=\;
        011011_{\color{gray} 2}
        \\
    =
        \underbrace{\texttt{00}}_{\text{I}} \; 
        \dots \;
        \underbrace{\texttt{00}}_{\text{I}} \;
        \underbrace{\texttt{01}}_{\text{X}} \;
        \underbrace{\texttt{10}}_{\text{Y}} \;
        \underbrace{\texttt{11}}_{\text{Z}} .
\end{gather*}
We choose to label the Pauli strings and bitstrings according to Convention Box~\ref{box:BitEndianness}, and denote the base of the numeral via gray subscripts for clarity.
The reason there are seemingly unnecessary identity Paulis on the left is because, depending on the bit-length of the chosen integer type, bits in excess of $2n$ will remain zero.  
We ensure optimal efficiency by choosing an integer precision inferred from the number of qubits in the user's simulated circuit, without compromising the compile-time bounds. This is done with type templating: At the time of instantiation of a \code{PauliString} or \code{PauliSum} object, we check whether suitable built-in integer types exist and otherwise generate the next higher precision type. \texttt{Julia}'s multiple dispatch will just-in-time (JIT) recompile any propagation-related function which is specialised upon the type, to run as fast as with a built-in, hardcoded integer type. All propagated Pauli strings will use the same precision of integer, avoiding type instability.

The representation of Pauli strings as unsigned integers is adopted by other simulators such as \texttt{QuEST}~\cite{jones2019quest}, \texttt{Stim}~\cite{gidney2021stim}, as well as the related \texttt{PauliStrings.jl}~\cite{loizeau2025quantum} and Ref.~\cite{beguvsic2024real}. Some works maintain \textit{two} unsigned integers which each represent or ``mask" the location of $X$ and $Z$ operators in a Pauli string. The location of a $iY = -X\cdot Z$ operator is consequently indicated by a 1-bit in the same location of both masks. This representation bears closer similarity to the design choices in the stabilizer formalism~\cite{aaronson2004improved}.
Both representations satisfy all our previous considerations and, as we will discuss in the next section, enable rapid processing of Pauli strings with bitwise operations.
We focus on the single integer representation for the remainder of this paper since this is adopted by PauliPropagation.jl.

\subsection{Processing a Pauli string}\label{sec:bit-operations}

\begin{datastructbox}[label={box:bitwiseops}]{Bitwise Operations}
Bitwise operations are defined by their operation directly upon the bits of a number. Particularly relevant to processing binary-encoded Pauli strings are: AND (\code{\&}), OR (\code{|}), XOR (\code{$\oplus$}), left-bitshift (\code{<<}) and  right-bitshift (\code{>>}). For our purposes, their main utility is that they simultaneously operate upon every corresponding pair of bits of the operands.
We can in-fact uniquely define the above bitwise operations by their actions upon operands $a,b$:
\begin{align*}
    &a = 0011_{\color{gray} 2} = 3_{\color{gray} 10} & & \hspace{-.7cm} a \;\code{\&}\; b = 0001_{\color{gray} 2}  \\
    &b = 0101_{\color{gray} 2} = 5_{\color{gray} 10} & & \hspace{-.7cm} a \;\code{|}\; b = 0111_{\color{gray} 2} \\
    & & & \hspace{-.7cm} a \;\code{$\oplus$}\; b = 0110_{\color{gray} 2} \, .
\end{align*}

AND, OR and XOR are useful for ``masking"; i.e, retaining, setting to one, or flipping specific bits respectively. To do so, one prepares another integer $m$ (the mask) with one (``set") bits at the locations of the targeted bits. For example, 
the mask $m = 0 \dots 01001_{\color{gray} 2}$ targets the \textit{first} and \textit{fourth} bits.
The integer $a\,\code{\&}\,m$ retains only the targeted bits of $a$, $a\,\code{|}\,m$ has additional ones at the target indices, and $a\,\code{$\oplus$}\,m$ flips all targeted bits in $a$.
\\[.2cm]
Preparing a mask is done with the left-bitshift operation, $a \,\code{<<}\, n$, which simply left-shifts the set bits of $a$ by the specified number of positions $n$. E.g.
\begin{align*}
    0011_{\color{gray} 2} \;\code{<<}\; 1 &= 0\textbf{11}0_{\color{gray} 2}
    \\
    0011_{\color{gray} 2} \;\code{<<}\; 3 &= \textbf{1}000_{\color{gray} 2}.
\end{align*}
Note, the leftmost $n$ bits of the $m$ in $a$'s full binary representation are discarded. Otherwise
\begin{equation*}
    a \; \code{<<} \; n \;\equiv\; a \times 2^n \, \, \, \text{for} \, \, \, a < 2^{m-n} \, .
\end{equation*}
The mask $m = 0 \dots 01101_{\color{gray} 2}$ can be prepared via
$$
m =
    (1_{\color{gray} 10} \,\code{<<}\,0) \,\code{|}\,
    (1_{\color{gray} 10} \,\code{<<}\,2) \,\code{|}\,
    (1_{\color{gray} 10} \,\code{<<}\,3).
$$
\end{datastructbox}

Encoding Pauli strings as binary integers is not only memory efficient, but permits us to process them using so-called \textit{bitwise} operations. Such low-level operations, illustrated in Under~The~Hood~Box~\ref{box:bitwiseops}, correspond one-to-one with CPU instructions and are processed in a dedicated arithmetic logic unit (ALU). Requiring a single clock cycle, they are extremely efficient -- the fastest among arithmetic instructions (such as the addition of two integers), which are again orders of magnitude faster than floating-point operations (such as the addition of two decimals). This is regardless of the number of represented Pauli operators in the string, provided that the integer type fits within a single CPU register.
In this section, we highlight several operations upon Pauli strings that are implemented by \ppjl{}.
These are similarly spirited but distinct from the optimisations recently added to \texttt{PauliStrings.jl} which performs analogous schemes on two distinct masks. 

\medskip

\emph{Preparing a Pauli string.}
The integer $0$ with bits $\code{00\dots00\,00}_{\color{gray}2}$ represents the all-identity Pauli string. Preparing a string where the $i$-th Pauli is \textit{not} the identity requires overwriting the bits at indices $2i$ and $2i-1$ with the decimal value $1,2,3$ (for $X$, $Y$, $Z$ respectively). We do this by \textit{bit-shifting} the decimal value, as described in Under~The~Hood~Box~\ref{box:bitwiseops}, by \textit{twice} the index. For example,
\begin{equation*}
    Y_3 
    \equiv \code{00\,\textbf{10}\,00\,00}_{\color{gray}2}
    \;=\;
    \code{10}_{\color{gray}2} \,\code{<<}\, 4
    \;=\;
    \code{2}_{\color{gray}10} \,\code{<<}\, 4\,.
\end{equation*}
In general, the Pauli string with a single non-identity Pauli $\sigma\in\{1,2,3\}$ on site $i$ is computed by $\sigma_{\color{gray}10} \;\code{<<}\; 2(i-1)$. Preparing a Pauli string with \textit{multiple} non-identity Paulis involves \textit{masking} their individual bitmasks using the bitwise OR operator, also described in Under~The~Hood~Box~\ref{box:bitwiseops}. For example
\begin{equation*}
     Z_1Y_3 
     \equiv \code{00\,\textbf{10}\,00\,\textbf{11}}_{\color{gray}2}
    \;=\;
    (2_{\color{gray}10} \,\code{<<}\, 4) \, \code{|} \,
    (3_{\color{gray}10} \,\code{<<}\, 0).
\end{equation*}

\emph{Setting a Pauli.}
Modifying the $i$-th Pauli in a Pauli string instantiated with integer \code{P} requires overwriting the bits at indices $2i$ and $2i-1$. When $\code{P}=0$, this can be done using the bitwise OR operator as above, but otherwise requires first \textit{unsetting} the targeted bits to zero. This is achieved using the \textit{bitwise negation} operator (\code{$\sim$}) which unconditionally flips all bits.
Consider 
$$m = 00 \dots \overbrace{11}^{2i, 2i-1} \dots 00 = 11_{\color{gray}2} \code{ << } ( 2(i-1) ),$$ 
which masks the targeted bits. The integer $ \code{$\sim$} m= 11 \dots 00 \dots 11 $ masks \textit{every other} than the targeted bits. As such $\code{P} \code{ \& } (\code{$\sim$} m)$ retains all ones of \code{P} \textit{except} at the targeted bits which are overwritten to zero; this is equivalent to overwriting the $i$-th operator with identity $I$. The desired Pauli ($\sigma \in \{0,1,2,3\}$) can then be set using bitwise OR as before,
\begin{gather*}
    \code{P}' =
    (\code{P \& } ( \code{$\sim$} m) )
        \code{\,|\,} 
    (\sigma \,\code{<<}\, ( 2(i-1) ) \, .
\end{gather*}

\medskip

\emph{Extracting a Pauli.} 
Consider a Pauli string instantiated with an unsigned integer \code{P}. The $i$-th Pauli operator $\code{P}^{(i)} \in \{0,1,2,3\}$ (which encodes respective symbols $I,X,Y,Z$) is constituted by the two bits at indices $2i$ and $2i-1$. We can extract the Pauli by bit-shifting this pair to the lowest two positions, then \textit{masking} them through the bitwise AND operation; see Under~The~Hood~Box.~\ref{box:bitwiseops} for an elaboration. The individual Pauli is given by
    \begin{align}
        \code{P}^{(i)} =
        (\code{P >> } 2(i-1)) \code{ \& } 11_{\color{gray}2},
    \end{align}
where \code{>>} is a \textit{right}-bitshift (the opposite of the left shift).

\medskip

\medskip

\emph{Products of Paulis.} Obtaining the product of two Pauli strings is a common subroutine of Pauli propagation, e.g., when applying Pauli rotations.
We recall that the product of two single-qubit Paulis $P \in \{X, Y, Z\}$ is given by
\begin{equation}
    P_\alpha P_\beta = \delta_{\alpha \beta} I + i \epsilon_{\alpha \beta \gamma} P_\gamma,
\end{equation}
where $\epsilon_{\alpha \beta \gamma} $ is the Levi-Civita tensor.
It follows that when two Pauli strings are encoded as unsigned integers \code{P1} and \code{P2}, then the encoded product \code{P3} (without prefactor) is simply given by bitwise XOR:
\begin{equation}
    \code{P3} \;\propto\; \code{P1 $\oplus$ P2}.
\end{equation}
For example, $XY \cdot ZI  = -\mathrm{i} \, YY$ and $\code{0110}_{\color{gray}2} \oplus \code{1100}_{\color{gray}2} = \code{1010}_{\color{gray}2}$. 
However, the prefactor of \code{P3} is also needed by Pauli propagation to obtain the coefficients of the final Pauli sum and evaluate the ultimate inner-product.
We currently evaluate a cached Levi-Civita tensor for each pair of bits in \code{P1} and \code{P2}, necessitating a loop over the sites affected by the Pauli product. Note that a constant-time bitwise evaluation of the sign of the \textit{commutator} of two Pauli strings is known~\cite{loizeau2025quantum}.

\medskip

\emph{Indexing into look-up tables}. A core subroutine of Pauli propagation is mapping a \textit{subset} of  $\mathcal{O}(1)$ sites in a Pauli string to one or more other substrings. This is beneficial, for example, when applying general PTMs representing circuit operations. While sometimes performed with hardcoded logic, such as when implementing a Pauli rotation's PT map as per Eq.~\eqref{eq:PauliRotationPtmAsCommutator}, output Pauli substrings are often instead obtained using \textit{look-up tables}. Consider a function $f$, e.g., here corresponding to a Clifford gate, that acts on $k$ qubits and thus maps $k$-qubit  substrings to another. That is, we have
\begin{equation}
    f: \PC^{(k)} \rightarrow \PC^{(k)}.
\end{equation}
When $k$ is tractably small, we can pre-compute $f(\code{P})$ for all $\code{P} \in \{0,\dots,2^k-1\}$, store the outputs in an array $\vec{f}$, and thus can replace future evaluations of $f$ with the lookup $\vec{f}[\code{P}+1]$. That is, we utilise that our Pauli substrings are instantiated as unsigned integers to ``directly address" $\vec{f}$, offsetting by $+1$ due to \texttt{Julia}'s indexing convention. 

For example, consider the PTM of the Hadamard gate:
\begin{equation}
\begingroup
\renewcommand*{\arraystretch}{1} 
    \text{PTM}_{H} = 
    \begin{pmatrix} 1&0&0&0 \\ 0&0&0&1 \\ 0&0&-1&0 \\ 0&1&0&0 \end{pmatrix}.
\endgroup
\end{equation}
While the action of $\text{PTM}_{H}$ on the Pauli string $X$  can be computed via matrix-vector multiplication, e.g.,
\begin{equation}
\text{PTM}_H \cdot (0,1,0,0)^T = (0,0,0,1)^T \equiv Z,
\end{equation}
it is faster to access a pre-computed lookup array
\begin{equation}
    \vec{f}_H \coloneqq
    [ I, Z, -Y, X ],
    \hspace{.8cm}
    \vec{f}_H[1 + 1] = Z \, ,
\end{equation}
where we use the fact that $X$ is instantiated by the unsigned integer $\code{P}=1$, and offset by $+1$ since indexing from one.

\medskip

\emph{Checking commutation.} It is critical to quickly determine whether a Pauli string commutes with another, given that this check is critical to whether Pauli rotations induce path branching within ``hot loops" (see Eq.~\ref{eq:PauliRotationPtmAsCommutator}). Two strings commute only if an \textit{even} number of their individual Paulis \textit{anti}-commute, which we can quickly determine using $\mathcal{O}(1)$ bitwise operations.

Consider single-qubit Paulis $a$ and $b$, respectively instantiated by two-bit integers with bits $(\code{a}_2,\code{a}_1)$ and $(\code{b}_2,\code{b}_1)$. Since single-qubit Paulis only commute when identical or one is the identity, it turns out that two single-qubit Paulis $a$ and $b$ \textit{anti}-commute when $c=1$, where
\begin{align}
c \coloneq
(\code{a}_1 \code{ \& } \code{b}_2) \code{ $\oplus$ } 
(\code{b}_1 \code{ \& } \code{a}_2).
\label{eq:bit_cond_for_single_paulis_to_anticommute}
\end{align}
This can be understood since $c$ is only ever $0$ (implying commutation) when the XOR operands are equal, occurring when $a=b$ or one of $a$ or $b$ is identity (both AND operands include a $0$ and so return $0$, yielding $0 \oplus 0 = 0$).

The above quantity is for a single site, but we can compute it for every corresponding pair of Paulis among many-qubit Pauli strings \code{A} and \code{B} by use of specialized bit masks.
We prepare (at compile-time) an alternating mask
\begin{equation}\label{eq:mr}
    m_R=01\dots 0101_{\color{gray}2},
\end{equation}
such that $\code{A} \code{ \& } m_R$ isolates every second bit of $\code{A}$; retaining only the \textit{right} bit of each individual Pauli within, interspaced by zero bits, i.e.,
$$
\code{A}_R \; \coloneqq \;
\code{A} \code{ \& } m_R \; \equiv \; (\dots, \, 0, \, \code{A}_3, \, 0, \, \code{A}_1).
$$
The mask $m_L=10\dots 1010_{\color{gray}2} = m_R \code{ << } 1$ similarly retains all \textit{left} bits, i.e.,
$$
\code{A}_L \; \coloneqq \;
\code{A} \code{ \& } m_L \; \equiv \; (\dots, \, \code{A}_4, \, 0, \, \code{A}_2, \, 0).
$$
We evaluate $\code{B}_R$ and $\code{B}_L$ similarly, then bit-shift them by one so that their non-zero bits of interest align with those of $\code{A}_L$ and $\code{A}_R$ respectively.
Therefore, the expression
$$
C \coloneqq
\left(
    \code{A}_R \code{ \& } (\code{B}_L \code{ >> } 1)
\right)
\oplus
\left(
    \code{B}_R \code{ \& } (\code{A}_L \code{ >> } 1)
\right)
$$
simultaneously evaluates the left-hand-side of Eq.~\eqref{eq:bit_cond_for_single_paulis_to_anticommute} for all Paulis within $\code{A}$ and $\code{B}$. The resulting integer
$$
C
\equiv (\dots, 0, \, c_3, \, 0, \, c_2, \, 0, \, c_1)
$$
consists of bits $c_i=1$ indicating that the $i$-th individual Paulis anti-commute.
The strings \code{A} and \code{B} therefore commute when $C$ contains an even number of one bits, which can ordinarily be computed in a \textit{single} bespoke CPU instruction such as the inbuilt \code{count\_ones()} in \code{Julia}.

\medskip

\emph{Computing Pauli weight.} Counting the number of non-identity Paulis in a Pauli string, necessary when performing Pauil weight truncation, is equivalent to counting the number of Pauli bit-pairs that are not $\code{00}_{\color{gray}2}$. Consider a single-qubit Pauli $\code{b}$ with bits 
$\code{b} = (\code{b}_L,\code{b}_R)$.
The expression \code{b$_L$ | b$_R$} is only $0$ when both bits are zero and so $\code{b}$ is the identity. Consider now a many-qubit Pauli \code{B} whereby \code{B >> 1} aligns the \textit{left} bit of the Pauli bit pairs with the \textit{right} bits:
\begin{align*}
    \code{B} &= (\dots, \, \code{B}_3, \, \code{B}_2, \, \code{B}_1)
        \\
    \code{B >> } 1 &= (\dots, \, \code{B}_4, \, \code{B}_3, \, \code{B}_2) \, .
\end{align*}
As such, the bits of \code{B|(B>>1)} indicate (with a set-bit) where neighbouring (and overlapping) pairs of bits contain at least one set-bit. 
We finally isolate every second bit (one from each non-overlapping neighbouring pair) using the alternating mask $m_R$ in Eq.~\eqref{eq:mr}. Hence we are left with the bitwise expression
\begin{align*}
    W = m_R \code{ \& } (\code{B} \code{ | } (\code{B} \code{>>} 1)).
\end{align*}

\medskip

\emph{Computing XY weight.} Counting the number of $X$ or $Y$ Paulis (excluding $Z$) in a Pauli string follows an almost identical approach to above, except that the bitwise OR operation is replaced with an XOR. This is because $X = \code{01}_{\color{gray}2}$ and $Y = \code{10}_{\color{gray}2}$ have different first and second bits, while $I = \code{00}_{\color{gray}2}$ and $Z = \code{11}_{\color{gray}2}$ have repeated bits. Counting, for example, YZ weight is even simpler, because one only needs to count how many bit pairs carry \code{1} on the left bit.

\subsection{Representing a Pauli sum}\label{sec:representing_pauli_sum}

\begin{datastructbox}[label={box:paulisum_datastrucs}]{Data structures}

We here summarise several data structures relevant to Pauli propagation. Let $m[i]$ signify the $i$-th memory address and white and gray boxes indicate filled and empty memory slots respectively.

\medskip

\emph{Arrays} store a fixed number of directly addressable items contiguously in memory.

\vspace{-3mm}

    \begin{center}
    \begin{tikzpicture}
      \def\boxwidth{.6}
      \def\boxheight{.6}
    
      \foreach \i in {1,...,10} {
        \pgfmathsetmacro\x{(\i - 1) * \boxwidth}
        
        \draw[fill=white] (\x, 0) rectangle ++(\boxwidth, \boxheight);
        
        \node at (\x + 0.5*\boxwidth, 0.5*\boxheight) {$v_{\i}$};
        
      }
    
    \node at (0.5*\boxwidth, \boxheight+0.3) {\small $m[1]$};
    \node at (0.5*\boxwidth + 1.6*\boxwidth, \boxheight+0.3) {\small $\dots$};
    \node at (9.5*\boxwidth, \boxheight+0.3) {\small $m[10]$};
      
    \end{tikzpicture}
    \end{center}

\emph{Lists} permit appending items in amortised $\OC(1)$, inserting items at specific locations in $\OC(N)$, and can be implemented by an array with extra allocated memory. When exhausted, an entirely new array with additional capacity must be allocated and copied into, in $\mathcal{O}(N)$ time. In \code{Julia}, a 1D \code{Array} curiously follows this definition.

\vspace{-3mm}

    \begin{center}
    \begin{tikzpicture}
      \def\boxwidth{.6}
      \def\boxheight{.6}
    
      \foreach \i in {1,...,7} {
        \pgfmathsetmacro\x{(\i - 1) * \boxwidth}
        
        \draw[fill=white] (\x, 0) rectangle ++(\boxwidth, \boxheight);
        
        \node at (\x + 0.5*\boxwidth, 0.5*\boxheight) {$v_{\i}$};
        
      }

      \foreach \i in {8,...,10} {
        \pgfmathsetmacro\x{(\i - 1) * \boxwidth}
        
        \draw[fill=lightgray] (\x, 0) rectangle ++(\boxwidth, \boxheight);
        
        
      }
    
    \node at (0.5*\boxwidth, \boxheight+0.3) {\small $m[1]$};
    \node at (0.5*\boxwidth + 1.6*\boxwidth, \boxheight+0.3) {\small $\dots$};
    \node at (6.5*\boxwidth, \boxheight+0.3) {\small $m[7]$};
     \node[color=gray] at (6.5*\boxwidth + 1.45*\boxwidth, \boxheight+0.3) {\small $\dots$};
    
    \node[color=gray] at (9.5*\boxwidth, \boxheight+0.3) {\small $m[10]$};
      
    \end{tikzpicture}
    \end{center}

\medskip

\emph{Linked lists} can forego direct indexing by storing values alongside the memory address $p$ of the next value. This allows $\OC(1)$ insertion and deletion, but makes memory \textit{non-contiguous}, sabotaging caching performance and parallel enumeration.

\vspace{-3mm}

    \begin{center}
    \begin{tikzpicture}
      \def\boxwidth{.6}
      \def\boxheight{.6}

      \def\i{1}
      \pgfmathsetmacro\x{(\i - 1) * \boxwidth}
        \draw[fill=white] (\x, 0) rectangle ++(\boxwidth, \boxheight);
        \draw[fill=white] (\x + \boxwidth, 0) rectangle ++(\boxwidth, \boxheight);
        \node at (\x + 0.5*\boxwidth, 0.5*\boxheight) {$v_{1}$};
        \node at (\x + 1.5*\boxwidth, 0.5*\boxheight) {$p_{2}$};
        \node at (\x + 0.5*\boxwidth, \boxheight + 0.3) {\small $m[p_1]$};

      \def\i{4}
      \pgfmathsetmacro\x{(\i - 1) * \boxwidth}
        \draw[fill=white] (\x, 0) rectangle ++(\boxwidth, \boxheight);
        \draw[fill=white] (\x + \boxwidth, 0) rectangle ++(\boxwidth, \boxheight);
        \node at (\x + 0.5*\boxwidth, 0.5*\boxheight) {$v_{2}$};
        \node at (\x + 1.5*\boxwidth, 0.5*\boxheight) {$p_{3}$};
        \node at (\x + 0.5*\boxwidth, \boxheight + 0.3) {\small $m[p_2]$};

      \def\i{8}
      \pgfmathsetmacro\x{(\i - 1) * \boxwidth}
        \draw[fill=white] (\x, 0) rectangle ++(\boxwidth, \boxheight);
        \draw[fill=white] (\x + \boxwidth, 0) rectangle ++(\boxwidth, \boxheight);
        \node at (\x + 0.5*\boxwidth, 0.5*\boxheight) {$v_{3}$};
        \node at (\x + 1.5*\boxwidth, 0.5*\boxheight) {$p_{4}$};
        \node at (\x + 0.5*\boxwidth, \boxheight + 0.3) {\small $m[p_3]$};

      \foreach \i in {3, 6, 7, 10} {
        \pgfmathsetmacro\x{(\i - 1) * \boxwidth}
        \draw[fill=lightgray] (\x, 0) rectangle ++(\boxwidth, \boxheight);
      }

    \end{tikzpicture}
    \end{center}

\emph{Sorted lists} improve search from $\OC(N)$ to $\OC(\log(N))$. Insertion is $\OC(N)$ or $\OC(\log(N))$ when implemented with arrays or links, resp.
\medskip

\emph{Hash tables} are commonly used to implement \textit{dictionaries}. They store values of \texttt{(key,value)} pairs in a larger fixed-size array, at addresses informed by \textit{hashing} the keys with a many-to-one function $h$.
Assuming the ``load factor" (number of values divided by array capacity) is kept small, insertion is amortised $\OC(1)$.
When the load factor grows too large, the array is resized and all items are re-inserted at cost $\OC(N)$.

\vspace{-3mm}

    \begin{center}
    \begin{tikzpicture}
      \def\boxwidth{.6}
      \def\boxheight{.6}

      \def\i{2}
      \pgfmathsetmacro\x{(\i - 1) * \boxwidth}
        \draw[fill=white] (\x, 0) rectangle ++(\boxwidth, \boxheight);
        \node at (\x + 0.5*\boxwidth, 0.5*\boxheight) {$v_{3}$};
        \node at (\x + 0.5*\boxwidth - 0.2, \boxheight + 0.3) {\small $m[h(k_3)]$};

      \def\i{4}
      \pgfmathsetmacro\x{(\i - 1) * \boxwidth}
        \draw[fill=white] (\x, 0) rectangle ++(\boxwidth, \boxheight);
        \node at (\x + 0.5*\boxwidth, 0.5*\boxheight) {$v_{1}$};
        \node at (\x + 0.5*\boxwidth + 0.2, \boxheight + 0.3) {\small $m[h(k_1)]$};
        
      \def\i{7}
      \pgfmathsetmacro\x{(\i - 1) * \boxwidth}
        \draw[fill=white] (\x, 0) rectangle ++(\boxwidth, \boxheight);
        \node at (\x + 0.5*\boxwidth, 0.5*\boxheight) {$v_{5}$};
        \node at (\x + 0.5*\boxwidth, \boxheight + 0.3) {\small $m[h(k_5)]$};
        
      \def\i{10}
      \pgfmathsetmacro\x{(\i - 1) * \boxwidth}
        \draw[fill=white] (\x, 0) rectangle ++(\boxwidth, \boxheight);
        \node at (\x + 0.5*\boxwidth, 0.5*\boxheight) {$v_{4}$};

        \node at (\x + 0.1, \boxheight + 0.3) {\small $\dots$};
        
      \def\i{9}
      \pgfmathsetmacro\x{(\i - 1) * \boxwidth}
        \draw[fill=white] (\x, 0) rectangle ++(\boxwidth, \boxheight);
        \node at (\x + 0.5*\boxwidth, 0.5*\boxheight) {$v_{2}$};

      \foreach \i in {1, 3, 5,6, 8} {
        \pgfmathsetmacro\x{(\i - 1) * \boxwidth}
        \draw[fill=lightgray] (\x, 0) rectangle ++(\boxwidth, \boxheight);
      }

    \end{tikzpicture}
    \end{center}

\end{datastructbox}

\begin{datastructbox}[label={box:hpc}]{HPC considerations}

Different implementations of the same algorithm can have wildly different performances depending on the code's use of control flow and memory movement.
Here we summarise some considerations relevant to Pauli propagation.

\medskip

\emph{Branching} of the control flow due to \code{if} statements is fine in most contexts but can damage performance when inside hot loops. This is because conditional instructions interfere with the CPU's use of instruction-level parallelism (ILP) which may resort to ``branch prediction". When a CPU incorrectly guesses the condition outcome, it must undo speculatively executed instructions, interrupting execution.

\medskip

\emph{Caching} copies a subset of main memory into a smaller cache to speed up future accesses. A modern CPU has progressively larger caches as intermediaries to RAM, which are updated \textit{in batch}; one ``cache line" at a time. The main memory only needs to be consulted when the CPU requests data not already in a cache (a `miss'). Irregular access to data, like out-of-order indexing of an array, encourage expensive cache misses. HPC code accesses data in a predictable, contiguous manner.

\medskip

\emph{Multithreading} distributes a task between concurrent cores of a CPU(s). Efficient parallelisation requires the task is non-trivial (to outweigh thread invocation costs), can be homogeneously divided between cores, and that the divided tasks are mostly independent.

\end{datastructbox}

The previous sections motivated our representation of individual Pauli strings as unsigned integers. But Pauli propagation evolves rapidly growing \textit{weighted sums} of Pauli strings, necessitating the design of a \textit{container} data structure to store a large ensemble of the above discussed types, each coupled with a coefficient.
This section considers several choices of canonical data structures (introduced in UTH~Box~\ref{box:paulisum_datastrucs}) as motivated by the below list of considerations, and the performance considerations introduced in UTH~Box~\ref{box:hpc}.
We note that this list focuses on {merging-BFS} simulations, the main focus of \ppjl{} currently, and may differ for other variants of Pauli propagation. 
Below, we take $N$ to be the number of Pauli strings presently in the container, and we will often discuss \textit{amortised} asymptotic scaling, which can be understood as the \textit{worst-case averaged} over many (towards infinite) repetitions.

\begin{enumerate}[i]
    \item \textit{Memory efficient}:
    Propagation sees the ensemble of evolving Pauli strings rapidly grow, requiring the simultaneous storage of an extreme number of Pauli strings - easily in the billions. Due to the small memory footprint of each string represented as an unsigned integer, a non-negligible memory overhead \textit{per string} will multiplicatively exacerbate the memory costs.
    This encourages use of a \textit{contiguous memory} data structure with a zero amortised overhead per element, such as an array or hash table. 

    \item \textit{Compact}:
    There are $4^n$ unique Pauli strings with $n$ sites. When each is stored as an $n$-bit unsigned integer (ignoring storage of the coefficients), collectively storing all possible strings (like a full-state simulator) would require $n\, 4^n$ bits and quickly become prohibitive. For example, $n=16$ qubits would require at least $8\,$GB to even represent the Pauli strings, not to speak of the $64\,$GB for their real-valued coefficients. Yet, we seek to push simulation far beyond this regime, to circuits of hundreds of qubits which require simultaneous storage of a tractable but unpredictable subset of all possible strings.
    As such, we require that our ensemble data structure is \textit{compact}: that it can store a small fraction of all possible elements in a space without allocating memory for the full space.
    This eliminates ``directly addressed" structures which map Pauli strings one-to-one with memory locations, like used in full-state simulation.

    \item \textit{Dynamically sized}: 
    It is typically hard to predict the total number of Pauli strings that will need to be propagated at the beginning of a simulation. As such, we cannot reliably pre-allocate our container, and must permit our ensemble to grow in tandem with our attempts to reduce costs using truncation. This requires that our container can be freely dynamically sized (like a linked list with $\OC(1)$ insertion), else \textit{infrequently} and \textit{efficiently} re-allocated to a larger, static size when necessary (like an array with $\OC(N)$ reallocation).
    It discourages use of non-linked structures where re-allocation requires re-insertion of every element in non-constant time, such as trees implemented with static arrays.

    \item \textit{Efficient search}:
        A core subroutine of Pauli propagation via merging-BFS is \textit{searching} the container for a particular Pauli string, in order to merge with it. Since this is performed almost every time an operator is applied to a Pauli string, it is essential that this search is fast.
        Our use of unsigned integer primitives to represent Pauli strings enables us to hash or compare Paulis in $\OC(1)$ time, enabling amortised $\OC(1)$ and $\OC(\log(N))$ searches inside hash tables and non-linked sorted lists respectively.
    
    \item \textit{Efficient insertion, modification and deletion}: 
        It is clear a rapidly growing ensemble should be stored in a container which supports efficient insertion of new elements. 
        Yet our needs are greater. Pauli propagation sees the corpus of contained Pauli strings \textit{change} frequently.
        As such, runtime is dominated by the costs of repeatedly enumerating and modifying the billions of contained strings. This might incentivise a data structure supporting in-place mutation, though this tends to be at odds with our other requirements. Thus,  we might instead demand it is efficient to \textit{remove} and \textit{insert} Pauli strings, discouraging the use of non-linked sorted lists with $\OC(N)$ penalties, and encourages use of hash tables or trees with amortised $\OC(1)$ and $\OC(\log(N))$ costs. Efficient removal is also crucially important for the truncation of Pauli strings.

    \item \textit{Efficient iteration}:
        Pauli propagation requires repeatedly iterating over or enumerating the Pauli sum to process every Pauli string it contains under the action of an operation. It is straightforward to enumerate in $\mathcal{O}(N)$ time using an array, linked-list, hash table or tree, though their runtimes can differ greatly. This is because only arrays store their items in contiguous memory (see UTH~Box~\ref{box:paulisum_datastrucs}), supporting fast caching, parallel enumeration, and enabling vectorisation.
        Hash tables, for example, are multiplicatively slower to enumerate than arrays, as informed by the ratio of filled slots called the \textit{load factor}. Trees and linked structures can be slower still due to storing items at unpredictable memory locations.

    \item \textit{Efficient union}: 
        It is not straightforward to parallelise the processing of some of the mentioned data structures. We can sometimes work around this by partitioning our structure into multiple, smaller instances, process each independently and concurrently, before re-merging them. This is worthwhile when the processing time outweighs the partitioning and merging costs, and encourages use of a data structure which is efficiently \textit{unioned}.

\end{enumerate}

Combining all considerations, we have so far established the hash table and arrays to be very appropriate structures. They have different strength and weakness profiles, and one may be superior over the other depending on the quantum circuit, the task, or the computational hardware.

\subsubsection*{Hash Tables}

Hash tables are how native dictionaries in the \code{Julia} language are implemented. Our keys are Pauli strings which, since instantiated with unsigned integers, are hashable in $\OC(1)$ time\footnote{
The diligent reader may wonder how an unsigned integer with at least $2n$ bits can be hashed in $\OC(1)$ time. 
Indeed the ultimate number of CPU clock cycles to perform a single bitwise instruction technically grows \textit{linearly} with the bits in the operands, once the CPU register size has been exceeded.
In practice the linear scaling is so small as to be insignificant, especially when considering the SIMD nature of CPU registers, hash-table implementations which cache hashes of keys, and sub-linear hash functions. In such cases, we can say the hash-table lookup time is effectively $\OC(1)$ when amortised over many lookups.
}.
%
%
%
%
The stored values are the associated coefficients of the strings in the represented Pauli sum. For example, sum 
$
S = c_{P_1} P_1 + c_{P_2} P_2 + c_{P_3} P_3
$
may be stored as:

\begin{center}
\begin{tikzpicture}
\HashTableDiagramSingle
\end{tikzpicture}
\end{center}
where $h(P_i)$ denotes the hash function called upon the integer representation of Pauli string $P_i$. Beware: this sketch is a simplification. Each value is actually stored alongside its corresponding key (the associated Pauli string) as is necessary since the hash function is generally non-invertible and key-equality checks will be common.

Inserting, removing and searching for Pauli strings are all amortised $\OC(1)$, permitting rapid modification of strings under the action of an operator. For example, $S$ above may be modified to $S' = c_{P_1'} P_1 + c_{P_4} P_4 + c_{P_5} P_5 + c_{P_6} P_6$, resembling:

\begin{center}
\begin{tikzpicture}
\HashTableDiagramOverviewGate
\end{tikzpicture}
\end{center}

The next section elaborates on this process, which in fact makes use of \textit{two} hash tables.
Meanwhile, parallel processing is possible by using a \textit{concurrent} hash table (making contentious actions ``atomic"), or by division into separate, concurrently modified hash tables before a final, serial union.

There are a few drawbacks to choosing hash tables.
Firstly, the aforementioned operations are efficient when the load factor is small, i.e., when the number of empty hash table slots is low. Typically, a load factor of  $\approx 0.5$ is said to be desireable, necessitating that the internal array capacity is at least \textit{twice} the number of Pauli strings contained within. Since a \texttt{Julia} hash table quadruples in capacity when the load factor exceeds $0.75$, the internal array can be up to \textit{five times} larger than necessary to store the strings contiguously.

Secondly, enumerating the hash table to process Pauli strings is suboptimal. This is because either the redundantly large internal array with empty entries is iterated over, or a separate list of keys is kept, requiring invoking superfluous re-hashing and control flow branching. These issues may be allayed by caching key hashes (sometimes called ``tokens") or by hybridisation with arrays; facilities offered by \texttt{Julia}'s \texttt{Dictionaries.jl} library.

And thirdly, many implementations of hash maps are not \textit{thread safe}, i.e., inserting elements is often sequential and not easily parallelizable. This can be solved via \textit{locking} of hash map segments, or specialized lock-free data structures. Most general implementations in \code{Python} or \code{Julia}, however, remain serial. As such, there remain alternative choices of data structure to represent the Pauli sum which are optimal for different situations.

\ppjl{} presently balances these considerations by instantiating its \texttt{PauliSum} type as a hash table in the form of \code{Julia}'s \code{Base.Dict} dictionary.  
This offers asymptotically favorable support for all circuit families, with simple and fast single-threaded defaults even for custom gates.

\subsubsection*{Arrays}

Pauli sums can also be represented via two arrays, one array containing all Pauli strings (e.g., \code{Array\{UInt32\}} for $\leq 16$ qubits) and one array containing their coefficients (often  \code{Array\{Float64\}}).
When using a single thread, some elementary operations may be slower and others faster than with hash tables/dictionaries, but our design choices for arrays aim to maximize parallelization on CPU or GPU. This is particularly great news for running on HPC clusters (which we would like to use for their larger memory), because computations will be significantly sped up by their many CPU threads.

Arrays are highly advantageous for circuits consisting predominantly of non-branching gates such as Clifford operations or Pauli noise channels. This is typical, e.g., in quantum error correction circuits~\cite{knill2005quantum}. Because Cliffords map Pauli strings one-to-one, the number of Pauli strings is conserved, and no merging is necessary, thus relaxing our requirements for fast insertion, deletion and lookup.
The contiguous data structure enables rapid enumeration and, more importantly, \textit{in-place mutation}. 
A Pauli string can be modified by merely overwriting its array element, which is considerably faster than removal and re-insertion into a hash table despite the asymptotic equivalence.
Modifying all Pauli strings also becomes \textit{embarrassingly parallel}, and is straightforwardly accelerated with several CPU threads or GPUs.

One may wonder how branching gates and merging strings are efficient in this data structure. The key insight is that the arrays can be \textit{sorted} by the numeric value of the Pauli string integer. Comparison-based sorting will take time $\OC(N \log N)$, but there exist algorithms such as \textit{Radix sort} that sort integers in time $\OC(N)$ with non-negligible prefactors. Searching the sorted Pauli array to determine where to merge or insert new elements then takes time $\OC(\log N)$. This is the approach taken in SPD~\cite{beguvsic2023fast,beguvsic2024real}. The naive $\OC(N)$ cost of insertion can be avoided by collecting the $k$ new Pauli strings in a smaller, temporary array which is merged in-batch to the full ensemble array in $\OC(N+k)$. 

Another approach (indeed the one we take) is to \textit{pre}-allocate and \textit{over}-allocate all arrays. By sizing all arrays to an explicit capacity $M \geq N$, we amortise the $\OC(N)$ resizing events and associated garbage collection over several gates, only kicking into action when a branching operation would result in over $M$ (unmerged) Pauli strings. 

To coordinate threads and perform massively parallel work, we leverage paradigms from the HPC and GPU communities, i.e., we write all subroutines as \textit{kernels}. A kernel is a function that is defined to act on an array at index \code{i} without interfering with how it acts on index \code{i+1}. This is necessary to allow all CPU and GPU threads to work independently of each other.

Our most frequent subroutine is commonly called \textit{stream compaction} or simply a \textit{filter} operation. Let us exemplify two important variants of this subroutine for truncating and merging a Pauli sum. We define \code{P} and \code{C} as the arrays carrying the Pauli strings and coefficients, and \code{P2} and \code{C2} their pre-allocated, auxiliary counterparts. We further allocate a boolean flag array \code{F} and an index array \code{I} of the same length. All loops and parallel operations on these arrays are written with the \code{AcceleratedKernels.jl} library that compiles all kernels to CPU or GPU, depending on which device the incoming arrays are allocated.

\medskip

\emph{Truncate.} Presume the function \code{iskept()} computes whether a term in the Pauli sum is kept, i.e., evaluates to 1 if kept and is 0 otherwise. We then perform the following three steps:
\begin{enumerate}
    \item Flag the elements that are not truncated.\\ 
    \code{for i = 1...N:}\\
    \hspace*{4mm}\code{F[i] = iskept(P[i], C[i])}
    \item Perform a parallel cumulative sum over \code{F} and write the results into \code{I}, \code{cumsum(F) $\rightarrow$ I}.
    \item Write the untruncated elements.\\
    \code{for i = 1...N:}\\
    \hspace*{4mm}\code{if F[i] == true}: \\
    \hspace*{8mm}\code{P2[I[i]] = P[i]}\\
    \hspace*{8mm}\code{C2[I[i]] = C[i]}
\end{enumerate}
These steps will place the untruncated terms in \code{P2} and \code{C2}, where conveniently \code{I[N]} denotes the new number of valid terms, i.e., the new $N$. Any residual or ``trash'' terms in the valid region of the arrays from previous iterations are completely overwritten, as will be the remaining terms in \code{P} and \code{C} in the next iteration. It suffices to relabel \code{P2} and \code{C2} as the main arrays and proceed with the propagation. 

Crucially, the pre-allocated flag array allows threads to independently check the truncation condition, and the index array carrying the cumulative sum allows threads to independently transfer un-truncated Pauli strings and coefficients into the correct slot of $\code{P2}$ and $\code{C2}$, respectively. With a suitable implementation of a parallel cumulative sum, the entire procedure is \emph{thread safe}.  

\medskip

\emph{Merge.} We start by performing a sort permutation of \code{P} and \code{C} by the integer-valued entries of \code{P}. Importantly, this causes duplicate strings to be next to each other. Then perform the following three steps:
\begin{enumerate}
    \item Flag the beginning of a new domain.\\
    \code{for i = 1...N:}\\
    \hspace*{4mm}\code{if i == 1:}\\
    \hspace*{8mm}\code{F[i] = true}\\
    \hspace*{4mm}\code{else:}\\
    \hspace*{8mm}\code{F[i] = (P[i] != P[i-1])}
    \item Perform a parallel cumulative sum over \code{F} and write the results into \code{I}, \code{cumsum(F) $\rightarrow$ I}.
    \item Move the unique strings and accumulate their coefficients.\\
    \code{for i = 1...N:}\\
    \hspace*{4mm}\code{if F[i] == true}: \\
    \hspace*{8mm}\code{P2[I[i]] = P[i]}\\
    \hspace*{8mm}\code{C2[I[i]] = C[i]}\\
    \hspace*{8mm}\code{j = i+1}\\
    \hspace*{8mm}\code{while (j $\leq$ N) and (P[j] == P[i]):}\\
    \hspace*{12mm}\code{C2[I[i]] += C[j]}\\
    \hspace*{12mm}\code{j += 1}
\end{enumerate}
Now \code{P2} and \code{C2} represent the merged Pauli sum. On GPUs, the inner \code{while} loop in step 3 is likely suboptimal if terms can be duplicated dozens or hundreds of times, but this can be solved by several sweeps with finite-sized loops over the inner index \code{j}.

\begin{figure}
    \centering
    \includegraphics[width=1\linewidth]{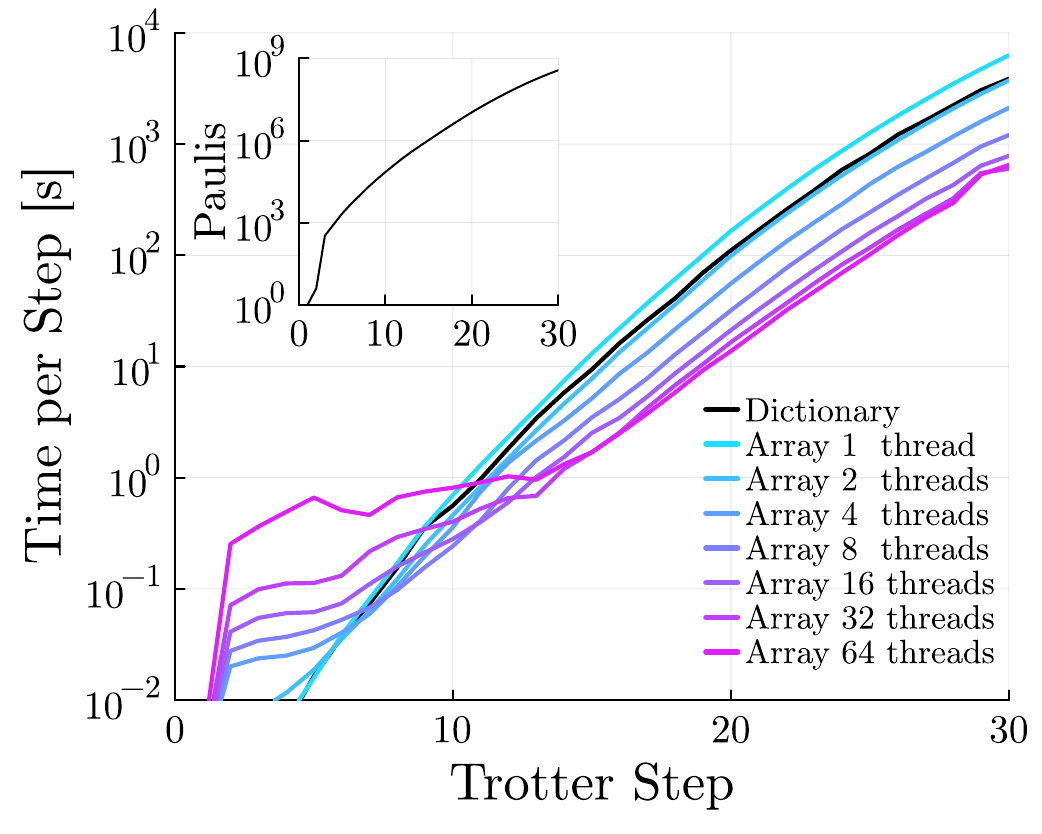}
    \caption{\textbf{Benchmarking multithreading capabilities in \ppjl{}}. We again consider the Trotterized dynamics of a $6\times6$ tilted-field Ising model with $dt=0.05$, 132 Pauli rotations per step, the $Z_{21}Z_{22}$ observable, a coefficient cutoff of $2^{-20}$, and no Pauli weight truncation. Multi-threaded array-based propagation here achieves up to an order of magnitude speedup over single-threaded dictionary-based propagation. We refer to Fig.~\ref{fig:convergence} for the expectation values.}
    \label{fig:threadtiming}
\end{figure}

Our library \ppjl{} provides array-based capabilities starting version \code{0.7} using the \texttt{VectorPauliSum} type (given that \code{Vector} in \code{Julia} is an alias for a 1D \code{Array}). Fig.~\ref{fig:threadtiming} demonstrates the performance improvement we can attain for quantum circuits consisting exclusively of Pauli rotations on an Intel Xeon Platinum 8360Y CPU node. Circuits including Clifford gates would be sped up significantly more, yet we observe up to an order of magnitude improvement using $\sim32$ threads even for Pauli rotations. This corresponds to applying a gate to a physical observable (not a random Pauli sum) with $10^8$ Pauli strings and pre-allocated memory in approximately 1 second, including merging and truncations. As is common in parallel computing, we observe strongly diminishing returns or even performance degradation when splitting the work over too many threads. Note also that local machines may have fewer and faster threads, in turn reducing the gap between single-threaded and multi-threaded propagation.

A downside of this array-based approach is that it is more challenging to write correct and efficient code. This is particularly true when targeting GPUs because their performance comes at the cost of additional constraints. For example, Pauli strings on many qubits (in our case $>32$) can no longer be represented by a single long integer due to GPU hardware architecture. Instead, they need to be split into chunks of \code{UInt32} or \code{UInt64}. Our GPU acceleration is still in an early stage, and future versions of \ppjl{} will improve on it. However, when adhering to all architectural constraints, even custom extensions to our library compile to multi-threaded CPU or GPU code via \code{Julia}'s multiple dispatch and the \code{AcceleratedKernels.jl} library. Future engineering efforts by the community will also inform us whether more powerful parallelization approaches on CPU exist (consider, e.g., Ref.~\cite{broers2025scalable}), and how GPU acceleration can be fully leveraged.

\subsubsection*{Trees}
Another interesting choice of data structure is a \textit{tree} where each node stores an individual Pauli and a scalar, and the path from root to a leaf node encodes a Pauli string. A string's coefficient is the product of the scalars among its constitutent nodes. For example, the string $(ab) XX + (ac) XY + d IZ$ might be stored as:
\begin{center}
    \begin{tikzpicture}

      \tikzset{
        circled node/.style={circle,draw, inner sep=0, minimum size=1.5em,text height=.8em,text depth=.25em},
      }

      \tikzset{
          shortened arr/.style={->,shorten >=2pt} 
      }
    
        \node[circled node] (A) at (0,0) {$*$};
        \node[circled node] (B) at (-1,-1) {X};
            \node at (-1.5,-1) {$a$};
        
        \node[circled node] (C) at (1,-1) {I};
            \node at (0.5,-1) {$1$};

        \node[circled node] (D) at (-1.5, -2) {X};
            \node at (-2,-2) {$b$};
        
        \node[circled node] (E) at (-0.5, -2) {Y};
            \node at (0,-2) {$c$};

        \node[circled node] (F) at ( 1.5, -2) {Z};
            \node at (1,-2) {$d$};

        \draw[shortened arr] (A) -- (B);
        \draw[shortened arr] (A) -- (C);
        \draw[shortened arr] (B) -- (D);
        \draw[shortened arr] (B) -- (E);
        \draw[shortened arr] (C) -- (F);

    \end{tikzpicture}
\end{center}
The tree has depth $n$ (the number of qubits), and the $i$-th layer contains all Paulis of the $i$-th site. Strings are searched, inserted or removed in $\mathcal{O}(n)$ regardless of the number stored, and enumeration of all $N$ leaf nodes remains $\mathcal{O}(N)$ though will be practically slower than enumerating an array.
Modifying a node's Pauli or scalar simultaneously updates all weighted strings encoded by all descendant nodes. 
Such a structure (without scalars) has been used in algorithms to decompose dense matrices into Pauli strings~\cite{koska2024tree}, 
but could also be used in Pauli propagation to update many Pauli strings \textit{in batch} through tree \textit{transformations}.
For example, a single-qubit Clifford gate upon the leftmost Pauli requires modifying only the first layer of nodes, avoiding processing all contained strings. Many-qubit and non-Clifford gates however require more complicated traversals and transformations, with performance dependent upon the structure of the tree and the distance of the affected site indices to its root. The use of trees was reported in Ref.~\cite{shao2024simulating} although only as a means of enumerating the strings, and was not used for accelerating the simulation of gates.

\subsection{Implementing a gate}\label{sec:implementing_gates}

Recall from Theory Box~\ref{box:PTM} that the action of an $n$-qubit channel $\mathcal{E}$ upon a Pauli string can be captured by the operator's Pauli Transfer Matrix.
For a generic $k$ qubit channel, computing this $4^k\times 4^k$ matrix takes $\OC(k 16^k)$ time~\cite{hantzko2024pauli} and its subsequent operation upon a Pauli sum in any representation takes at least $\OC(16^k)$. This is monstrously slow and memory inefficient.
Such a method fails to leverage that for most canonical channels, the PTM is sparse, few-branching (i.e. contains few non-zero identities in any given column) and known at compile-time. 
Here we will show how many gates and channels of interest to the quantum computing community offer optimised, specialised simulation.

An essential motivation for the representations of \code{PauliString} and \code{PauliSum} described above was the speed to modify a Pauli string under the action of a circuit operator.
We here describe precisely \textit{how} a \code{PauliSum} is efficiently modified by the family of operators presented at a higher-level in Sec.~\ref{sec:gates&ops}. Throughout, we take \code{P} to be the unsigned integer instantiating the Pauli string $P$ which is being operated upon, with corresponding coefficient $\code{c}_\code{P} \in \mathbb{R}$. 
We note that to modify the stored Pauli strings we find it convenient to maintain \textit{two} persistent Pauli sums. The specific usage of the auxiliary Pauli sum on which data structure was chosen to represent it, as well as on the circuit operator applied.

\medskip

\emph{Pauli noise} is implemented by a diagonal PTM irrespective of the number of decohering qubits. 
For example, single-qubit depolarising noise on qubit $i$ with error probability $p$ scales Pauli strings with any non-identity Pauli on site $i$ by a factor of $\lambda=4p/3$, leaving the Pauli string itself unchanged. We can efficiently modify the coefficient of $P$ without control flow branching via an expression like
\begin{center}
\code{
    $
        \code{c}_\code{P} \rightarrow 
        \code{c}_\code{P} \times
        (1 - \lambda \times (\code{P}_i \ne 0)).
    $
}
\end{center}
Similarly, dephasing noise updates the coefficient if \code{$(\code{P}_i \neq 0) \code{ \& } (\code{P}_i \neq 3)$}, and many-qubit Pauli noise channels merely replace the conditional with slightly more complicated (but potentially $\OC(1)$ bitwise) logic.
The coefficients of the Pauli strings can be updated by directly mutating the enumerated Pauli sum, avoiding use of the auxiliary Pauli sum and associated memory-movement costs.
All coefficients are independently modified in this manner, presenting an ideal scenario for parallelisation. This can, for example, be sketched as follows (the auxiliary table is not used in this case of Pauli noise but is included in the sketch because it will be used for other gates).

\begin{center}
\begin{tikzpicture}
\HashTableDiagramPauliNoise
\end{tikzpicture}
\end{center}

\medskip

\emph{Clifford gates} are also non-branching but may modify $P$ to a distinct Pauli string $Q$ while updating only the \textit{sign} of the coefficient. Namely,
\begin{center}
\code{ $
    \code{P} \rightarrow \code{Q},
        \;\;\;\;
    \cP \rightarrow \cQ = \pm \cP.
$ }
\end{center}
The precise update only depends on the Paulis on the sites the Clifford gate acts on and is determined via lookup tables. Since a Clifford PTM is invertible, a set of unique Pauli strings are mapped to a set of unique output strings; no merging of strings occurs. This permits independent, embarrassingly parallel processing of the strings (if performed in-place, e.g., with arrays). For hash tables, we insert all new strings into the (initially empty) auxiliary hash table which is thereafter taken as the primary hash table.

\begin{center}
\begin{tikzpicture}
\HashTableDiagramClifford{}  
\end{tikzpicture}
\end{center}

\medskip

\emph{Pauli rotations} of angle $\theta$ around the Pauli string $G$, $R_G(\theta) = e^{-i\theta G/2}$, are either $1$ or $2$-branching depending on whether $P$ and $G$ commute. Recall Eq.~\eqref{eq:PauliRotationPtmAsCommutator}:
\begin{equation*}
    R_G(\theta)
    [P] = \begin{cases}

    P & \text{if } [P,G] = 0\\
    \cos(\theta)P + \sin(\theta)Q & \text{else}
    \end{cases}
\end{equation*}
where $Q = P\cdot G$, modulo the coefficient. 
Both $Q$ and whether $[P,G] = 0$ can be found in $\OC(1)$-many bitwise operations as described in Sec.~\ref{sec:bit-operations}.

Since $R_G(\theta)[P]$ always includes $P$, we can modify its coefficient in place and decide where to place $Q$. For hash tables, we insert $Q$ into the (initially empty) auxiliary hash table. Once enumeration is complete, we union the hash tables.
This ensures the strings inserted into the auxiliary hash table are unique, relaxing the need to check for collisions and merging which otherwise introduces control flow branching.
Our implementation employs further tricks to reduce memory movement and superfluous re-allocating the hash tables.

\begin{center}
\begin{tikzpicture}
\HashTableDiagramPauliRotation{}  
\end{tikzpicture}
\end{center}

For array data structures and our \code{VectorPauliSum} type, we adapt the stream compaction subroutine outlined in Sec.~\ref{sec:representing_pauli_sum}. Inside the following pseudo-code, we again denote \code{P} and \code{C} as the main arrays carrying the Pauli strings and their coefficients, \code{P2} and \code{C2} their respective auxiliary arrays, \code{F} the boolean flag array, and \code{I} the index array. Define \code{paulirotationprod()} to compute the new Pauli string and its real-valued coefficient (see Eq.~\eqref{eq:PauliRotationPtmAsCommutator}). We then apply a Pauli rotation generated by Pauli string \code{G} with parameter \code{theta} as follows:
\begin{enumerate}
    \item Flag the elements that anticommute.\\
    \code{for i = 1...N:}\\
    \hspace*{4mm}\code{F[i] = !commutes(P[i], G)}.
    \item Perform a parallel cumulative sum over \code{F} and write the results into \code{I}, \code{cumsum(F) $\rightarrow$ I}.
    \item Ensure sufficient capacity.\\ 
    \code{if N+I[N] > length(P):}\\
    \hspace*{4mm}determine new capacity, e.g., \code{M = 2*(N+I[N])}\\
    \hspace*{4mm}resize \code{P, P2, C, C2, F, I} to length \code{M}
    \item Write new string and adapt coefficients.\\
    \code{for i = 1...N:}\\
    \hspace*{4mm}\code{if F[i] == true}:\\
    \hspace*{8mm}\code{Q, s = paulirotationprod(P[i], G)}\\
    \hspace*{8mm}\code{P[N+I[i]] = Q}\\
    \hspace*{8mm}\code{C[N+I[i]] = C[i] * s * sin(theta)}\\
    \hspace*{8mm}\code{C[i] = C[i] * cos(theta)}
\end{enumerate}
Step 3 ensures that all new terms can be written into the main arrays without additional memory movement involving the auxiliary arrays. Furthermore, the joint resize guarantees that all arrays are ready for subsequent processing, like merging or truncating.

\medskip

\emph{Custom operations} beyond those natively supported by \ppjl{} can be efficiently implemented, provided their (potentially parametrized) action is known at compile-time. The package defines user-overloadable functions (like \code{apply()}) to specify the action of a custom operation upon a Pauli string. Such functions are automatically invoked during simulation, leveraging \texttt{Julia}'s multiple dispatch, yet abstract away the complications of merging Pauli strings and truncation.
\ppjl{} also permits overloading increasingly sophisticated functions (like \code{applytoall!()}) in an ``opt-in" fashion, to incorporate performance considerations like memory movement and type-stability. 
This offers a trade-off between performance and code complexity.
Elegantly, these very functions are leveraged by the natively supported gates.

Arbitrary operations known only at runtime as few-qubit computational-basis matrices or Kraus operators can also, in principle, be efficiently implemented. 
If the gate is fixed, it can directly be turned into a \code{TransferMapGate}, which computes their PTMs via Eq.~\eqref{eq:definition_of_ptm}, converts it into a look-up map, and will then be sequentially applied to all Pauli strings. If instead the gate is parametrized, the unitary and its PT map can be computed once per gate (not once per Pauli string).
Alternatively, one could precompute their PTMs by use of a symbolic algebra library. In that case, it is essential to symbolically simplify the PTM 
using prior knowledge, e.g., the realness of the parameters, since the density of the PTM can be crucial for efficiency.
Pre-computing the Pauli transfer map in this way is expensive, but may drastically improve the runtime of the propagation. 
This technique is leveraged by \texttt{QuESTlink} functions such as \texttt{CalcPauliTransferMap}~\cite{jones2020questlink}, and is a planned future facility of \ppjl{}.

\medskip

\emph{Further optimisations} remain possible. For example, truncation of the propagated \code{PauliSum} via the strategies presented in Sec.~\ref{sec:truncations} presently involves periodically enumerating all contained Pauli strings and removing some from the hash table. This occurs after each gate, and enables simple, modular code with separated concerns.
Alas, it is suboptimal; it invokes expensive re-enumeration of the full \code{PauliSum}, doubling memory-movement costs. It also sees the temporary storage of new Pauli strings which are immediately removed by truncation, incurring superfluous hashing and potentially hash table re-allocation costs.
 We could instead truncate earlier, moving logic into the \texttt{apply()} functions to avoid ever storing truncated strings.
Such a scheme complicates the code and custom user truncations, and might only be worthwhile for hand-picked truncations and gate sets.

\subsection{Calculating the overlap} \label{sec:calculating_overlap}

The final step of Pauli propagation is to compute the \textit{overlap} of the propagated Pauli sum with another object - like a state or observable - in order to compute an expectation value.
Below, let $S$ be the final $n$-qubit Pauli sum evolved from an initial observable $P_\text{obs}$ under the circuit $\ECdag$ in the Heisenberg picture, i.e.
\begin{equation}\label{eq:EvolvedObs}
    S = \ECdag(P_\text{obs}) = \sum_P c_P P\,.
\end{equation} 
Note that since \textit{unnormalised} $n$-qubit Pauli strings do not form an ortho\textit{normal} basis (consider $\Tr[I^{\otimes n}]=2^n$), the norm-squared of a Pauli string is $2^n$ in lieu of unity:
\begin{align}
    {\|P\|_2}^2
    = \langle P, P \rangle
    = \Tr[P P] 
    = 2^n,
\end{align}
having adopted the Hilbert-Schmidt inner product.
As such, the coefficient of a string $P$ in the Pauli sum $S$ is obtained by scaling the inner-product of $P$ and $S$.
\begin{align}
    c_P = \frac{1}{2^n} \langle P, S \rangle = \frac{1}{2^n} \Tr[P S].
\end{align}
The form of the output overlap depends on the goal of the simulation.

\medskip

\emph{Correlation functions}. A commonly studied quantity in quantum physics is the correlation between a given observable $Q$ and its time evolved version $\mathcal{E}(Q)$. That is, the quantity 
\begin{equation}
    \frac{1}{2^n}\Tr[Q \mathcal{E}(Q)] = c_Q,
\end{equation} 
where $Q$ is a Pauli string. To compute such correlation functions it suffices to lookup the coefficient of $Q$, i.e., $c_Q$, in the data structure representing the Pauli sum $\mathcal{E}(Q)$. 
Under our use of a hash table, lookup is $\OC(1)$ time and no summation is necessary.
In \ppjl{} code, this is performed as
\begin{center}
\code{
    \cQ = getcoeff(S, Q)
}
\end{center}
where \code{S} is a \code{PauliSum} object and \code{Q} is either the integer representation of $Q$ or a \code{PauliString} object.
This is equivalent to the generic approach
\begin{center}
\code{
    val = scalarproduct(S, Q)
}
\end{center}

\medskip 

\emph{State expectation values}. 
Overlaps with quantum states $\rho$, 
\begin{equation}
\langle P_{\text{obs}} \rangle 
=
\Tr[ \mathcal{E}(\rho) P_{\text{obs}} ]
=
    \Tr[\rho S] = \sum_P c_P \Tr[\rho P] \, ,
\end{equation}
where $S$ is the propagated observable as defined in Eq.~\eqref{eq:EvolvedObs},
can be evaluated in a variety of ways depending on the form of $\rho$.

Among the simplest is when $\rho$ is a tensor product of single qubit \textit{stabilizer states}, i.e., when it can be written as $\rho = \bigotimes_{i=1}^n \left( \frac{I + s_i\sigma_i}{2}\right)$ where $\sigma_i\in(X, Y, Z)$ is a non-identity Pauli on site $i$ and $s_i=\pm1$ is a sign. For example, $\sigma_i=Z, s_i=+1, \forall i$ yields the all-zero state $|0\rangle\langle0|$. Then, only if the Pauli on site $i$ is $I$ or $\sigma_i$ does that term contribute, and $c_P$ can be added to the running sum. In pseudo-code this may read:\\
\smallskip

\noindent\text{\hspace{10mm}}\code{val = 0.0  }\\
\text{\hspace{10mm}}\code{for ($\cP$, P) in pairs(S):}\\
\text{\hspace{14mm}}\code{sign = 1}\\
\text{\hspace{14mm}}\code{for i in 1...n:}\\
\text{\hspace{18mm}}\code{if $\code{P}_i$ == $\sigma_i$: }\\
\text{\hspace{22mm}}\code{sign *= $s_i$}\\
\text{\hspace{18mm}}\code{elseif $\code{P}_i$ != I: }\\
\text{\hspace{22mm}}\code{sign = 0}\\
\text{\hspace{14mm}}\code{val += sign * $\cP$}
\smallskip

In the end, \code{val} is the computed expectation value. Notice that only Pauli strings contribute with their coefficients when they contain $I$ or $\sigma_i$ on all sites $i$. If any Pauli is different, the Pauli string is \textit{orthogonal} to the state $\rho$ and does not contribute.

In the common case of the all-zero state only Pauli strings consisting of $I$ and $Z$ Paulis contribute which can be checked with $\OC(1)$ bit operations. In \ppjl{} this is implemented via the inbuilt function
\begin{center}
\code{
    val = overlapwithzero(S).
}
\end{center}
In many cases, this will be the overlap function of choice when simulating quantum circuits. Similarly, we provide the \code{overlapwithplus()} for $\rho = |+\rangle\langle+|$ and \code{overlapwithcomputational()} for an arbitrary computational basis state $\rho = |x\rangle\langle x|$.

\medskip

If $\rho$ can be fully written down in the Pauli basis due to a low number of qubits or if it is a high-temperature state, then the overlap can be computed straightforwardly with one important caveat. Assume now that \code{rho} is a \code{PauliSum} object capturing the Pauli basis decomposition of $\rho$, then the function call
\begin{center}
\code{
    val = overlapwithpaulisum(rho, Q)
}
\end{center}
returns the correct expectation value. Note that this function calls the \code{scalarproduct()} function and multiplies the result by \textit{an otherwise missing factor of} $2^n$. This is because it is almost always more convenient for our code to function as if we worked with \textit{normalized} Pauli operators for which $\Tr[P^2]=1$ holds -- something that only comes into play when explicitly representing quantum states.

\medskip

More generally, for quantum states $\rho$ that are neither stabilizer states nor efficiently representable in Pauli basis, the overlap can still be computed if the expectation values $\Tr[\rho P] = \langle P \rangle_\rho$ can be evaluated. This may be possible if the state is represented in another classical form, for example, as a tensor network, or if the state lives on a \textit{quantum device}. In such cases, direct measurements or randomized protocols like classical shadows, can be used to estimate the Pauli expectation values $\langle P \rangle_\rho$, yielding 
\begin{equation}
     \Tr[\rho S] = \sum_P c_P \Tr[\rho P] =  \sum_P c_P \langle P \rangle_\rho \,.
\end{equation}
Then a \code{PauliSum} object \code{rhoP} storing the dictionary with pairs \code{ $P$ => $\langle P \rangle_\rho$} can handily be overlapped with the evolved Pauli sum via
\begin{center}
\code{
    val = scalarproduct(rhoP, Q) .
}
\end{center}
In this manner, Pauli propagation proves to be a framework with surprising flexibility in how it can be used and combined with other approaches.

\section{Outlook}

\subsection{What is the niche of Pauli propagation?}

\begin{theorybox}[label={box:magic}]{Magic and hardness of classical simulation in the Pauli basis}
Pauli propagation methods are efficient when the Pauli spectrum of $O$ is concentrated on few Pauli terms -- a condition that is sufficient but not always necessary.
A common way to formalize this notion is through the concept of \emph{magic}~\cite{dowling2024magic}. 
As a preliminary step, we observe that the squared coefficients of $O$ sum up to one, and therefore they can be interpreted as probability distribution, known also as \emph{characteristic distribution} 
\begin{align}
    \sum_{\alpha}c_\alpha^2 = 2^{-n}{\norm{O}_2^2} = 2^{-n}{\norm{P}_2^2} = 1.
\end{align}
A common metric of the ``spreadness'' of 
a distribution $p(x)$ is its \emph{collision probability} $\sum_x p^2(x)$.
In particular, the collision probability of the characteristic distribution of $O$ is known as the \emph{generalized Pauli purity}:
\begin{align}
    P^{(2)}(O) \coloneqq \sum_{\alpha} c^4_\alpha. 
\end{align}
A related quantity of interest is the \emph{operator stabilizer entropy} (OSE)~\cite{dowling2024magic}, defined as $\mathcal{M}^{(2)}(O) \coloneqq - \log\{P^{(2)}(O)\}$.
We say that $O$ exhibits ``high magic'' if $P^{(2)}(O)$ is exponentially small in system size.
In such regimes, approximating $O$ by truncating its small coefficients can be prohibitive, as the corresponding $\ell_1$-norm 
is exponentially large
\begin{align}
    \sum_{\alpha} \abs{c_\alpha} \geq \frac{\left({\sum_{\alpha} c^2_\alpha}\right)^\frac{3}{2}}{\left(\sum_{\alpha} c^4_\alpha \right)^\frac{1}{2} } 
    = \frac{1}{\sqrt{{P}^{(2)}(O)}} \geq \exp\left(n\right),
\end{align}
where the first step follows from 
Lemma\ 8 in Ref.\ \cite{nietner2023average}.
As most quantities to be estimated using Pauli propagation scale non-exponentially, we see that the $\ell_1$-norm cannot provide a non-trivial error estimates when the magic is high. 
 
\end{theorybox}

\paragraph*{Magic.}
\begin{figure}
    \centering
    \includegraphics[width=1\linewidth]{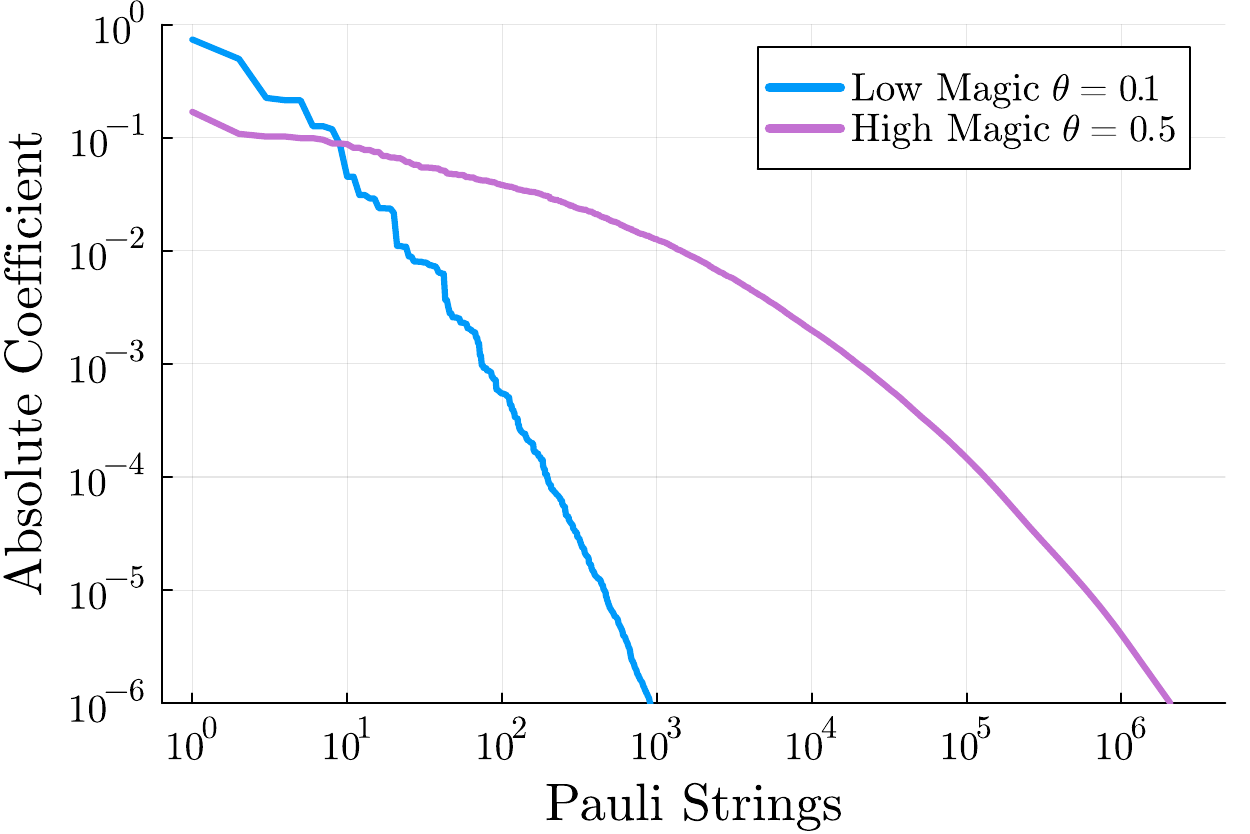}
    \caption{\textbf{Manifestation of magic in Pauli propagation simulation.} We propagate a single-site Pauli in the middle of a $6\times 6$ grid through a quantum circuit that corresponds to eight layers of the Trotterization of the tilted-field Ising Hamiltonian, i.e., eight repetitions of RX, RZ, and RZZ Pauli rotations. Importantly, all Pauli rotation angles are set to a single value $\theta$, with $\theta=0.1$ in one case and $\theta=0.5$ in another. The latter induces higher magic, or in other words, creates Pauli strings with coefficients whose magnitude is more spread out. This typically results in harder simulation, because more Pauli strings need to be kept for the same accuracy. We note that the total number of Pauli strings would be the same in both cases if none were truncated, but for practical purposes, we employed a coefficient truncation of $10^{-7}$.}
    \label{fig:magic}
\end{figure}

Pauli propagation is most directly limited by the growth of the number of Pauli terms in the propagation tree. This is quantified by \textit{magic}~\cite{chitambar2019quantum, dowling2024magic} as discussed in Theory Box~\ref{box:magic}.
If the magic generated is low, then the number of Pauli strings with non-negligible coefficients will be small, and Pauli propagation equipped with exclusively coefficient truncation will be extraordinarily fast and precise at simulating not only expectation values but the full propagated object. On the other hand, if magic is high, then coefficients are more spread over a large number of Pauli strings, and for in the worst case (i.e., some initial states), one would have to keep propagating a large fraction of all Pauli strings for precise estimates. See Fig.~\ref{fig:magic} for an example of how the value of Pauli rotation angles can drastically change the difficulty of a simulation as measured by the number of Pauli strings above certain absolute coefficient thresholds.
It remains an open question whether there exist physically interesting low magic but high entanglement circuits whereby Pauli propagation methods could prove very advantageous over popular tensor network methods, or to what extent Pauli propagation approaches can be successful in simulating high magic circuits.

\medskip

\paragraph*{Noise and open quantum systems.}
Several key factors contribute to non-unitary dynamics being potentially more promising applications for Pauli propagation. One quite naive argument is that, by working in the Pauli basis, we are already paying the cost of open quantum system simulation. This is exemplified by the significantly increased cost of exact density matrix simulation compared to pure state wavefunction simulation. More concretely, however, noise can \textit{reduce} magic. Pauli noise strictly reduces magic, and even non-unital noise like amplitude damping tends to effectively reduce magic in unstructured circuits. Working with open quantum systems may also enable working directly with quantum states as mixed states can be sparse in the Pauli basis. An extreme example would be the maximally mixed state, which can be represented with a single Pauli string (the identity), suggesting that high-temperature physics might be amenable to efficient Pauli propagation simulation.

\medskip

\paragraph*{Expectation values in typical random or scrambling circuits.}

While magic can be a useful quantity for bounding worst-case errors and runtime, the difficulty of many simulations does not strictly correlate with magic. Namely, for many initial states, their overlap with most Pauli strings can be negligible. If it can be deduced which Pauli strings are most likely to contribute to expectation values during propagation, expectation values can be significantly more efficient than magic alone would indicate. This intuitive claim is formalized in Ref.~\cite{angrisani2024classically} which shows that for typical quantum circuits, expectation values can be estimated to a reasonable approximation with polynomial resources. In other words, high magic implies worst-case hardness and is a necessary but not sufficient condition for average-case hardness. For one, this suggests that a wide range of parametrized quantum circuits used for variational algorithms can be simulated when randomly initialized~\cite{cerezo2023does,bermejo2024quantum}. It also implies that the scrambling behavior may lend certain chaotic quantum systems to be simulated using Pauli propagation methods~\cite{roberts2017chaos,belyansky2020minimal,geller2022quantum}. 

That being said, in our experience, simulating non-trivial high-magic circuits beyond a certain precision remains hard -- particularly in quantum dynamics. In current Pauli propagation approaches, there is no form of spacial factorization like in tensor networks that may enable working in exponential spaces with polynomial resources. Currently, the only way that exponential Pauli strings can contribute to a simulation is by explicitly propagating exponentially many Pauli strings. It thus remains to be seen to what extent the guarantees enjoyed by Pauli propagation for typical random circuits translate into practical benefits for physically motivated problems. 

\medskip

\paragraph*{Local observables and operator lightcones.}
\begin{figure}
    \centering
    \includegraphics[width=1\linewidth]{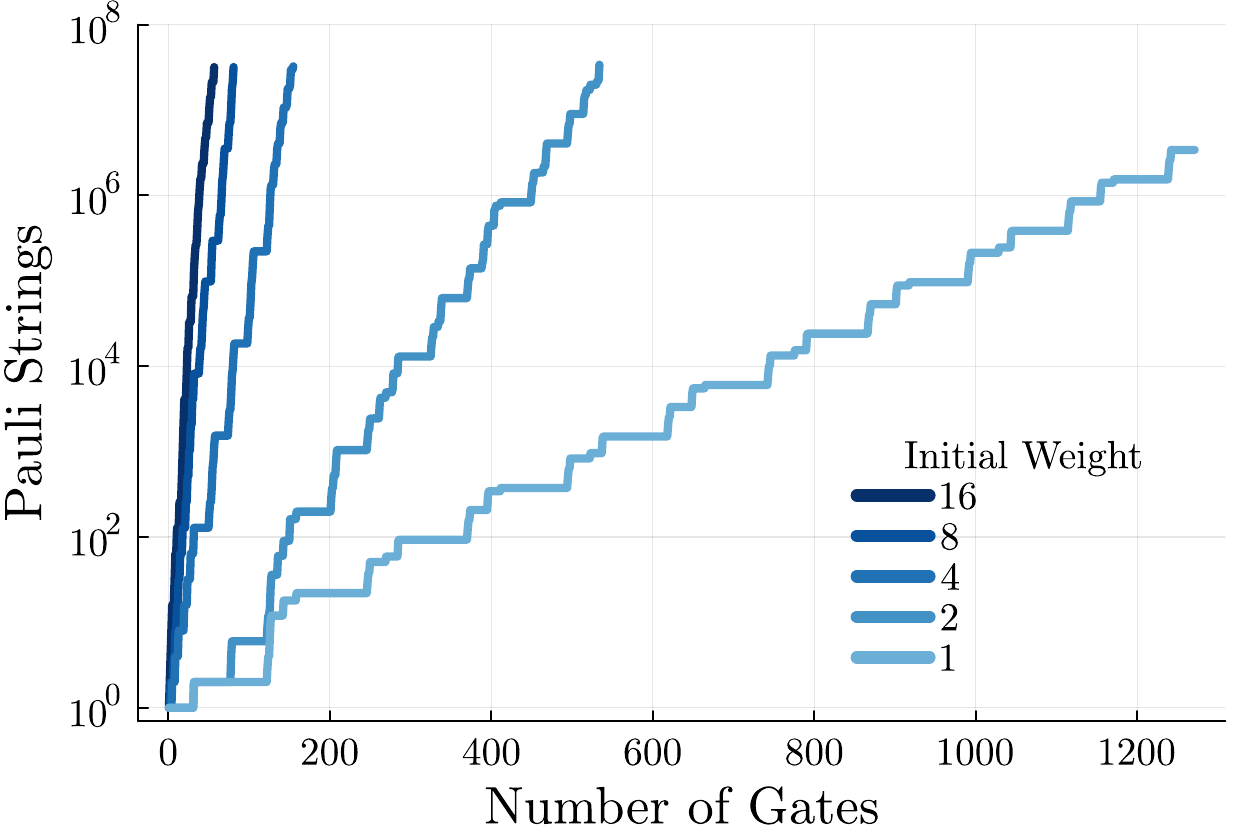}
    \caption{\textbf{Effect of operator lightcones.} We demonstrate how the Pauli weight of the observable (sometimes called \textit{bodyness}) affects the growth in number of Pauli strings in a simulation. The setup is a 32-qubit system with RX-RZ-RX rotations per qubit and RYY entanglers on a 1D periodic bricklayer ring. The non-identity Paulis are chosen at random and spaced out equally on the Pauli string, e.g., weight 16 is ZIYIYIZI\dots. If the non-identity Paulis were neighboring, the number of Pauli strings would still grow faster than with fewer Paulis, but likely not as fast as spaced-out Paulis.}
    \label{fig:lightcone}
\end{figure}
One feature that can make the Heisenberg simulation of quantum circuits substantially simpler than the Schrödinger evolution of states is the so-called \textit{operator lightcone} or \textit{entanglement lightcone}. It describes which operations in a circuit \textit{can} contribute to certain observable expectation values at all (called the \textit{hard} lightcone), or which operations \textit{predominantly} contribute (called the \textit{effective} lightcone). This effect is very visible for geometrically local observables and topologies, and low Pauli weight observables in general, where the number of active sites with non-identity components only grows slowly. Gates outside the hard lightcone could simply be removed from the circuit. This is often done in the Schrödinger picture but Pauli propagation natively skips over those gates when they commute with all current Pauli strings. However, some gates contribute only weakly based on their position in the circuit or because of their parameter. Suitable coefficient truncations will capture the most significant effect of such gates, which may thus not come at the full expected computational cost. An example demonstrating the drastic effects of Pauli weight of the initial observable can be viewed in Fig.~\ref{fig:lightcone}.

\medskip

\paragraph*{Circuit topology.}
Pauli propagation in its current form is flexibly adaptable to any circuit topology, i.e., to which qubits are connected via multi-qubit gates. It is in principle not strictly harder to simulate all-to-all gates, such as Pauli rotations with global generators. In practice, however, the topology strongly affects how quickly Pauli strings spread into the exponentially-sized space, with the operator lightcone, and consequently magic, increasing more quickly in highly connected systems. In contrast to tensor networks, current Pauli propagation approaches are not biased towards or against certain topologies, implying that in cases where tensor networks work well, they likely outperform Pauli propagation, but in cases where they do not, Pauli propagation may have a practical advantage.
Given the impressive performance of MPS for simulating a wide range of 1D and quasi-1D systems, we do not expect Pauli propagation to yield a better time-accuracy trade-off there. However, in 2D and 3D systems, for example, as explored in Ref.~\cite{beguvsic2024real}, or in more exotic long-range connected systems, this may be possible. For a more detailed theoretical discussion of the role of topology for Pauli propagation see Box~\ref{box:geom}.

\begin{theorybox}[label={box:geom}]{Entanglement lightcone and geometric dimension}
Consider a \textit{local} Pauli operator \( P \) and a quantum circuit \( U = U_L U_{L-1} \dots U_1 \) consisting of \( L \) layers. Each layer \( U_j \) consists of non-overlapping two-qubit gates arranged on a \( D \)-dimensional topology. A straightforward lightcone argument shows that the Heisenberg-evolved operator 
\[
U^\dagger P U
\]
has support only on those qubits within a lightcone whose size scales as \(\mathcal{O}(L^D) \). As a consequence, one could perform statevector simulation on these \( \mathcal{O}(L^D) \) qubits and compute $\Tr\!\left[P U \rho U^\dagger\right] $
exactly with computational resources scaling as 
\[
\exp\!\left(\mathcal{O}(L^D)\right).
\]
Pauli propagation and certain other methods operating in the Heisenberg picture will automatically adapt to such lightcones because 
gates outside the lightcone act trivially on identity Paulis. Truncations based on, for example, coefficient magnitude or Pauli weight additionally give rise to the concept of  \textit{effective} lightcones focusing on the most significant contributions to the expectation values. These dominant contributions often spread more slowly than the so-called \textit{hard} lightcone containing all qubits where any non-identity Pauli can exist. 
\end{theorybox}

\medskip
\paragraph*{Operator dynamics.}
One concrete potential application for Pauli propagation methods is for studying operator spreading and operator hydrodynamics\cite{nahum2017quantum, von2018operator}. To learn about how operators spread in quantum systems, tensor networks have been a popular choice of computational framework~\cite{rakovszky2022dissipation}. While MPS has been applied to analyze coarse-grained behaviors such as operator spreading in one-dimensional systems, Pauli propagation captures the fine-grained evolution of individual Pauli strings at a microscopic level. 
In this context, Pauli propagation offers a direct algorithm for computing the dynamics of quantum circuits, particularly in near-Clifford, shallow-depth, or noisy settings. Its precision and flexibility in simulating individual Pauli string evolution make it particularly valuable in verifying and benchmarking current quantum simulations and experimental measurements of operator dynamics across various platforms, such as superconducting qubits~\cite{mi2021information, kim2023evidence}, trapped ion systems~\cite{joshi2020quantum}, and Rydberg atom arrays~\cite{bluvstein2021controlling}.

Where tensor networks may, in some cases, allow to simulate for longer times or to higher precision, they need to be queried to retrieve information about individual Pauli strings. This is technically \textit{efficient}, but is practically slower than a single look-up in the Pauli sum. More importantly, however, propagated Pauli sums transparently display which Pauli strings are present and with which coefficients. This could be imitated by sampling Pauli strings from a tensor network, but may be less practical. 

\medskip
\paragraph*{Repeated evaluations.}  
As discussed in Sec.~\ref{sec:surrogate}, Pauli propagation surrogates aim to trade an initial computational overhead for a potentially drastic speed-up for new parameters. There are a lot of nuances that make such an algorithm challenging to realize in practice, but one could imagine practitioners compiling a surrogate once, for example as part of a particular experimental setup, then the extremely fast re-evaluation is an integral part of the data processing pipeline. At this current stage, the surrogate implementation in PauliPropagation.jl is far from ideal (though significantly better than demonstrated in Ref.~\cite{rudolph2023classical} and able to simulate quantum convolutional neural networks at scale until 1024 qubits~\cite{bermejo2024quantum}). But this is a possibility that appears truly unique for propagation methods. 

\medskip

\paragraph*{Hybridization with quantum devices.} 
Finally, and going hand-in-hand with the transparency of which Pauli strings are contained in an evolved observable, Pauli propagation can straightforwardly be combined with quantum devices within the simulation of any quantum circuit. Following the framework of classical-enhanced quantum computation from Ref.~\cite{cerezo2023does}, a quantum device merely needs to perform Pauli measurements of the initial state or a partially evolved state to estimate the expectation values in Eq.~\eqref{eq:expectation_value}. The hybrid framework is provably efficient in the same cases where Pauli propagation itself is provably efficient~\cite{lerch2024efficient}, but given the empirical success of Pauli propagation, it stands to reason that this framework can be pushed beyond classical and beyond current quantum limits with clever algorithm engineering. A proof-of-principle realization was recently presented in Ref.~\cite{fuller2025improved}. We believe such hybrid approaches are a promising avenue for eking advantage out of moderate-scale non-fault-tolerant quantum devices~\cite{preskill2018quantum, zimboras2025myths}.

\subsection{Avenues for further developments}
Pauli propagation is a new classical simulation tool for quantum physics and quantum computing practitioners. Despite only recently entering the stage, Pauli propagation shows promise for becoming the computational framework of choice for some tasks. A certain dose of optimism is crucial to make use of up-and-coming technologies, but this is not to downplay the many ways in which Pauli propagation can be improved. Mirroring the structure of this manuscript, these range from high to low level. 

Starting at the low level, the main challenges -- perhaps unsurprisingly -- are to increase the speed of computation and lift the absolute constraint imposed by finite memory. We already support multithreaded CPU simulation and limited GPU acceleration via the \code{VectorPauliSum} type. They heavily rely on sorting and writing back and forth between pre-allocated memory. While any performance increases are generally desirable, with our code available through \ppjl, one likely runs out of memory before they run out of time. This is exacerbated by GPUs commonly having less memory than HPC CPU nodes. In other words, with current propagation approaches GPUs will be faster at running out of memory, unless applied to tasks that are more time- than memory-limited (e.g., learning and optimization tasks). Larger-scale quantum simulations may require the use of propagation methods beyond pure merging-BFS (see Sec.~\ref{sec:tree-traversal}), for example, deterministic or randomized depth-first components. Both GPU-accelerated and non-memory-limited propagation strategies will in the future be supported via \ppjl.

On the middle level, there are many questions surrounding how to best utilize prior knowledge about the circuits to be simulated. Pauli weight truncation, for example, can be seen as a truncation tailored to random quantum circuits~\cite{angrisani2024classically} which is significantly less effective in structured circuits. A Majorana Propagation method has recently been developed for the simulation of Fermionic systems~\cite{miller2025simulation,alam2025fermionic,alam2025programmable,d2025majorana}. This method is substantially more suited to simulating the long Pauli strings that arise naturally in Fermionic simulations.
Symmetric merging strategies, whereby Pauli operators that behave identically due to symmetry properties of the circuit are merged to a single propagated string, can further reduce the memory cost of simulation~\cite{loizeau2025quantum,teng2025leveraging}.

Beyond manifesting physical understanding, there are many algorithmic and conceptual tricks to be played for performing Pauli propagation more efficiently. One such trick is what we call \textit{meet-in-the-middle} simulation of expectation values, where the initial and final objects are propagated in the Schrödinger and Heisenberg pictures, respectively. This replaces one depth $L$ simulation by two depth $L/2$ simulations, in turn halving the exponent of the computation, but with the additional overhead of calculating the inner product of the two half-evolved objects. In the case of spin correlation functions (see the enumeration in Sec.~\ref{sec:framework}), both parts can be simulated via Pauli propagation with very efficient inner product computation, but in principle, any two computational methods (classical or quantum) can be combined~\cite{beguvsic2023fast, lerch2024efficient,fuller2025improved}. Other conceptual innovations could, for example, exploit the unitarity of certain gates and iteratively compute or correct Pauli sum coefficients via backward and forward evolution, or find suitable extrapolations for expectation values via variable noise rates or truncation parameters. More exploratory would be to utilize Pauli propagation for sampling quantum states, which could be done via the sampling-to-expectation-value reduction described in Ref.~\cite{aharonov2022polynomial}, but it has, to our knowledge, never been realized at scale. Crucially, recent works appear to indicate that propagation methods can be very proficient at learning ground state preparation circuits or estimating the ground state energies of chemistry or condensed matter Hamiltonians~\cite{angrisani2024classically,miller2025simulation,li2025dual,lin2025utility,shrikhande2025rapid}. Many more exciting innovations are possible and merely require practitioners to try them out. 

Finally, on a high level, we need to better understand how Pauli propagation and similar methods can be useful to practitioners. What are the applications where such methods stand above the rest? Perhaps it appears dissatisfying that our work cannot provide a definitive answer to this question, but we hope to have pointed out our current understanding of strengths and weaknesses. One point we would like to stress here is that current Pauli propagation codes are implemented on a very low computational level, using basic bit operations and simple data structures. This means that in its current form, Pauli propagation is exceptionally efficient in cases that it is well-equipped to handle, while suffering in other cases where more sophisticated algorithms (e.g. tensor networks) likely have the upper hand. That being said, it is possible that future variations of propagation methods exceed what we perceive as their current limitations, finding even broader application, and we are hope that our package \ppjllink{} can play a role in these developments.

\addtocontents{toc}{\string\tocdepth@munge}

\begin{acknowledgments}
The authors would like to acknowledge Tomislav Begušić, Joseph Tindall, Matteo D'Anna, Julia Gerecke, Julia Guignon, and Su Yeon Chang for helpful comments on this manuscript.
The authors would like to acknowledge Patrick Rall, Nikita Astrakhantsev, Joey Tindall, Lukas Devos, Marc Drudis, and Aaron Miller for insightful discussions.
MSR acknowledges funding from the 2024 Google PhD Fellowship and the Swiss National Science Foundation [grant number 200021-219329].
TJ acknowledges support of the NCCR MARVEL, a National Centre of Competence in Research, funded by the Swiss National Science Foundation (grant number 205602). YT acknowledges support from NCCR spin, a National Centre of Competence in Research, funded by the Swiss National Science Foundation (grant number 565785). AA and ZH acknowledge support from the Sandoz Family Foundation-Monique de Meuron program for Academic Promotion. 
\end{acknowledgments}

\section*{Data Availability}
All simulations in this manuscript were performed using the open-source library PauliPropagation.jl~\cite{paulipropagation_jl}. All data can be replicated on version \texttt{v0.7.3} using the descriptions provided in the text, and is available upon request.

\bibliography{quantum,extra}

\section*{Controversies}

This work sits at the intersections of quantum information theory, quantum computing, high-performance classical computing, and software development.
In lieu of an appendix, this \textit{controversies} section documents the process behind many of the work's decisions which compromise the conventions of the various fields.

\appendix

\section{Pauli normalisation} \label{app:pauli_normalisation}

The \ppjl{} library effectively propagates observables in the \textit{unnormalised} Pauli basis. That is, the coefficients of the terms in a Pauli sum  $\EC[P] = \sum_{j} c_j P_j$
are stored simply as $c_j $. We could have instead explicitly worked in a normalized Pauli basis $\mathsf{P}_i = 2^{-\frac{n}{2}} P_i$, as appeared in the definition of the PTM (Theory~Box~\ref{box:PTM}) or numerous theoretical works using the Pauli path formalism~\cite{aharonov2022polynomial, angrisani2024classically}. Had we done so, we might then have worked with Pauli sums of the form
\begin{equation}
    \EC [P] = \sum_{j} 2^{\frac{n}{2}} c_j \mathsf{P}_i , 
\end{equation}
and so stored $2^{\frac{n}{2}} c_j$ instead of $c_j$.
Choosing to work without normalization is akin to non-dimensionalisation in continuous-space numerical modelling~\cite{roache1998verification}.

As a consequence, on viewing our Pauli strings as \textit{vectors}, our propagated basis is \textit{not} ortho\textit{normal}, as mentioned in Sec.~\ref{sec:calculating_overlap}. This sometimes necessitates diligence when evaluating quantities naturally expressed as Hilbert--Schmidt inner products. We saw, for example, that the \code{scalarproduct()} function computes the inner product between the \textit{coefficients} of two Pauli sums\footnote{Or, at the risk of further confusing the reader, it can equivalently be viewed as computing the Hilbert Schmidt inner product between two Pauli sums of normalized Pauli strings.}. However, to compute the Hilbert Schmidt inner product of a state and a Pauli sum, i.e., $\sum_{j} c_j \Tr[\rho P_j] $, we need to multiply \code{scalarproduct()} by $2^n$ to account for the fact that $\Tr[I] = 2^n$. This subtlety is hidden from a typical user who can work with the \code{overlapwithpaulisum(rho, Q)} function that already includes this $2^n$ factor.

\section{Indexing and endianness}

As noted in Convention~Box~\ref{box:BitEndianness}, this manuscript has juggled multiple endian conventions. We notated binary numbers with little-endian (least significant bit on the right, indexing from right to left), prevalent among programmers.
This matches the ubiquitous positional notation used to denote numerals in \textit{any} base, including decimal.
\begin{equation}\label{eq:bitendian}
\begin{aligned}
    {\color{blue}1}
    {\color{cyan}1}
    {\color{LimeGreen}0}
    {\color{Dandelion}1}_{
        \color{gray}2
    } 
    &=
    {\color{blue}1}\times 2^{3} + 
    {\color{cyan}1}\times 2^{2} + 
    {\color{LimeGreen}0}\times 2^{1} + 
    {\color{Dandelion}1}\times 2^0
    \\
    = \hphantom{11} 
    {\color{WildStrawberry}1}
    {\color{Lavender}3}_{
        \color{gray} 10
    } &=
    {\color{WildStrawberry}1} \times 10^1 +
    {\color{Lavender}3} \times 10^0
\end{aligned}
\end{equation}

We chose however to \textit{index} bits and digits from \textit{one} in lieu of the alternate, natural choice of \textit{zero}. As such, the bit at index $1$ of 
$
    {\color{blue}1}
    {\color{cyan}1}
    {\color{LimeGreen}0}
    {\color{Dandelion}1}_{
        \color{gray}2
    } 
$
is
$
{\color{Dandelion}1}
$.
This matches the indexing convention of the \texttt{Julia} language within which \ppjl{} is implemented, but clutters our bitwise logic (presented in Sec.~\ref{sec:bit-operations}) with otherwise superfluous $-1$ index offsets. 

In contrast, we chose to employ \textit{big}-endian (indexing from left to right) more familiar to the QIT community when notating Pauli strings, such that  
\begin{equation}\label{eq:qitendian}
    Z X \equiv Z \otimes X \equiv Z_1 X_2.
\end{equation}
But consider that a Pauli string is a base-four numeral with digit symbols $\{I,X,Y,Z\}=\{0,1,2,3\}$; an equivalence exploited by our instantiation of a string as an unsigned integer in Sec.~\ref{ssec:Paulis}. We have thus adopted a distinct endianness for base-four numerals versus base-two and base-ten. 

Further, consider that sites of a Pauli string correspond one-to-one with qubits in a circuit. This necessitates that our kets are also denoted in big-endian, i.e.
\begin{equation}
Z_1 X_2  \ket{a,b}
\equiv
(Z \otimes X) \ket{a,b} 
\equiv 
Z\ket{a} \otimes X \ket{b}
\end{equation}
where $\ket{a}$ has index $1$ and $\ket{b}$ has index $2$. \textit{Qubit} states are consequently also big-endian ($\ket{01} = \ket{0}_1 \ket{1}_2$), but this conflicts with the \textit{bit} endianness in Eq.~\eqref{eq:bitendian}. This tension is the unfortunate consequence of our attempts to use the most natural and common conventions of quantum information theorists and computer scientists, namely the conventions in Eq.~\eqref{eq:qitendian} and Eq.~\eqref{eq:bitendian}, respectively.

\section{Programming language}

Pauli propagation simulation is an interesting case study. It is logic-heavy and yet when running at scale, it becomes memory-bandwidth-bound. Much of its logic is irreducible to numerical routines which can be dispatched to existing accelerated libraries; instead, the hot loops perform custom, nuanced processing. Since its algebra is highly scientific, so too must be its reader; it must be maintainable by the scientific community, and so be written in a language which is high-level and easy to use. Most importantly, to compete with quantum hardware providers the code needs to run \emph{very fast}.

These attributes necessitate its implementation in a programming language which is both simple yet fast. Ideally, it should not delay the scientific programmer with long compilation times - yet it must make excellent use of compiler optimisations so that its high-level syntax does not compromise performance. It should further empower the use to extend the existing \ppjl{} code base with project-specifc adaptations that can be engineered to be equally performant. These considerations made us arrive at \textit{Julia}. That being said, we are aware that \code{Python} is the language of choice for a large fraction of our readers. In fact, we encourage \code{Python} wrappers for \ppjl{}, and invite the community to contribute!

\end{document}